\documentclass[aps,prd,twocolumn,superscriptaddress,showpacs,showkeys,nofootinbib]{revtex4-1}
\usepackage{graphicx}
\usepackage{dcolumn}
\usepackage{amssymb}
\usepackage{amsmath}
\usepackage{amsfonts} 
\usepackage{amsbsy}
\usepackage{color}
\usepackage{rotating}
\usepackage[english]{babel}
\usepackage{soul}
\usepackage{xcolor}
\usepackage[T1]{fontenc}
\usepackage[utf8]{inputenc}
\usepackage{hyperref}
\hypersetup{
colorlinks=true,
linkcolor=blue,
filecolor=magenta,
urlcolor=cyan,
citecolor=cyan
} 
\usepackage{amssymb,amsmath,epsfig}
\usepackage{eurosym}
\usepackage{amsfonts}
\usepackage{array}
\usepackage{changes}
\usepackage{amsthm}
\usepackage{bm}
\usepackage{palatino}

\newcommand{\be}{\begin{equation}}
\newcommand{\ee}{\end{equation}}
\newcommand{\bea}{\begin{eqnarray}}
\newcommand{\eea}{\end{eqnarray}}

\def\0{{\sst{(0)}}}
\def\1{{\sst{(1)}}}
\def\2{{\sst{(2)}}}
\def\3{{\sst{(3)}}}
\def\4{{\sst{(4)}}} 
\def\5{{\sst{(5)}}}
\def\6{{\sst{(6)}}}
\def\7{{\sst{(7)}}}
\def\8{{\sst{(8)}}}
\def\sst#1{{\scriptscriptstyle #1}}

\DeclareMathOperator{\sech}{sech}

\usepackage{mathpazo}
\usepackage{supertabular}
\usepackage{graphics}
\usepackage{color}
\usepackage{graphicx}

\begin{document}

\title{Generalized Uncertainty Principle Impact on Nonextensive Black Hole Thermodynamics}

\date{\today}

\author{Ilim \c{C}imdiker}
\email{ilim.cimdiker@phd.usz.edu.pl}
\affiliation{Institute of Physics, University of Szczecin, Wielkopolska 15, 70-451 Szczecin, Poland}
\author{Mariusz P. D\c{a}browski}
\email{mariusz.dabrowski@usz.edu.pl}
\affiliation{Institute of Physics, University of Szczecin, Wielkopolska 15, 70-451 Szczecin, Poland}
\affiliation{National Centre for Nuclear Research, Andrzeja So{\l}tana 7, 05-400 Otwock, Poland}
\affiliation{Copernicus Center for Interdisciplinary Studies, Szczepa\'nska 1/5, 31-011 Krak\'ow, Poland}
\author{Hussain Gohar}
\email{hussain.gohar@usz.edu.pl}
\affiliation{Institute of Physics, University of Szczecin, Wielkopolska 15, 70-451 Szczecin, Poland}


\begin{abstract} 
The effect of the generalized uncertainty principle (GUP) on nonextensive thermodynamics applied to black holes, as well as the sparsity of the radiation at different temperatures associated with each nonextensive entropy, is investigated. We examine the R\'enyi, Tsallis-Cirto, Kaniadakis, Sharma Mittal, and Barrow entropies, temperatures, and heat capacities and show that, in each case, due to GUP corrections, the temperature and entropy have finite values, implying that the final state of the black hole is a remnant at the end of the evaporation process and that the sparsity of the radiation for massless bosons at each temperature depends on the mass of the black hole. We also find that GUP reduces the value of the sparsity profile for each case as compared to the sparsity parameter at Hawking temperature, which is always constant throughout the evaporation. 
\end{abstract}
\maketitle
\section{Introduction}
\label{intro}
Black holes emit radiation due to the Hawking evaporation process, and therefore, there is an established concept of Hawking temperature \cite{Hawking:1974rv} and Bekenstein entropy \cite{Bekenstein:1973ur} connected with the black hole horizon. The black hole evaporation process operates within the purview of quantum field theory, and one of its more intriguing aspects may be that it appears to indicate a non-unitary evolution, which gives rise to the well-known issue of the information loss paradox \cite{Hawking:1976ra,Chen:2014jwq,Unruh:2017uaw}. Black holes behave like thermodynamic objects, and the laws of black hole thermodynamics \cite{Bardeen:1973gs,Gibbons:1976pt,Hawking:1982dh,Hawkingnew1,Hawking:1975vcx} are analogous to the conventional thermodynamic laws. The thermodynamics of black holes have been extensively studied and used in a variety of cosmological and gravitational applications \cite{Jacobson:1995ab,Verlinde:2010hp,Padmanabhan:2009kr,Kubiznak:2014zwa,Cvetic:2010jb,Caldarelli:1999xj,Cai:2005ra,Davies:1977bgr,Dolan:2012jh,Easson:2010av}.

Entropy measures how difficult it is for an outside observer to get information about the underlying structure of the system. This is a clear reflection of the macroscopic features that result from the quantum statistical mechanics that govern the behavior of quantum microstates. For the case of black holes, there is no definition of Bekenstein entropy in quantum statistical mechanics and it only relies on Hawking's area theorem \cite{PhysRevLett.26.1344}, therefore, it would be required to have a complete theory of quantum gravity in order to fully comprehend the origin of this entropy and the nature of microstates in the case of black holes. In its absence, we rely on the definition of Bekenstein entropy for black holes. For the case of a Schwarzschild black hole with mass $M$, the Hawking temperature $T_H$ and Bekenstein entropy $S_B$ are given by \cite{Hawking:1974rv,Bekenstein:1973ur}
\begin{equation}
T_H= \frac{\hbar \kappa}{2\pi k_B c}~~,  ~~S_B=\frac{k_Bc^3A}{4G\hbar}~~, 
 \label{TH}
\end{equation}
where $\hbar$, $G$, $k_B$, and $c$ are the reduced Planck constant, the Newton gravitational constant, the Boltzmann constant, and the speed of light, respectively. The area $A$ of the event horizon is defined as $A=4\pi r_h^2$ in the above equation (\ref{TH}), where $r_h=2GM/c^2$ is the Schwarzschild radius and $\kappa=c^4/4\pi G M$ is the surface gravity defined on the event horizon of the Schwarzschild black hole.

Gibbs statistical mechanics is based on two key hypotheses: that entropy is extensive and that internal energy and entropy follow the additive composition rule. All thermodynamic relations in Gibbs statistical mechanics are defined in light of these presumptions.
It is very important to differentiate extensivity and additivity of a thermodynamic quantity in general (for more comprehensive discussion, see Refs. \cite{soton29487,Swendsen2011,Mannaerts_2014}. Assume two independent systems $A$ and $B$ with an ensemble of configurational possibilities $\Omega_A$ and $\Omega_B$ and corresponding probabilities $P_A$ and $P_B$. Consider $AUB$ now, with $P_{AUB}$ being the probability and $\Omega_{AUB}$ being the set of possibilities. Because of the systems' independence, $P_{AUB}=P_AP_B$. Therefore, if $S(A+B)\equiv S(P_AP_B)=S(P_A)+S(P_B)\equiv S(A)+S(B)$, then an entropy functional $S(P)$ is said to be additive. In order to define extensivity, we will use Tsallis' definition of extensive entropy, which states that if a system's total number of microstates, $\Omega$, is proportional to its number of particles or degrees of freedom, the entropy is extensive.
For instance, the Gibbs entropy is defined as $S_G(N)=k_B \ln \Omega (N)\propto N$, where $N$ is the total number of particles or degrees of freedom in the system. Keep in mind that extensive entropy can be nonadditive. In Gibbs thermodynamics, entropy is defined as extensive because it scales with the size of the system. This definition does not capture its full significance, and is not stated with full mathematical rigour: what does it mean to ‘scale’? What is meant by ‘size’ ? Is it the volume? Mole number? Both? In order to understand the definition of extensive variables more clearly, we define a function $f$, the fundamental relation of thermodynamic variables ($X_0$, $X_1$,$X_2$,...,$X_k$) such that $X_0=f(X_1,X_2,...,X_k)$. Here, $f$ is homogeneous first order function of $X_1$,$X_2$,...,$X_k$ when $f(aX_1,aX_2,...,aX_k)=af(X_1,X_2,...,X_k)$ for every positive real numbers $a$ for all $X_1$, $X_2$, ...$X_k$. The thermodynamic variables $X_i$ can be the energy $U$, entropy $S$ and mole number $N$ and expressing $f$ in differential form will give the first law of thermodynamics. For example, in Gibbs thermodynamics, the fundamental relation $f$ for the entropy $S$ can be written as $S=f(U,V,N)$ for an ideal case and and $f(aU,aV,aN)=af(U,V,N)$, hence $S$ is extensive. In terms of the scaling symmetry of the fundamental relation, the geometric framework provides a precise way of defining what extensive variables are. In a nutshell, we will say that a set of thermodynamic variables is "extensive" when the first-order homogeneous property is imposed on the fundamental relation.  This way, we avoid ambiguity in the word 'size,' as well as claims that volume and mole number are 'obviously' extensive, as seen frequently in discussions of extensivity.

 Nonextensive statistical mechanics, such as Tsallis nonextensive statistical mechanics \cite{Tsallis:1987eu,TSALLIS1998534,tsallisbook,Abe_2001,Abe_2001a,Abe_2001c,Bir__2011,PhysRevE.67.036114,PhysRevE.83.061147,Parvan:2004vn}, is the outcome of removing the assumption of extensivity. The assumption of the extensive nature of entropy is connected to ignoring the long-range forces between thermodynamic sub-systems. Since the size of the system exceeds the range of the interaction between the system's components, Gibbs thermodynamics ignores these forces. Because of this, the total entropy of a composite system equals the sum of the entropies of the individual subsystems and entropy grows with the size of the system. However, long-range forces are important in various unique thermodynamic systems. For instance, if we think of a black hole as a $(3+1)$ dimensional object, it is vital to note that Bekenstein entropy scales with the area and is thus regarded as a nonextensive quantity \cite{Tsallis:2012js,Biro:2013cra,Czinner:2015ena,Czinner:2017bwc,Czinner:2015eyk,Czinner:2017tjq,Tsallis:2019giw}. Furthermore, because of the area scaling, Bekenstein entropy is nonadditive and follows a nonadditive composition rule $S_{12} = S_1 + S_2 + 2 \sqrt{S_1} \sqrt{S_2}$ (see e.g. \cite{Alonso-Serrano:2020hpb}), whereas Gibbs statistical mechanics or thermodynamics is based on the extensive and additive properties of the entropy.
  Therefore, Gibbs thermodynamics or statistical mechanics may not be the appropriate choice for studying the thermodynamics of black holes. In order to understand the nonextensive and nonadditive nature of Bekenstein entropy, several extensions \cite{Tsallis:1987eu,Renyi1,SM,sharma1975new,Kaniadakis:2002zz,Kaniadakis:2005zk,Barrow:2020tzx} of standard Gibbs thermodynamics have been applied to black holes and cosmological horizons \cite{Nojiri:2021czz,Nojiri:2022aof,Nojiri:2022sfd,Nojiri:2021jxf,Promsiri:2020jga,Promsiri:2021hhv,Tannukij:2020njz,Nakarachinda:2021jxd,Cimdiker:2022ics,Promsiri:2022qin,Nakarachinda:2022gsb,Saridakis:2020zol,Dabrowski:2020atl,Nojiri:2022aof,Nojiri:2022ljp,Komatsu:2016vof,Komatsu:2015nkb,Nunes:2015xsa,Liu:2022snq,Majhi:2017zao,Luciano:2021mto,DiGennaro:2022ykp,DiGennaro:2022grw,Asghari:2021bqa,Abreu:2022pil,SayahianJahromi:2018irq,Drepanou:2021jiv}. One of the main proposals is the Tsallis-Cirto's black hole entropy definition \cite{Tsallis:2012js}, which makes the black entropy extensive and compatible with the Legendre structure.  R\'enyi entropy \cite{Renyi1}, being a measure of entanglement, is another definition of entropy applied to black holes and cosmological horizons which is nonextensive, but additive (by assumption). There have been some other nonextensive forms of entropy suggested such as  the Sharma-Mittal entropy \cite{SM,sharma1975new} as a generalization of R\'enyi entropy, the Kaniadakis entropy \cite{Kaniadakis:2002zz} which takes inspiration from Lorentz group transformations and the Barrow entropy \cite{Barrow:2020tzx} which is based on a hypothetical fractal structure of black hole horizon as a result of quantum fluctuations.

Due to the prevalence of quantum gravity effects, it is anticipated that the semiclassical technique would fail during the last phases of Hawking evaporation. There is currently no satisfactory theory of quantum gravity that enables us to completely explain that regime, despite the development of several quite diverse proposals \cite{Carlip:2001wq,Konishi:1989wk,Adler:1999bu,Rovelli:1996dv,Meissner:2004ju,Scardigli:1999jh,Hossenfelder:2012jw}. Investigating the phenomenological consequences of an underlying theory of quantum gravity is one technique to explore the quantum gravity effects at those scales. The generalized uncertainty principle (GUP) \cite{Maggiore:1993kv,Scardigli:1999jh,Giddings:1992hh,Hossenfelder:2012jw} is one approach that has the benefit of being sufficiently generic to be compatible with several quantum gravity theories. The Bekenstein entropy and Hawking temperature of a black hole in its last phases of evaporation are modified within this framework \cite{Adler:1999bu}. Because of these modifications, black holes do not entirely evaporate during the evaporation process, and the final state of the black hole is a remnant of the order of Planck mass 

Sparsity \cite{Page:1976df,Page:1976ki,Page:1977um,Schuster:2018lmz,Gray:2015pma,Schuster:2019xvp,Paul:2016xvb,PhysRevD.97.044029,Alonso-Serrano:2018mfo,Ong:2018syk,Alonso-Serrano:2020hpb,Feng:2018jqf} is an important feature of Hawking radiation. It is defined as the average time between the emission of successive quanta over the timescales set by the energies of the emitted quanta. It was shown that Hawking radiation is very sparse during the black hole evaporation process \cite{Gray:2015pma}, which is one of the key characteristics that distinguish it from black-body radiation. However, it has been found that when GUP corrections are incorporated \cite{PhysRevD.97.044029,Alonso-Serrano:2018mfo,Ong:2018syk}, the sparsity decreases toward the late stages of evaporation. When nonextensivity is considered in the context of R\'enyi temperature \cite{Alonso-Serrano:2020hpb}, the R\'enyi radiation is initially not sparse, but as evaporation progresses, it begins to become sparse and eventually approaches the case of Hawking radiation. 

In this paper, we are interested in exploring the GUP modifications to the nonextensive entropies and corresponding thermodynamic quantities in R\'enyi, Tsallis-Cirto, Sharma-Mittal, Kaniadakis, and Barrow nonextensive statistics. Furthermore, the sparsity of the radiation is analyzed at different temperatures corresponding to different nonextensive entropies. 

The following is the outline of the paper.
In Sec. \ref{GUP}, we introduce the notion of GUP and apply it to the case of standard thermodynamic black hole quantities. In Sec. \ref{GUPvsNEE}, we introduce nonextensive entropies and accompanying nonextensive thermodynamic quantities, as well as GUP modifications to nonextensive black hole thermodynamics. Finally, in Sec. \ref{summary}, we summarize and discuss our findings. 

\section{GUP and Black Hole Thermodynamics}
\label{GUP}
\subsection{Generalized Uncertainty Principle}

One common aspect of several quantum gravity theories is that they all predict a minimum measurable length \cite{Hossenfelder:2012jw,Amati:1988tn}. For example, the notion of minimal length is defined in string theory as the string length \cite{Kempf:1994su,Konishi:1989wk}, in loop quantum gravity \cite{Rovelli:1996dv} it is the expectation value of the length operator, and this notion can also be developed by the phenomenological aspects coming from black hole physics \cite{Hossenfelder:2012jw}.
Because of the appearance of another minimum length at the Planck scale in various quantum gravity approaches, it has been proposed that the Heisenberg Uncertainty Principle (HUP) 
\be
\Delta x_0 \Delta p \geq \hbar, \hspace{0.2cm} {\rm or} \hspace{0.2cm} \Delta x_0 \sim \frac{\hbar}{\Delta p} 
\label{HUP}
\ee
where $\Delta x_0 $ and $\Delta p$ are position and momentum uncertainties can be modified when gravitational interaction is introduced. The simplest argument for the modification of HUP within the framework of Newtonian theory is that there is a gravitational acceleration $\vec{a}$ of an electron due to a photon of mass $E/c^2$ \cite{Adler:1999bu}, where $E$ is the photon energy and  $r$ is the photon-electron distance, which reads
\be
\vec{a} = \ddot{\vec{r}} = - \frac{G(E/c^2)}{r^2} \frac{\vec{r}}{r} ,
\label{accel}
\ee
and the interaction takes place in a characteristic region of length $L \sim r$ and in characteristic time $t \sim L/c$. Then, the velocity acquired by an electron $\Delta v$ is
\be
\Delta v \sim \frac{GE}{c^2 r^2} \frac{L}{c} , 
\label{velocity}
\ee
and the (extra due to gravity) distance $\Delta x_1$ it is shifted reads
\be
\hspace{0.5cm} \Delta x_1 \sim  \frac{GE}{c^2 r^2} \frac{L^2}{c^2} \sim \frac{G \Delta p}{c^3} = \frac{c \Delta p}{4F_{pl}} = l_p^2 \frac{ \Delta p}{\hbar},
\label{shift_x1}
\ee
where $l_p = \sqrt{G\hbar/c^3}$ is the Planck length, and $F_{pl} = c^4/4G$ is the Planck force (often called the maximum force in the context of general relativity) \cite{schiller2005general,BG2014,dabrowski2015abolishing,Ong:2018xna}. Extra uncertainty (\ref{shift_x1}) adds to the standard HUP uncertainty of position $\Delta x_0$ as in (\ref{HUP}) giving 
\be
\Delta x = \Delta x_0 + \Delta x_1 \sim \frac{\hbar}{\Delta p}+ l_p^2 \frac{ \Delta p}{\hbar},
\ee
leading to the generalized uncertainty principle (GUP)
\begin{equation}
    \Delta x \Delta p \geq \hbar \left(1 +\frac{l^{2}_p}{\hbar^2}(\Delta p)^2\right) \label{gupxx} .
\end{equation}

Taking an algebraic point of view, GUP can be derived from the deformed commutation relation between the position operator $\hat x$ and the momentum operator $\hat p$ such that
 \begin{equation}
\left[\hat{x},\hat{p}\right]=i\hbar f(\hat p),
\end{equation}
where $f(\hat p)$ is a general function of momentum operator $\hat p$ and there exist different proposed functions for $f(\hat p)$. In order to make the function $f(\hat p)$ compatible with (\ref{gupxx}), following the literature, we choose 
\begin{equation}
f(\hat p)=1+\alpha \frac{l_p^2}{\hbar^2} \hat p^2,
 \end{equation}
where we the introduce GUP parameter $\alpha$ -- a dimensionless parameter predicted to be of order of unity, but there are different (mostly upper) bounds on it from different experiments and observations \cite{Gao:2016fmk,Feng:2016tyt,Bosso:2018ckz,Gao:2017zch,Giardino:2020myz}. By introducing $\alpha$, the equation (\ref{gupx}), now, reads as
\begin{equation}
    \Delta x \Delta p \geq \hbar \left[1 +\alpha\frac{l^{2}_p}{\hbar^2}(\Delta p)^2\right]
    \label{gupx} .
\end{equation}

 \subsubsection{GUP Modified Hawking Temperature and Bekenstein Entropy}

 An interesting application of (\ref{gupx}) to black hole physics is the modification to the Hawking temperature, which can be derived by 
solving it for $\Delta p$, which gives
\begin{eqnarray}
 \Delta p = \Delta x  \frac{\hbar}{\alpha l^2_p }\left[1 \pm \sqrt{1- \frac{\alpha l^2_p }{(\Delta x)^2} }\right] .
 \label{Delta_p}
\end{eqnarray}
We consider the $''+''$ sign in (\ref{Delta_p}) as $\alpha \rightarrow 0$ limit yields the standard Heisenberg uncertainty principle, whereas the negative sign does not. Considering the minimum position uncertainty near the event horizon of the Schwarzschild black hole as $\Delta x = 2l_p=4GM/c^2$, where $l_p$ is taken as the Schwarzschild radius $r_h$, the GUP modified Hawking temperature $T_{GUP}$ reads 
\begin{eqnarray}
 T_{GUP}= \frac{m_p^2 c^2}{8 \pi k_B M} \left[\frac{4}{2+\sqrt{4-\alpha \frac{m_p^2}{M^2}}}\right].
 \label{Tgup}
\end{eqnarray}
By introducing a correction term due to GUP, $\mathcal{K}(\alpha, M)$, $T_{GUP}$ can be written in terms of $T_H$ and $\mathcal{K}$, such that
\be
T_{GUP} =T_H(M) \mathcal{K}(\alpha,M) ,
\label{Tgup_comp}
\ee
where the GUP correction term is defined as
\be
\mathcal{K}(\alpha,M) = \frac{4}{2+\sqrt{4- \alpha\frac{m_p^2}{M^2}}} .
\label{Kfactor}
\ee
This provides us with a more compact form of $T_{GUP}$, which will be used in the next sections for GUP modifications to the thermodynamic quantities. Note that we consider the case where $M^2\ge \alpha m_p^2/4$ to make the parameter $\mathcal{K}$ real valued function.

\begin{figure}[hbt!]
    \centering
    \includegraphics[scale=0.5]{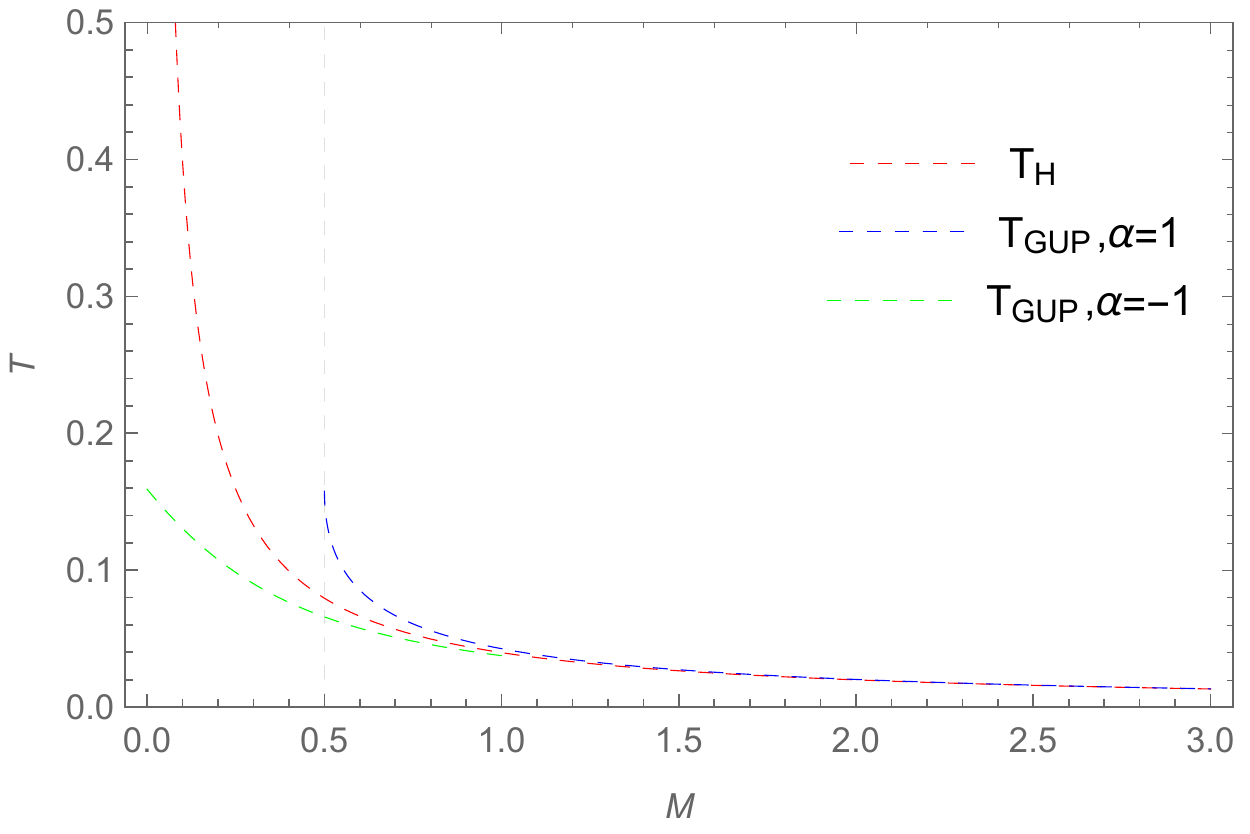}
    \caption{Temperature vs mass for the Hawking temperature $T_H$ and the GUP corrected temperature with positive and negative values of $\alpha$. Threshold with positive $\alpha$ for mass lies at the remnant mass $M_r^2=(\alpha/4) m_p^2$ (cf. formula (\ref{M_remnant})).}
    \label{fig:thvstgup}
\end{figure}

\begin{figure}[hbt!]
    \centering
    \includegraphics[scale=0.5]{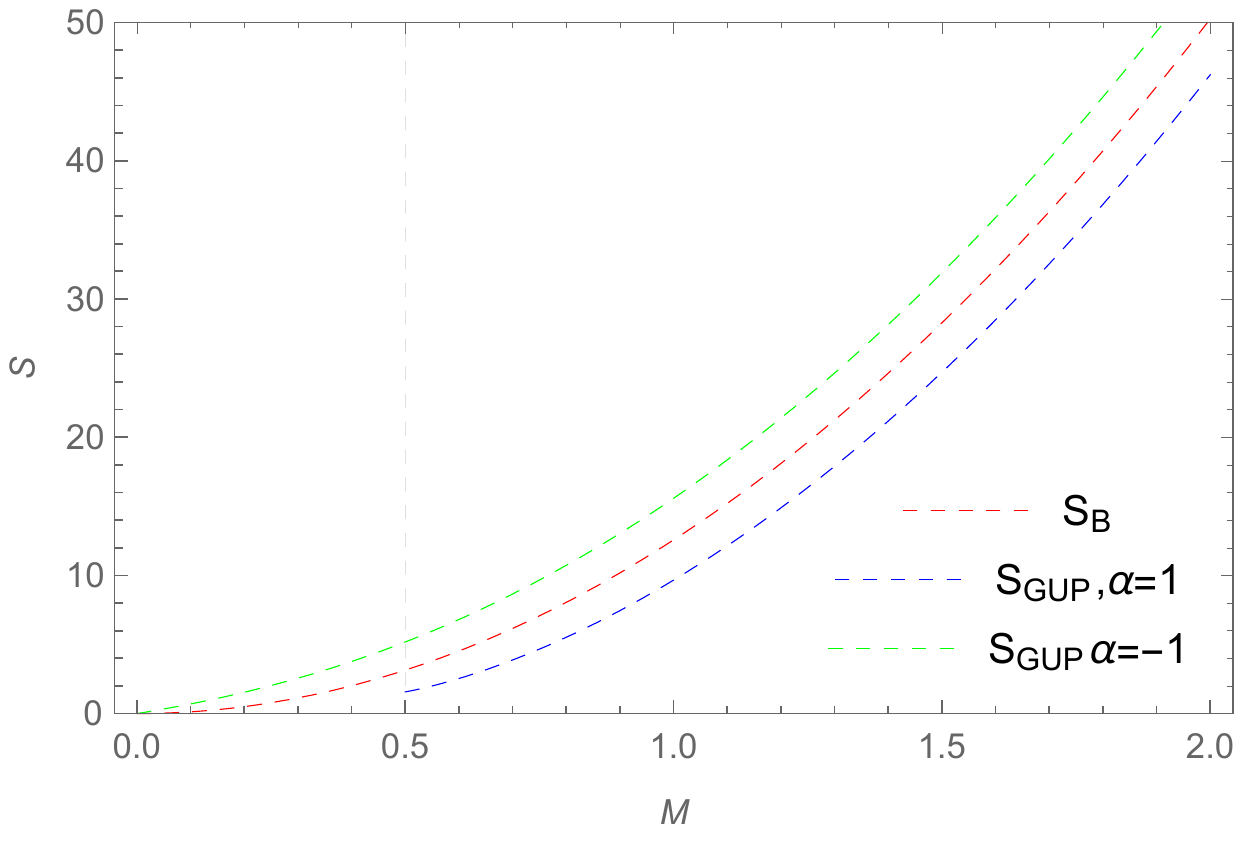}
    \caption{Entropy vs mass for the Hawking temperature and GUP corrected temperatures with positive and negative values of $\alpha$. The threshold for mass lies at the remnant mass given by $M_r^2=(\alpha/4) m_p^2$.}
    \label{fig:sbvssgup}
\end{figure}

Using the Clausius relation, the GUP modified Bekenstein entropy $S_{GUP}$ in terms of $S_B$ and the correction term $\mathcal{K}(\alpha,M)$ can be written as 
\begin{eqnarray}
S_{GUP}=\frac{S_{B}}{\mathcal{K}}-\frac{\alpha \pi k_B}{2} \ln{\left[\frac{4 M}{m_0 \mathcal{K}}\right]} , 
\label{Sgup}
\end{eqnarray}
where $m_0$ is a dimensionful constant of unit mass, which is introduced in order to make the logarithm dimensionless. In the limit $\alpha\rightarrow0$, the correction term $\mathcal{K}$ goes to one, and hence $T_{GUP}$ and $S_{GUP}$ reduce to $T_H$ and $S_B$. The plots of (\ref{Tgup}) and (\ref{Sgup}) are given in Figs. \ref{fig:thvstgup} and \ref{fig:sbvssgup}. Note that all the plots in the paper, unless explicitly stated, are given in natural units $\hbar = c = G =1$ and also with the GUP parameter $\alpha =1$.
For positive values of $\alpha$, the black hole evaporation stops when the mass of the black hole reaches some critical value of mass 
\be\label{M_remnant}
M_r=\frac{\sqrt{\alpha} m_p}{2}= \frac{l_p\sqrt{\alpha}}{2c^2}F_{pl},
\ee
which is called the black hole {\it remnant mass} and we introduce the Planck force $F_{pl}=c^4/G$ in above equation. Therefore, we can say that the final state of the black hole evaporation is a remnant having the mass $M_r$. In fact, without a well-defined quantum gravity theory, we cannot predict what happens if the mass of a black hole is smaller than this critical value. For the critical mass value $M_r$, the formulas (\ref{Tgup}) and (\ref{Sgup}) for $T_{GUP}$ and $S_{GUP}$, give the temperature $T_r$ and the entropy $S_r$ for the remnant as \cite{Alonso-Serrano:2020hpb} 
\be 
T_r = \frac{m_p c^2}{2\pi k_B \sqrt{\alpha}}, ~ S_r=\frac{\pi \alpha k_B}{2}\left[1- \ln{\left(\frac{ \sqrt{\alpha}m_p}{m_0}\right)}\right], 
\label{T_gup_remnant}
\ee
provided that  $\alpha >0$. For $\alpha< 0$ in (\ref{Kfactor}), we have a smooth correction function defined for all black hole mass values. In this case, the black hole continues to radiate slowly and yields an infinite lifetime \cite{Ong:2018syk}. When $M$ approaches zero, interestingly, the temperature is still finite, and for this case, in \cite{Jizba:2009qf}, it is referred to as a remnant with zero rest mass.

\subsubsection{GUP Modified Heat Capacity}

\begin{figure}[hbt!]
    \includegraphics[scale=0.5]{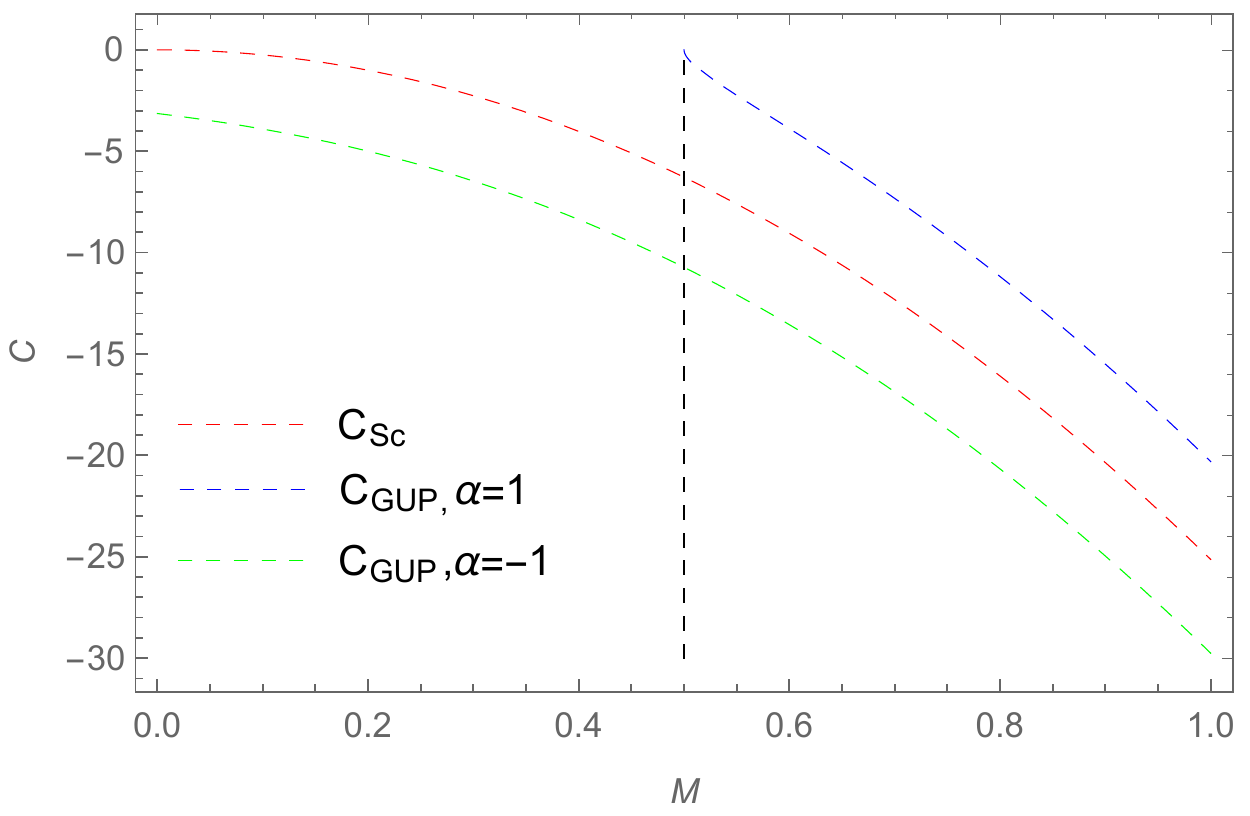}
    \caption{Specific heat capacity of the Hawking radiation for GUP corrected black holes. For positive $\alpha$, there is a remnant with zero heat capacity.}
    \label{fig:cgup}
\end{figure}

In order to investigate the GUP modifications to the heat capacity of a black hole with mass $M$, we use the definition of heat capacity $C$, which reads
\begin{eqnarray}
 C=-\frac{S'^2(M)}{S''(M)} ,
\label{SHeat}
\end{eqnarray}
where $S$ is the black hole entropy and prime and double prime denote the first and second derivative with respect to the mass $M$. For the case of Schwarzschild black hole, we have (denoting $C$ as $C_{Sc}$)
\begin{eqnarray}
 C_{Sc}=- 8 \pi k_B \frac{M^2}{m_p^2}, 
 \label{csc}
\end{eqnarray}
and we can see that it is negative for all mass values. This means that the Schwarzschild black hole is thermodynamically unstable. In order to introduce GUP corrections, we introduce the quantity 
\be 
\beta_{GUP}=\frac{1}{k_B T_{GUP}} ,
\ee 
which after using (\ref{Tgup}) gives 
\begin{eqnarray}
 \frac{ S_{GUP}'(M)}{k_B c^2}=\beta_{GUP}=\frac{\beta}{\mathcal{K}}\; ,
 \label{betaGUP}
\end{eqnarray}
where $\beta = 1/k_BT_H$ is the inverse Hawking temperature. Differentiating $\beta_{GUP}$ once more, and using equations (\ref{SHeat}) and (\ref{betaGUP}), we obtain the GUP modified heat capacity $C_{GUP}$, which can be written as (cf. Fig. \ref{fig:cgup})
\begin{eqnarray} \label{Cgup}
 C_{GUP}=C_{Sc}\left[\frac{2-\mathcal{K}}{\mathcal{K}^2}\right] .
\end{eqnarray}
This means that the GUP corrections still yield a negative heat capacity for $M>M_r$, and when the black hole mass approaches the critical mass $M_r$, we have $\mathcal{K}=2$ and interestingly, we get the zero heat capacity for the remnant. In such a case, from the thermodynamic point of view, a small amount of heat would then increase the temperature of the remnant by an infinite amount.

\subsubsection{GUP Modified Sparsity of Hawking Radiation}

One of the most important aspects of Hawking radiation is that it is extremely sparse as compared to black-body radiation. The sparsity can be defined by using the parameter $\eta$ \cite{Gray:2015pma,PhysRevD.97.044029,Alonso-Serrano:2020hpb},
\begin{eqnarray} \label{Eta}
  \eta = \frac{C}{g}  \left(\frac{\lambda^2_t}{ A_{eff}}\right) ,
\end{eqnarray}
where $C$ is a dimensionless constant associated with different physical cases \cite{Gray:2015pma}, $g$ is the spin degeneracy factor of the particle, $\lambda_t=2\pi \hbar c/k_BT$  is the thermal wavelength in terms of the temperature $T$ and 
\be
A_{eff}=27 A /4
\label{Aeff}
\ee
 is the effective area with $A$ being the horizon area for the case of black holes \cite{Page:1976df,Gray:2015pma}. For the Schwarzschild black hole, one can find the thermal wavelength $\lambda_t$ by taking $T=T_H=1/k_B\beta$ as 
\begin{eqnarray}
  \lambda_t=\frac{2 \pi \hbar c}{k_B T_H}= 2 \pi \hbar c \beta ,
  \label{lam_t}
\end{eqnarray}
and the sparsity profile  for massless bosons in the Hawking process yields \footnote{Here the sparsity profile $\eta_H$ does not represent the actual value for sparsity $\eta$ which will include spin degeneracy factor $g$ and $C$ a dimensionless constant which depends on the chosen time scale. Here we look for the qualitative behaviour of sparsity with respect to mass which depends on sparsity profile through temperature and area of the black hole.}
\begin{eqnarray}
\frac{\lambda^2_t}{ A_{eff}}=\eta_H = \frac{64 \pi^3 }{27}  \approx 73.38,
  \label{etahawking}
\end{eqnarray}
which does not depend on mass of the black hole. Note that for classical black body radiation, the value of $\eta$ is less than one. This implies that the sparsity parameter clearly differentiates the Hawking radiation from classical radiation. One can obtain the GUP effects on the sparsity by replacing the Hawking temperature with the GUP corrected temperature $T_{GUP}$ given by (\ref{Tgup}) \cite{PhysRevD.97.044029}. However, it is assumed that GUP also modifies the black hole horizon area \cite{PhysRevD.97.044029,Alonso-Serrano:2020hpb}. Thus, it is logical to take the effective area that GUP modifies. In fact, the GUP modifications to $A$ can be derived from the equation (\ref{Sgup}) by writing it as
\be
S_{GUP}= \frac{k_B c^3 A_{GUP}}{4\hbar G} ,
\ee
where the GUP modified area $A_{GUP}$ reads 
\be \label{Agup}
  A_{GUP}= \frac{A}{\mathcal{K}}- \alpha \pi l_p^2  \ln \left(\frac{16 A}{A_0 \mathcal{K}^2}\right) ,
\ee
and $A_0=16 \pi m_0^2 G^2/c^4$ is a constant having the dimension of area. Note that in \cite{Alonso-Serrano:2020hpb}, corrections are only in the first order of $\alpha$, while in the above equation (\ref{Agup}) the area is corrected to all orders in $\alpha$. Sparsity depends on the crossectional area of the body at the ray optics limit and the corresponding temperature of the body, which directly depends on the horizon area and the entropy associated with the body, respectively. Thus we heuristically obtain the GUP corrected sparsity profile by replacing $T$ by $T_{GUP}$ and $A$ by $A_{GUP}$ in the expressions for $A_{eff}$ in (\ref{Aeff}) and for $\lambda_t$ in (\ref{lam_t}). It then reads 
\begin{eqnarray}
  \eta_{GUP}= \frac{\eta_H}{\mathcal{K}^2}\left[\frac{A}{A_{GUP}}\right]\label{etagup}  .
\end{eqnarray}
Interestingly, GUP modified sparsity profile $\eta_{GUP}$, depends on the mass of the black hole and the GUP parameter $\alpha$. For the negative values of $\alpha$, the sparsity profile increases as $M$ goes to zero. For the positive values of $\alpha$, the sparsity parameter decreases below the values of sparsity for the Hawking radiation until it reaches the critical mass $M_r$. 
\begin{figure}
    \centering
    \includegraphics[scale=0.5]{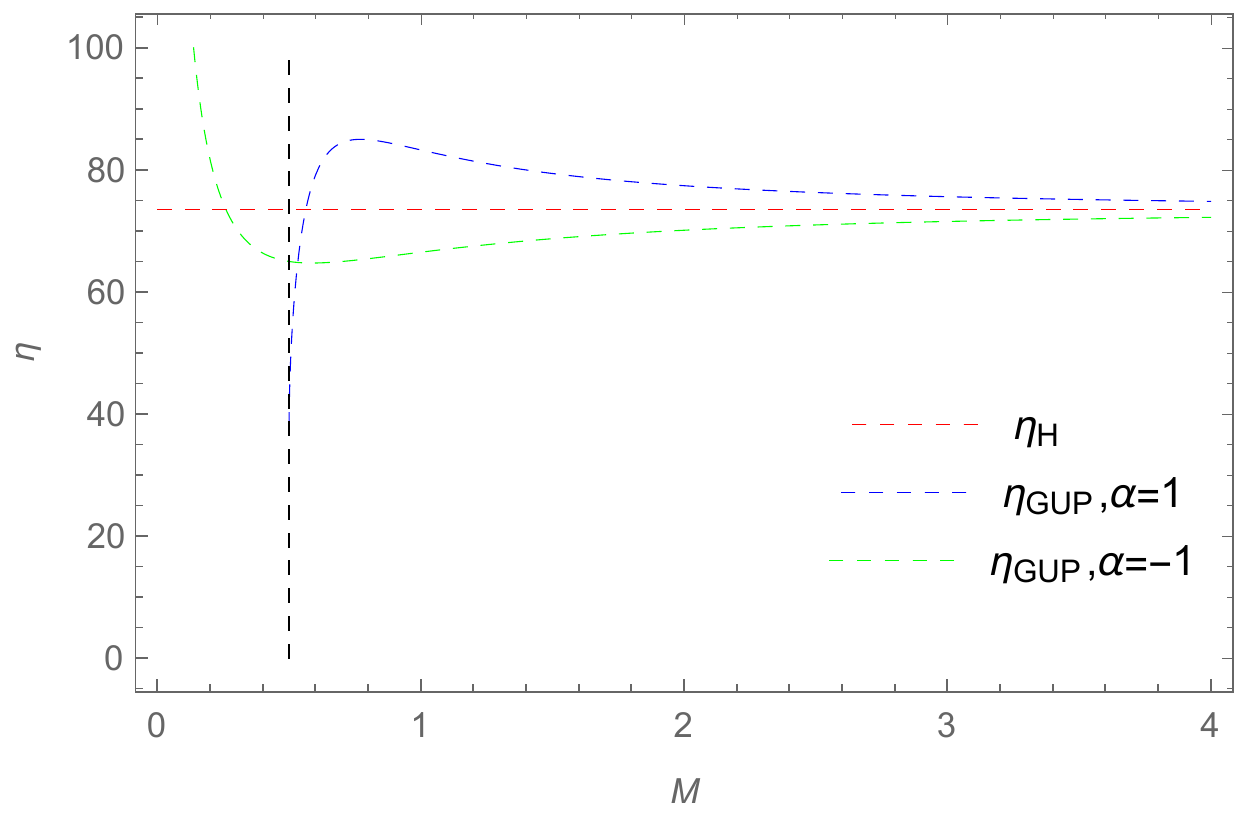}
    \caption{Sparsity profile of Hawking vs GUP corrected black holes in natural units. For positive values of $\alpha$, we observe that sparsity decreases when a black hole is near the final evaporation state.}
    \label{fig:sparsityhupgup}
\end{figure}
In Fig. \ref{fig:sparsityhupgup}, we can see that the GUP corrected sparsity profile is not a constant and it increases first before $M$ approaches $M_r$ for $\alpha>0$ and then it decreases to finite value when $M$ approaches to $M_r$. For the case of $\alpha<0$, first, it decreases, and then it goes to plus infinity when $M$ approaches zero. It is due to the fact that $A/A_{GUP}>1$ for $\alpha>0$ and $\eta_H/\mathcal{K}^2$ turns back the sparsity profile from a maximum value to a constant value, which is less than $\eta_H$. Therefore, we can clearly see the effects of GUP on sparsity due to $T_{GUP}$ and $A_{GUP}$ as depicted in Fig. \ref{fig:sparsityhupgup}. Similarly, $A/A_{GUP}<1$ for $\alpha<o$ and $\mathcal{K}$ goes to zero when $M$ approaches zero, therefore, sparsity decreases first, and then it goes to infinity. Note that in \cite{Ong:2018syk}, the GUP corrected area is not taken into account, therefore, there is no bump in the sparsity profile.

\section{GUP and Nonextensive Black Hole Thermodynamics}
\label{GUPvsNEE}
\subsection{Tsallis Nonextensive Entropy} 

Entropy plays a significant role in Gibbs thermodynamics or statistical mechanics. It is extensive and adheres to the additive composition rule. However, Gibbs statistical mechanics ignores long-range forces. Hence, there are some physical systems for which Gibbs thermodynamics cannot be the appropriate choice to apply \cite{tsallisbook} since they are subject to such forces. Important examples are some self-gravitating systems such as black holes, for which the forces are long-distance and play some significant role. For that reason Constantino Tsallis  in Refs. \cite{Tsallis:1987eu,tsallisbook} generalized the conventional Gibbs entropy for nonextensive systems in order to encompass and address this issue. Tsallis entropy $\mathcal{S}_{\mathcal{T}}$ was one of the earliest proposals to extend Gibbs entropy and the suggested new form of it reads 
\be
\mathcal{S}_{\mathcal{T}} =-k_B\sum_{i}[p(i)]^q\ln_qp(i), \label{S_T}
\ee 
where $p(i)$ is the probability distribution defined on a set of microstates $\Omega$, with the parameter $q$ determining the degree of nonextensivity, and we consider it positive to ensure the concavity of $S_q$. The q-logarithmic function $\ln_qp$ is given by
\be
\ln_qp=\frac{p^{1-q}-1}{1-q},
\ee 
where, in the limit $q \to 1$, Tsallis entropy $S_q$ given by (\ref{S_T}), reduces to Gibbs entropy $S_G$
\be
S_G=-k_B\sum_ip(i)\ln p(i). \label{SG}
\ee
In fact, the Tsallis entropy (\ref{S_T}) satisfies quite general, nonadditive composition rule of the following form
\begin{eqnarray}
\mathcal{S}_ {\mathcal{T} 12} = \mathcal{S}_{\mathcal{T}1}+\mathcal{S}_{\mathcal{T}2}+\frac{\lambda}{k_B} \mathcal{S}_{\mathcal{T}1} \mathcal{S}_{\mathcal{T}2} ,
\label{tsalliscomp}
\end{eqnarray}
for a composite system ''12'', made up of two subsystems ''1'' and ''2''. In above equation, we have defined a new nonextensivity parameter $\lambda = 1-q$. 

\subsection{R\'enyi Entropy}

The R\'enyi entropy \cite{Renyi1}, a measure of entanglement in quantum information that is additive and preserves event independence, is another important generalization of the Gibbs-Shannon entropy. It is defined as 
\be
S_R= k_B \frac{\ln \sum_i p^q(i)}{1-q}. \label{S_R}
\ee
It is important that $S_R$ can be written in terms of $S_{\mathcal{T}}$ by using the formal logarithm approach \cite{PhysRevE.83.061147}, and both entropies are related as follows 
\begin{eqnarray}
S_R=\frac{k_B}{\lambda}\ln[1+ \frac{\lambda}{k_B} \mathcal{S}_{\mathcal{T}}] .
\label{srenyi}
\end{eqnarray}
It is interesting to mention here that $S_R$ is the equilibrium entropy which corresponds to an equilibrium temperature $T_R$ defined from the equilibrium condition by maximizing the Tsallis entropy (\ref{tsalliscomp}), which is given by \cite{Cimdiker:2022ics}
\be
T_{R}=(1+\frac{\lambda}{k_B}S_{\mathcal{T}})\frac{1}{k_B\beta}. \label{T_R}
\ee
Here, $k_B \beta =\partial S_{\mathcal{T}}/\partial U$, where $U$ is the internal energy of the nonextensive system. 

\subsubsection{R\'enyi black hole Entropy and Temperature}

\begin{figure}[hbt!]
    \centering
    \includegraphics[scale=0.5]{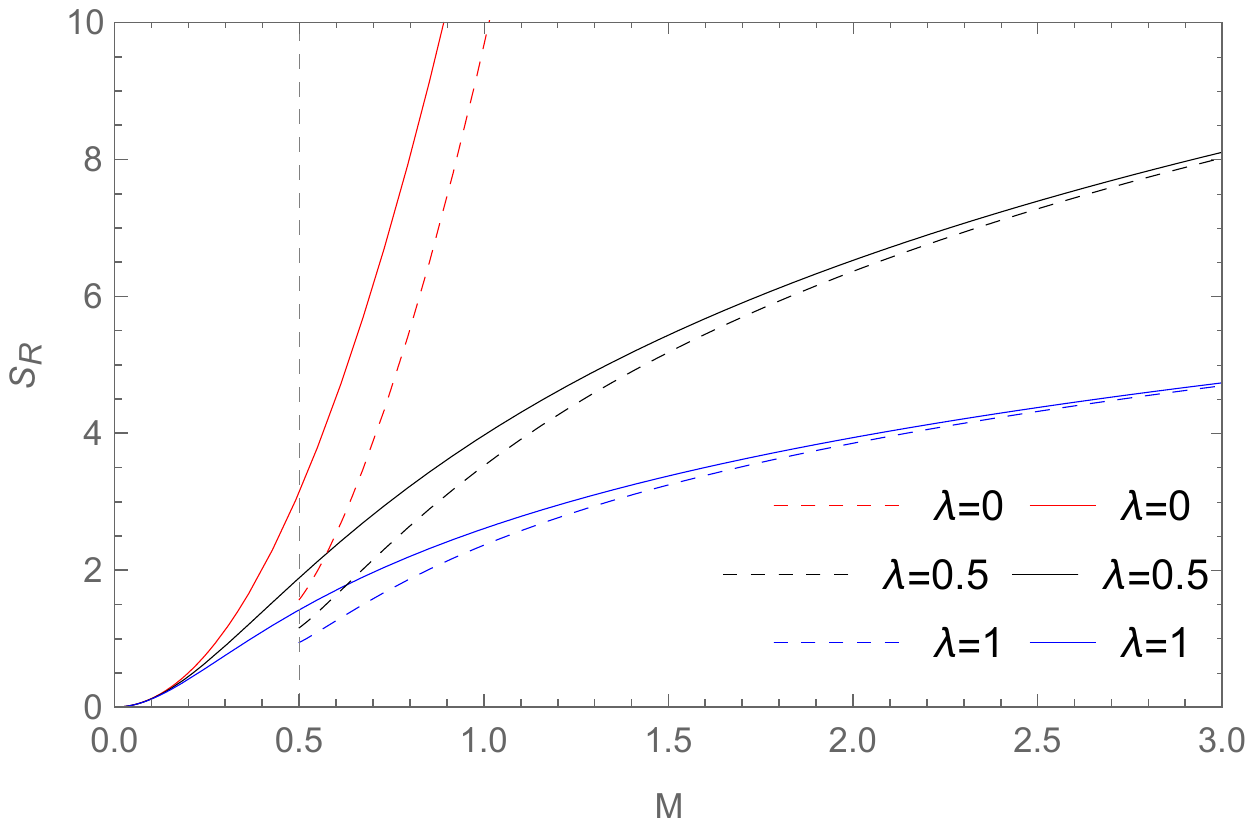}
    \caption{R\'enyi entropy $S_R$ of a black hole vs its mass $M$. Dashed lines represent GUP corrected cases, $\lambda\rightarrow0$ limit is the Bekenstein-Hawking case. }
    \label{fig:srenyi}
\end{figure}

\begin{figure}[hbt!]
    \centering
    \includegraphics[scale=0.5]{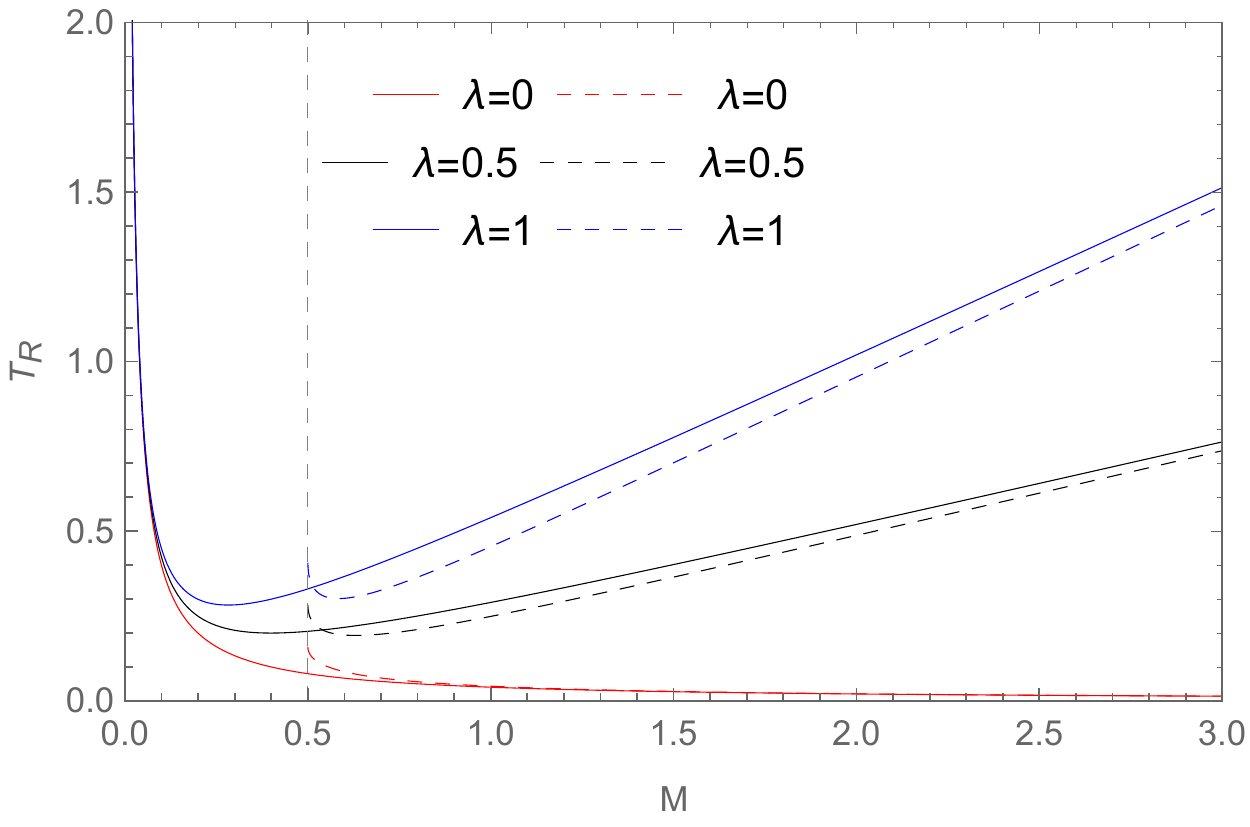}
    \caption{R\'enyi temperature $T_R$ of a black hole vs its mass $M$. Dashed lines represent GUP corrected cases, $\lambda\rightarrow0$ limit is the Bekenstein-Hawking case.}
    \label{fig:trenyi}
\end{figure}

For the case of a Schwarzschild black hole, assuming that the Bekenstein entropy $S_B$ is just the Tsallis entropy $S_{\mathcal{T}}$, and replacing internal energy $U$ with the mass of the black hole $M$ in equations (\ref{srenyi}) and (\ref{T_R}), the R\'enyi entropy can be defined on the horizon of a black hole as \cite{Biro:2013cra,Czinner:2015ena,Czinner:2015eyk,Czinner:2017tjq,Czinner:2017bwc}
\be \label{SR1}
S_R=\frac{k_B}{\lambda}\ln[1+ \frac{\lambda}{k_B}S_B], 
\ee
and the associated R\'enyi temperature reads 
\be
T_R=(1+\frac{\lambda}{k_B} S_B) T_H. \label{T_H1}
\ee
Furthermore, we can write down the GUP corrected R\'enyi entropy using GUP corrected Bekenstein entropy as follows \cite{Alonso-Serrano:2020hpb}  (cf. Fig. \ref{fig:srenyi})
\begin{eqnarray}
S_{Rgup}=\frac{k_B}{\lambda}\ln\left[1+\frac{\lambda}{k_B} (S_{GUP})\right] ,
\label{srgup}
\end{eqnarray}
and corresponding GUP modified R\'enyi temperature $T_{Rgup}$ can be written as (cf. Fig. \ref{fig:trenyi})
\begin{eqnarray}
T_{Rgup}= \left[1+\frac{\lambda}{k_B}(S_{GUP})\right]\mathcal{K} T_H.
\label{TRGup}
\end{eqnarray}
The R\'enyi entropy increases logarithmically (for $0<\lambda<1$), whereas the Bekenstein entropy ($\lambda\rightarrow0$) increases quadratically, as shown in Fig. \ref{fig:srenyi}. Furthermore, for the GUP corrections, the R\'enyi black holes do not completely evaporate; rather, evaporation stops at the critical mass $M_r$, leaving a remnant with finite entropy and temperature as the R\'enyi black hole's final state. 

Using (\ref{T_H1}) and (\ref{TRGup}), we can write the inverse R\'enyi temperature parameters, $\beta_R$ and $\beta_{Rgup}$, which will further be used in calculating the heat capacities, such that
\begin{eqnarray}
k_B \beta_R =\frac{S'_B(M)/c^2}{1+\frac{\lambda}{k_B} S_B}=\frac{k_B \beta}{1+\frac{\lambda}{k_B} S_B},
\label{betaR}
\end{eqnarray}
and the GUP-corrected inverse R\'enyi temperature reads
\begin{eqnarray}
k_B \beta_{Rgup} =\frac{S'_{GUP}(M)/c^2}{1+\frac{\lambda}{k_B} S_{GUP}}=\frac{k_B \beta_{GUP}}{1+\frac{\lambda}{k_B} S_{GUP}} .
\label{betaRGup}
\end{eqnarray}

One may determine the characteristic length scale $\mathcal {L_R}$ for $\lambda$ \cite{Promsiri:2021hhv,Promsiri:2020jga,Nakarachinda:2021jxd}, which reveals the impact of nonextensive parameter $\lambda$ in $S_R$ and $S_{Rgup}$, and in $T_R$ and $T_{Rgup}$. As a result, it can be concluded that below this characteristic length scale $\mathcal {L_R}$, the R\'enyi temperature behaves like $T_H$, and that above $\mathcal {L_R}$, the nonextensive effects increase and $T_R$ grows linearly with $M$. The precise value for the length scale is found in the following subsection.

\subsubsection{Heat Capacity for the R\'enyi black hole}

In order to investigate the thermodynamic stability of R\'enyi black holes, we define the heat capacity $C_R$ of the R\'enyi black hole as
\begin{eqnarray}
C_R=-\frac{S'{}^2_R(M)}{S''_R(M)} \label{cr1} .
\end{eqnarray} 
Inserting (\ref{betaR}) and (\ref{betaRGup}) into (\ref{cr1}), the heat capacity for the non-GUP case reads 
\be \label{cr2}
C_R=\frac{C_{Sc}}{1+\frac{\lambda}{k_B} S_B+\frac{\lambda}{k_B} C_{Sc}},
\ee
and for the GUP case, we have
\begin{eqnarray}
C_{Rgup}=\frac{C_{GUP}}{1+\frac{\lambda}{k_B} S_{GUP}+\frac{\lambda}{k_B} C_{GUP}} .
\end{eqnarray}
\begin{figure}[hbt!]
    \centering
    \includegraphics[scale=0.5]{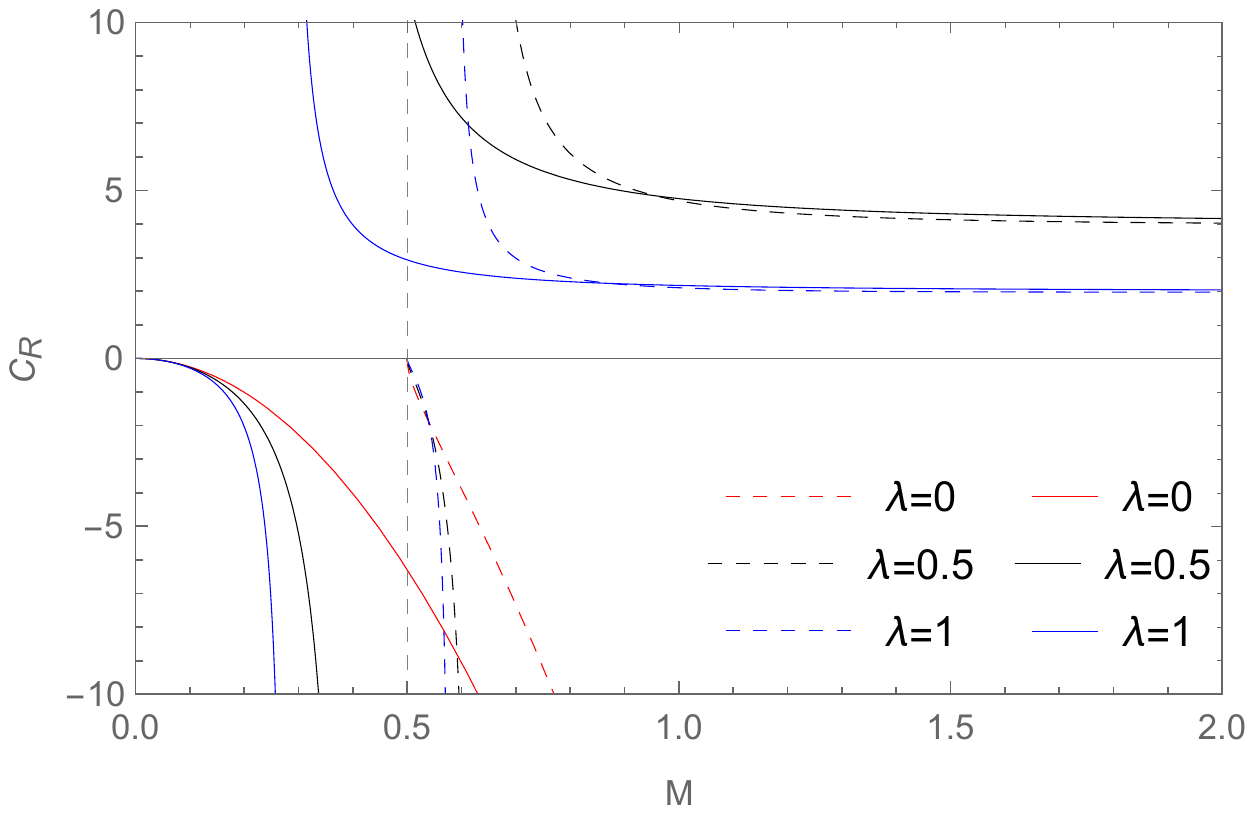}
    \caption{Heat capacity $C_R$ of a R\'enyi black hole vs its mass $M$. Dashed lines represent GUP corrected cases, $\lambda\rightarrow0$ limit is the Bekenstein-Hawking case.}
    \label{fig:crenyi}
\end{figure}
We plot the heat capacity in Fig. (\ref{fig:crenyi}), where we can see that $\mathcal{L}$ differentiates two regions for non-GUP and GUP cases. In order to understand the behavior of $C_R$ in both regions, we find  $\mathcal{L_R}$ in terms of $\lambda$ from the singular points of equation (\ref{cr2}) for the case Schwarzschild black hole. We find, for the non-GUP case
\be
\lambda= -\frac{k_B}{[S_B+C_{Sc}]}=\frac{m_p^2}{4\pi M^2}, \label{rleng1}
\ee
and for the GUP case, we have
\bea
 \lambda &=& -\frac{k_B}{[S_{GUP}+C_{GUP}]}  \\
 &\approx&  \frac{m_p^2}{4 \pi  M^2}+ \frac{3 \alpha  m_p^4}{64 \pi  M^4}+\frac{\alpha  m_p^4 \log \left(\frac{4 M}{m_p}\right)}{32 \pi  M^4} \nonumber
\label{rleng2}
\eea
by ignoring the higher order terms in $\alpha$. This means that for the non-GUP case, we define the mass scale
\be 
M_c=\frac{m_p}{2\sqrt{\pi \lambda}},
\ee
which differentiates the two regions and can be further used to define the characteristic length scale $\mathcal{L}_R$, which can be written as
\be
\mathcal{L}_R  =2l_p\sqrt{\pi \lambda},
\ee
where we have defined $\mathcal{L}_R= GM_c/c^2$. For the GUP case, we would expect the characteristic length scale $L_{Rgup} \approx LR + \alpha f (\lambda)$ by using equation (\ref{rleng2}), where $f$ is a function of the nonextensivity parameter $\lambda$. However, we can not solve it exactly, and it again shows the effects of $\alpha$ and $\lambda$ for the values of $M$ greater than the GUP corrected mass scale.
Interestingly, for the non-GUP case, the heat capacity is positive for the values greater than this scale, and below this scale, black holes have negative heat capacity. This means that black holes with higher masses than $M_c$ are thermodynamically stable and with masses lower than $M_c$, they are unstable. Note that, if we exclude quantum gravity effects, $\mathcal{L}_R$ should be greater than $l_p$. This puts a numerical constraint on the nonextensive parameter $\lambda > 1/4\pi$ and this can also be derived by considering $M_c>m_p$ by excluding the quantum gravity effects. In \cite{Promsiri:2021hhv,Promsiri:2020jga,Nakarachinda:2021jxd}, the authors derived this constraint as $\lambda > 1/\pi$ because they considered $\mathcal{L}_R=2GM_c/c^2$ as characteristic length scale for $\lambda$, where the extra $2$ in $\mathcal{L}_R$ is motivated by Schwarzschild radius $r_h=2GM/c^2$. We believe that the proper way to introduce the length or mass scale for $\lambda$ should be irrespective of the definition which is motivated by $r_h$. 

\subsubsection{Sparsity of the R\'enyi Radiation}

In order to calculate the sparsity profile of R\'enyi radiation, we replace $T$ with $T_R$ in (\ref{Eta}), and so the sparsity profile $\eta_R$ reads 
\begin{eqnarray} \label{EtaR}
\eta_R=\frac{\eta_H}{[1+\frac{\lambda}{k_B} S_B]^2} .
\end{eqnarray}
Replacing $T$ with $T_{Rgup}$ and using GUP modified area $A_{GUP}$ in equation (\ref{Eta}), the GUP modified sparsity profile $\eta_{Rgup}$ reads 
\begin{eqnarray}
\eta_{Rgup}=\frac{\eta_{GUP}}{ [1+\frac{\lambda}{k_B} S_{GUP}]^2} .
\end{eqnarray}
\begin{figure}[hbt!]
    \centering
    \includegraphics[scale=0.5]{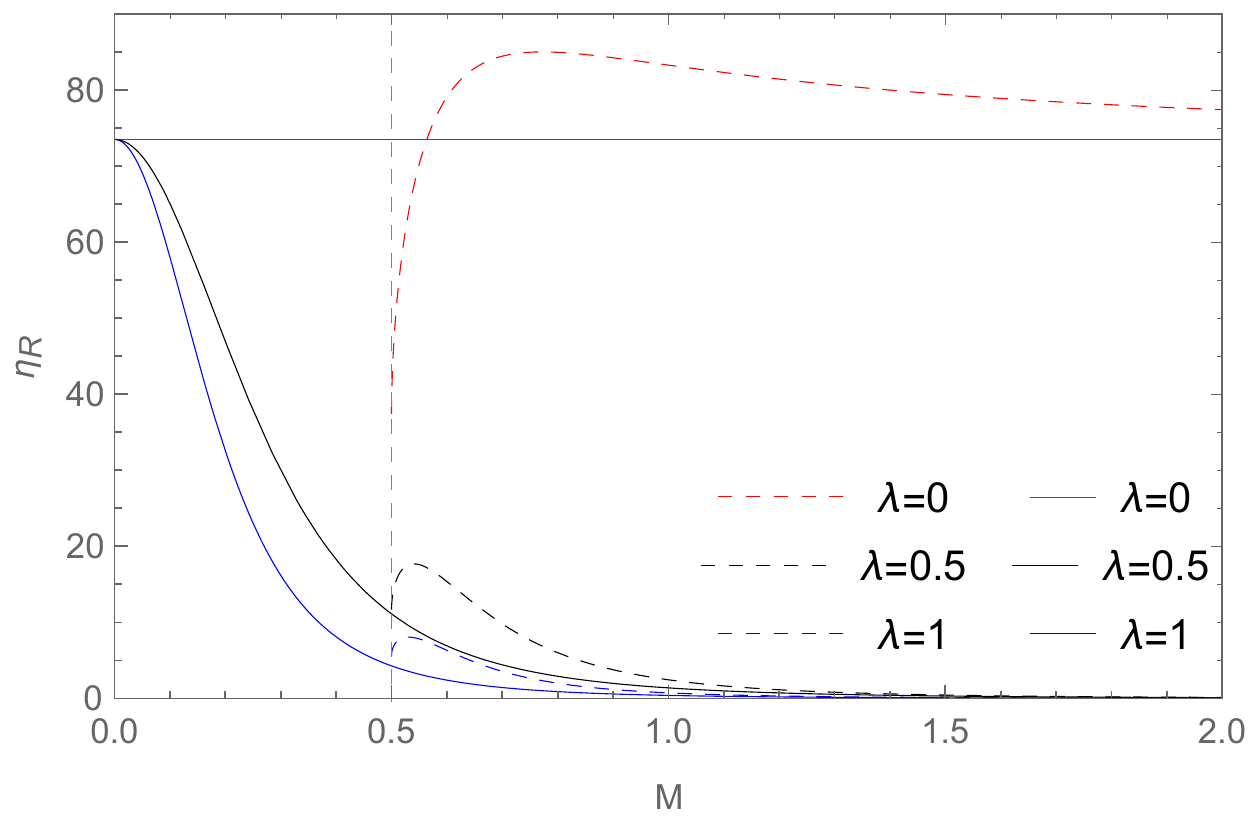}
    \caption{Sparsity profile $\eta_R$ of a R\'enyi blackhole vs its mass $M$. Dashed lines represent GUP corrected cases, $\lambda\rightarrow0$ limit is the Bekenstein-Hawking case.}
    \label{fig:etarenyi}
\end{figure}
From (\ref{EtaR}), we conclude that the sparsity profile $\eta_R$ depends on both the mass of the black hole and the nonextensivity parameter $\lambda$. From Fig. (\ref{fig:etarenyi}), we can easily see that the radiation is not sparse initially and then, at the final stages of the evaporation, the sparsity grows, reaching the value of $\eta_H$, when $M$ approaches to zero. For the GUP case, initially, the behavior of sparsity is similar to the non-GUP case, however, when $M$ approaches $M_r$, it has a finite value which is much less than the sparsity of Hawking radiation for the non-GUP and GUP cases. Again, we can see the {\it bump} before M reaches $M_r$, which is due to the effect of GUP corrections to the R\'enyi temperature and GUP corrections to the area. 

\subsection{Tsallis-Cirto Black Hole Entropy}

\begin{figure}[hbt!]
    \centering
    \includegraphics[scale=0.5]{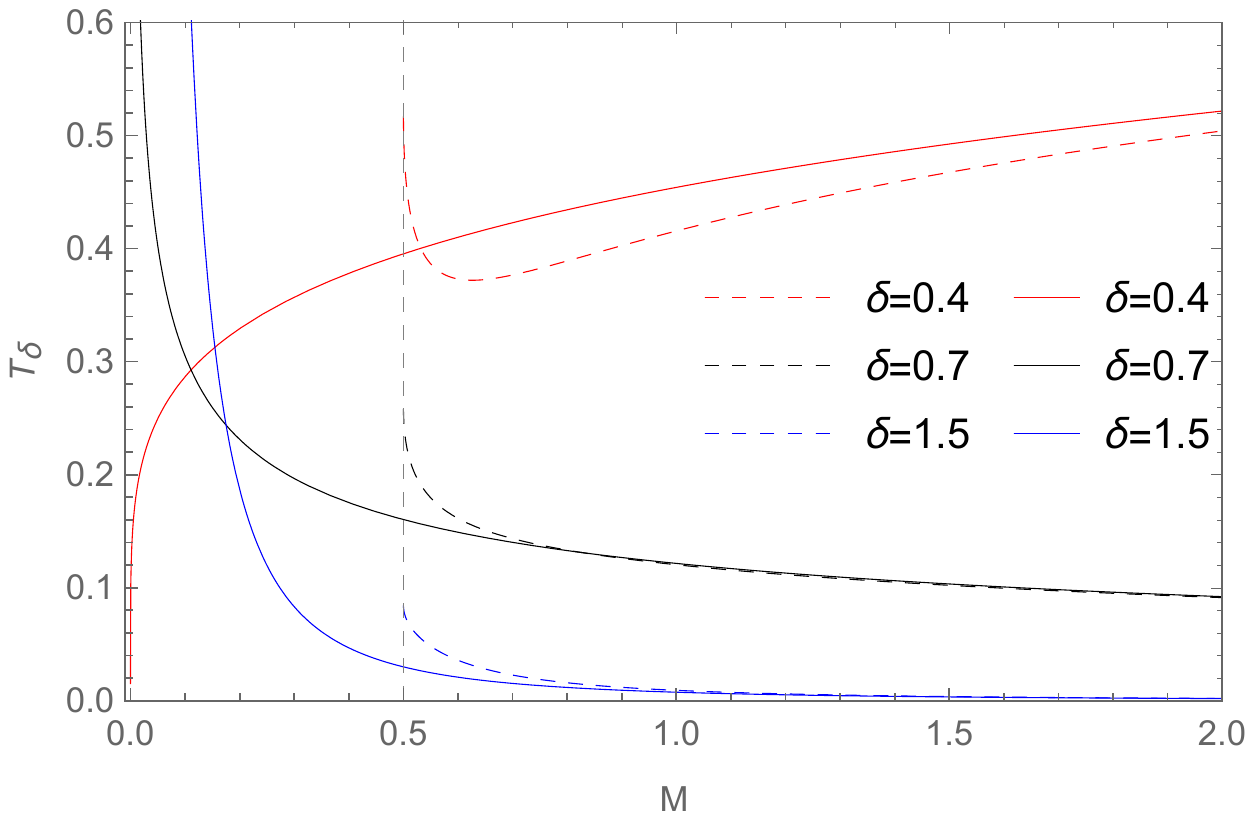}
    \caption{Temperature $T_{\delta}$ vs the mass $M$ for Tsallis-Cirto black hole entropy. Dashed lines correspond to a GUP case.}
    \label{fig:ttsallis}
\end{figure}

Tsallis-Cirto black hole entropy \cite{Tsallis:2012js} is based on key principles of Gibbs thermodynamics. First, the entropy must be extensive and additive, and second, the entropy and associated temperature for a thermodynamic system must satisfy the Legendre structure. As it was already said about the Bekenstein entropy in the Introduction, it violates a key principle of classical Gibbs thermodynamics and so new definitions of entropy and temperature for black holes are required in order to comply with the fundamental principles of thermodynamics in the case of $(3+1)$-dimensional black holes. Therefore, Tsallis and Cirto proposed the following entropy definition \cite{Tsallis:2012js,Tsallis:2019giw}. 
\begin{eqnarray}
\frac{S_{\delta}}{k_B}  = \left(\frac{S_{B}}{k_B} \right)^\delta ,
\label{deltas}
\end{eqnarray}
where $\delta >0$ is a real parameter and it follows the composition rule for a composite thermodynamic system, which is given by
\begin{equation}
    S_{\delta 12}=k_B\left[\left (\frac{S_{\delta 1}}{k_B}\right)^{1/\delta}+\left (\frac{S_{\delta2}}{k_B}\right)^{1/\delta}\right]^\delta. \label{tsallisdeltacomp}
\end{equation}
In this context, the $S_{B}$ is additive, and $S_{\delta}$ is nonadditive. For $\delta=3/2$, $S_{\delta}$ is proportional to the volume for the case of the Schwarzschild black hole, and so it is an extensive quantity. The corresponding Tsallis-Cirto temperature can be written by using the Clausius relation \cite{Cimdiker:2022ics}
\be
T_\delta=\frac{T_{H}}{\delta}\left(\frac{S_{B}}{k_B}\right)^{1-\delta} , \label{Tdelta}
\ee 
and it scales with $1/M^2$ for $\delta=3/2$, i.e., $T_\delta  \propto 1/M^2$, for the case of Schwarzschild black hole. 
GUP corrections to the Tsallis-Cirto black hole entropy can be obtained by the GUP corrected Bekenstein entropy $S_{GUP}$ given by (\ref{Sgup}) into (\ref{deltas}), which results in
\begin{eqnarray}
\frac{S_{\delta gup}}{k_B}  = \left(\frac{S_{GUP}}{k_B} \right)^\delta ,
\end{eqnarray}
and the corresponding GUP-modified Tsallis-Cirto temperature can be derived from the Clausius relation, giving
\be
T_{\delta gup}=\frac{T_{GUP}}{\delta}\left(\frac{S_{GUP}}{k_B}\right)^{1-\delta}.
\label{Tdeltagup}
\ee 
\begin{figure}[hbt!]
    \centering
    \includegraphics[scale=0.5]{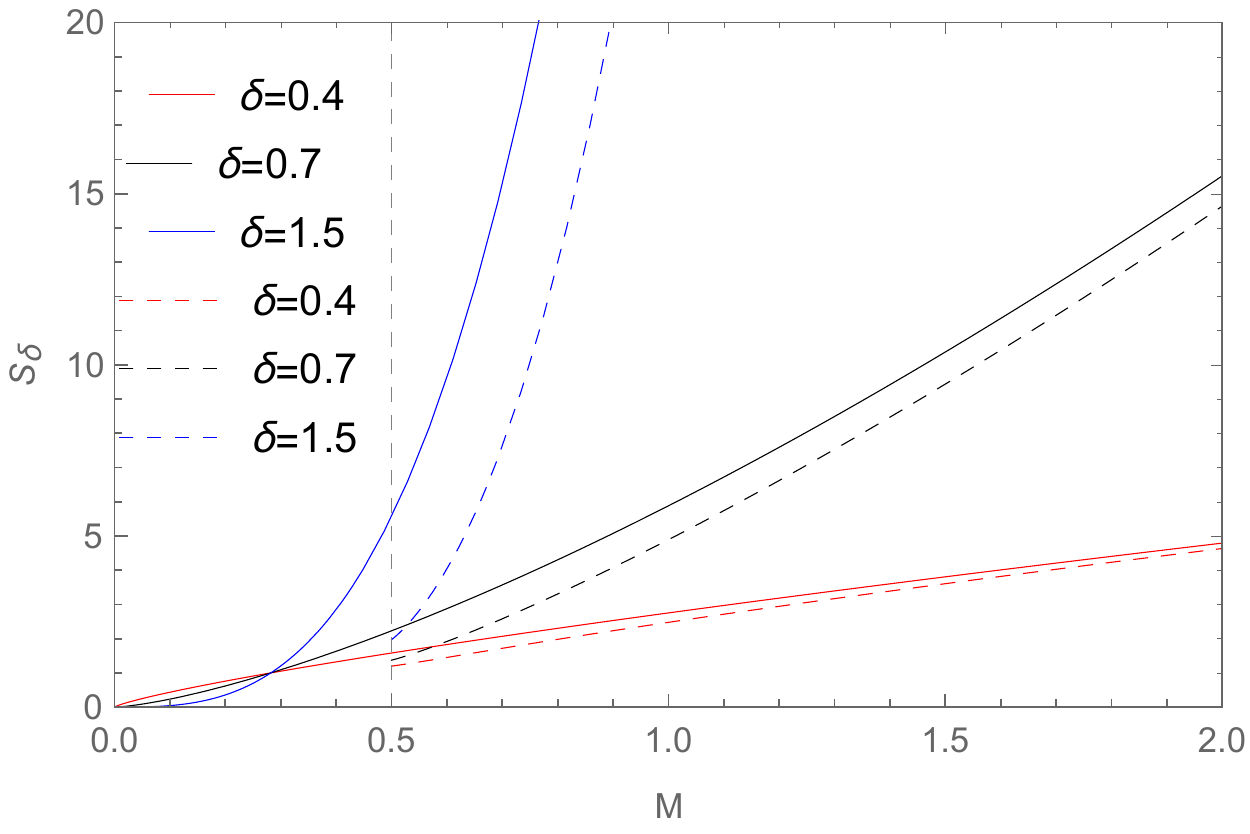}
    \caption{Tsallis-Cirto entropy $S_T$ of a black hole vs its mass $M$. Dashed lines represent GUP-corrected cases in this figure}
    \label{fig:stsallisdelta}
\end{figure}
From the Figs. (\ref{fig:stsallisdelta}) and (\ref{fig:ttsallis}), it shows that the evaporation process stops at the critical value $M_r$ for the Tsallis-Cirto case when GUP corrections are included. This means that the final state of the black hole for the Tsallis-Cirto case is also a remnant with finite entropy and temperature. Generally, for the non-GUP case,  the parameter $\delta$ plays a significant role. For $\delta > 1/2$, the Tsallis-Cirto entropy behaves similarly to Bekenstein entropy and increases as a power law of mass, whereas for $\delta < 1/2$, it increases with mass sub-linearly. For $\delta = 1/2$, the entropy depends linearly on mass, and in this case, Tsallis-Cirto temperature becomes constant. Furthermore, the behavior of the Tsallis temperature is similar to the Hawking temperature for $\delta > 1/2$ while for $\delta < 1/2$, the behavior is completely different for the non-GUP case and, interestingly, it behaves like R\'enyi temperature for the GUP-corrected case.
 Note that, unlike $\lambda$ parameter of the R\'enyi entropy, $\delta$ is {\it not associated} with the length scale for the non-GUP case. On the other hand, introducing GUP corrections to Tsallis-Cirto entropy, one can define a characteristic length scale for $\delta$ as well. 
 
\subsubsection{Heat Capacity for Tsallis-Cirto black holes}

Following the previous subsection, the heat capacity for the Tsallis-Cirto case can be written in terms of $C_{sc}$, and $S_{B}$
\be
C_\delta=C_{Sc}\left[\frac{S_B}{S_B- (\delta-1) C_{Sc}}\right],
\ee
where for the Schwarzschild black hole, we have $C_{Sc}=-2 S_B$. For $\delta = 1/2$, we have infinite heat capacity for all masses. For $\delta < 1/2$, we have positive heat capacity values and negative heat capacity for $\delta > 1/2$. This means that black holes are thermodynamically stable for $\delta<1/2$, and unstable for $\delta>1/2$.
\begin{figure}[hbt!]
    \centering
    \includegraphics[scale=0.5]{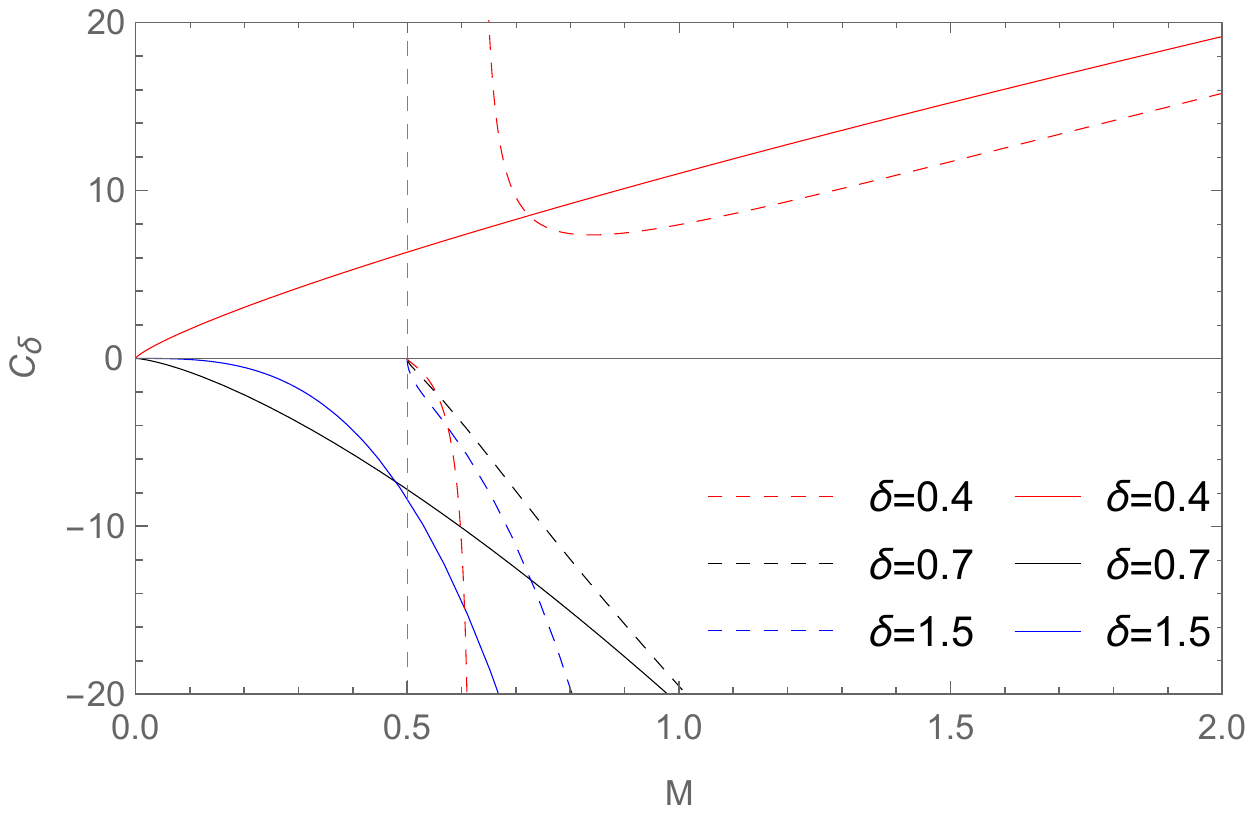}
    \caption{Heat Capacity $C_{\delta}$ for Tsallis-Cirto black hole entropy. Dashed lines correspond to a GUP case.}
    \label{fig:deltac}
\end{figure}
For the GUP corrections, we can write the GUP-corrected heat capacity as 
\begin{eqnarray}
 C_{\delta gup}&=&C_{GUP}\left[\frac{S_{GUP}}{S_{GUP}-(\delta-1) C_{GUP}}\right]  .\label{deltagup}
\end{eqnarray}
Note that from equations (\ref{Sgup}) and (\ref{Cgup}), we have $-2 S_{GUP}\neq C_{GUP}$, therefore, we can find an associated characteristic length scale $\mathcal{L}_{\delta gup}$ for the $\delta$ parameter, for which, we have two regions, which corresponds to positive and negative values of GUP corrected heat capacities. The length scale $\mathcal{L}_{\delta gup}$ can be found by using the singular points of the above equation (\ref{deltagup}) for $\delta$, which is given by
\begin{eqnarray}
\delta=\frac{S_{GUP}}{C_{GUP}}+1 . 
\label{ldelta}
\end{eqnarray}
One could solve the above equation (\ref{ldelta}) for mass $M$, which gives $\mathcal{L}_{\delta gup}$ as a function of $\delta$. However, it is analytically not possible. One may use the perturbative approach to solve the equation for $M$ and define the corresponding length scale or mass scale. From the Figs. (\ref{fig:stsallisdelta}) and (\ref{fig:deltac}), for $\delta<1/2$, and below $\mathcal{L}_{\delta gup}$, the  GUP corrected Tsallis-Cirto entropy behaves like $S_R$ and it gives positive GUP modified heat capacity for the GUP case. For values $\delta>1/2$, $\mathcal{L}_{\delta gup}$ does not exist as (\ref{ldelta}) yields imaginary numbers. Thus, it gives negative heat capacity, implying that GUP-corrected Tsallis black holes are thermodynamically stable for $\delta<1/2$, and unstable for $\delta > 1/2$.

\subsubsection{Sparsity of the Tsallis-Cirto Radiation}

By following the previous subsection, and using the Tsallis-Cirto temperature, we can write the sparsity profile $\eta_{\delta}$ for Tsallis-Cirto radiation as 
\begin{eqnarray}
\eta_\delta = \eta_H \delta^2 \left(\frac{S_B}{k_B}\right)^{2\delta-2} ,
\end{eqnarray}
and the GUP-corrected sparsity profile $\eta_{\delta gup}$, by using (\ref{Eta}) and (\ref{Tdeltagup}), it can be written as
\begin{eqnarray}
\eta_{\delta gup}= \eta_{GUP} \delta^2 \left(\frac{S_{GUP}}{k_B}\right)^{2\delta-2}  .
\end{eqnarray}
\begin{figure}[hbt!]
    \centering
    \includegraphics[scale=0.5]{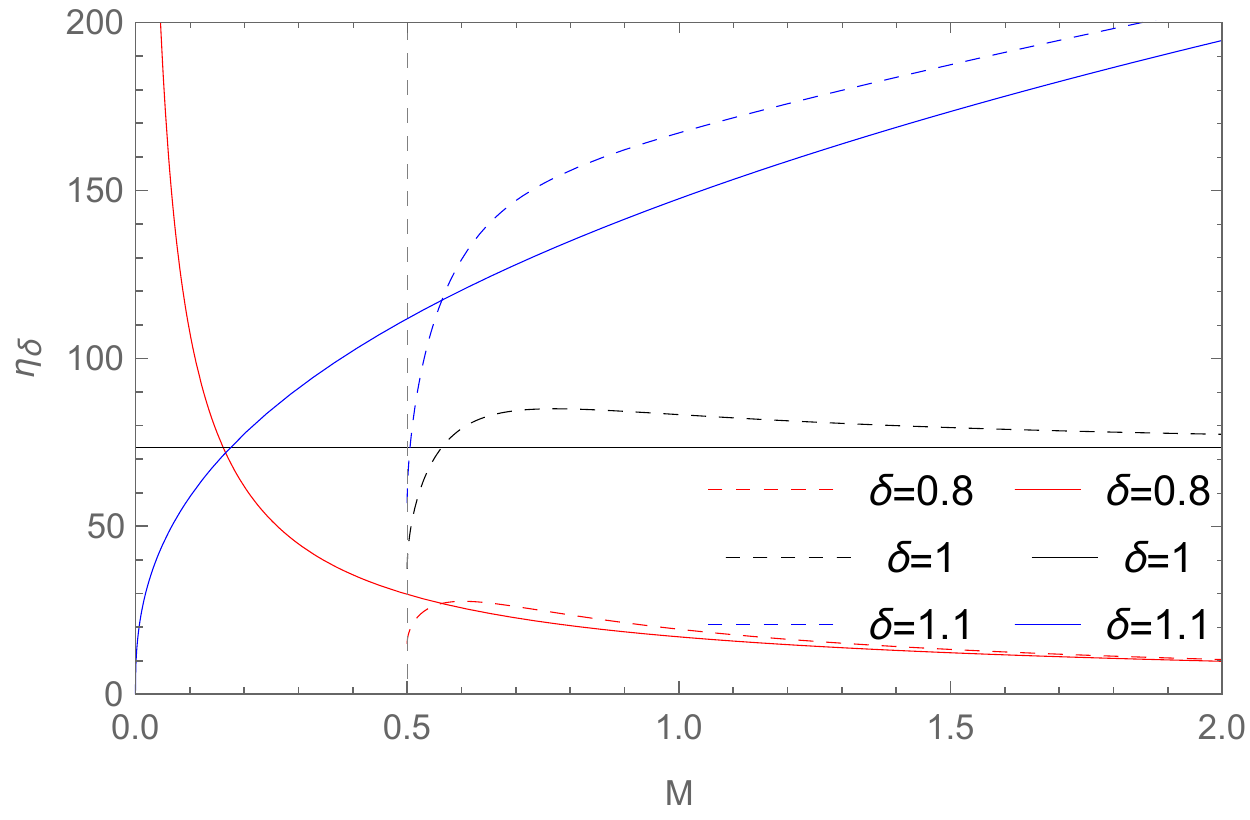}
    \caption{Sparsity profile $\eta_{\delta}$ for Tsallis-Cirto black hole entropy. Dashed lines correspond to a GUP case.}
    \label{fig:deltaeta}
\end{figure}
Fig. (\ref{fig:deltaeta}) depicts the sparsity profile vs. mass relationship. For the Tsallis-Cirto temperature, the sparsity scales with $M^{4\delta-4}$. Again, the value of $\delta$, significantly changes the behavior of the sparsity. It should be noted that the sparsity parameter is now affected by mass as well as $\delta$ and the GUP-parameter $\alpha$. In the non-GUP case, $\eta_{\delta}=\eta_H$ for $\delta=1$. When $\delta>1$, the value of $\eta_{\delta}$ is initially very high and approaches zero at the end of the black hole evaporation. This means that, initially, the Tsallis-Cirto radiation is highly sparse, and during the final stages of evaporation, it is not sparse at all. In this way, for $\delta < 1$, Tsallis-Cirto radiation is initially not sparse, but at the end of the evaporation, it is extremely sparse with the sparsity parameter infinite. For the GUP case, initially, the behavior is the same as for the non-GUP case, but when the mass approaches the order of Planck mass, i.e., the remnant mass $M_r$, the sparsity parameter decreases to some finite values for each case. Note that all these finite values of sparsity profiles are less than the standard sparsity profile $\eta_H$. 

\subsection{Sharma-Mittal Entropy}

\begin{figure}[hbt!]
    \centering
    \includegraphics[scale=0.5]{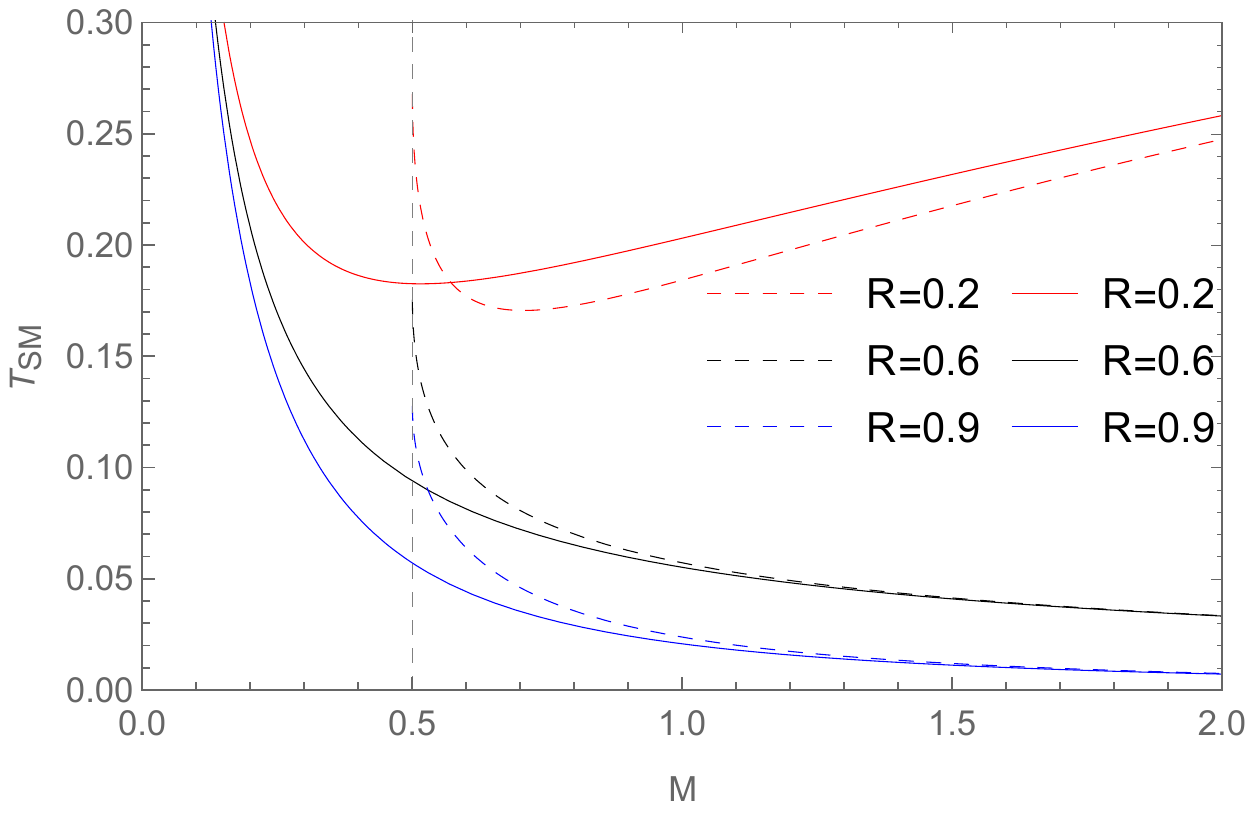}
    \caption{Sharma-Mittal temperature for $\lambda=0.7$. Dashed lines correspond to a GUP case.}
    \label{fig:tsm}
\end{figure}

Sharma-Mittal (SM) is an entropic form \cite{SM,MASI2005217} that {\it generalizes} the R\'enyi and Tsallis entropies. It is defined as
\begin{eqnarray}
    S_{SM}= \frac{1}{R} \left[\left(\sum_{i=1}^W p_i^{1-\lambda}\right)^{\frac{R}{\lambda}}-1 \right] \label{smpi}
\end{eqnarray}
where $R$ is another free parameter that is introduced in SM entropy.  Under the equiprobability condition of the states \cite{SayahianJahromi:2018irq}, the above equation (\ref{smpi}) reduces to
\begin{eqnarray}
S_{SM}=\frac{k_B}{R}\left[(1+\frac{\lambda}{k_B} S_T)^{R/\lambda}-1\right] ,
\label{SSM}
\end{eqnarray}
where $R\rightarrow\lambda$ limit yields the Tsallis entropy, and $R\rightarrow 0$ yields R\'enyi entropy. The Sharma-Mittal entropy obeys the same general nonextensive composition rule (\ref{tsalliscomp}). Assuming that the Bekenstein entropy $S_B$ is the same as the Tsallis entropy $S_{\mathcal{T}}$, we can write $S_{SM}$ for the case of a Schwarzschild black hole as
\begin{eqnarray}
S_{SM}=\frac{k_B}{R}\left[(1+\frac{\lambda}{k_B} S_B)^{R/\lambda}-1\right] ,
\end{eqnarray}
and replacing $S_{GUP}$ with $S_{\mathcal{T}}$ in equation (\ref{SSM}), the GUP corrected SM entropy $S_{SMgup}$ reads as
\begin{eqnarray}
S_{SMgup}=\frac{k_B}{R}\left[(1+\frac{\lambda}{k_B} S_{GUP})^{R/\lambda}-1\right] .
\label{ssmgup}
\end{eqnarray}
The corresponding temperatures can be found by using the Clausius relation, as 
\begin{eqnarray}
T_{SM}=T_H (1+\frac{\lambda}{k_B} S_B)^{1-\frac{R}{\lambda}}\label{tsm} ,
\end{eqnarray}
and the GUP corrected SM temperature $T_{SMgup}$ reads as 
\begin{eqnarray}
T_{SMgup}=T_{GUP} (1+\frac{\lambda}{k_B} S_{GUP})^{1-\frac{R}{\lambda}} .
\label{tsmgup}
\end{eqnarray}
 We can now define the inverse temperature parameters for GUP and non-GUP cases by using the above equations (\ref{tsm}) and (\ref{tsmgup}), which are given, for the non-GUP case, as
\begin{eqnarray}
 \beta_{SM}=\frac{S'_{SM}}{k_B c^2} = \beta (1+\frac{\lambda}{k_B} S_B)^{\frac{R}{\lambda}-1}  ,
\end{eqnarray}
and for the GUP case, as
\begin{eqnarray}
\beta_{SMgup}=\frac{S'_{SMgup}}{k_B c^2} = \beta_{GUP} (1+\frac{\lambda}{k_B} S_{GUP})^{\frac{R}{\lambda}-1}.\;\;
\end{eqnarray}
\begin{figure}[hbt!]
    \centering
    \includegraphics[scale=0.5]{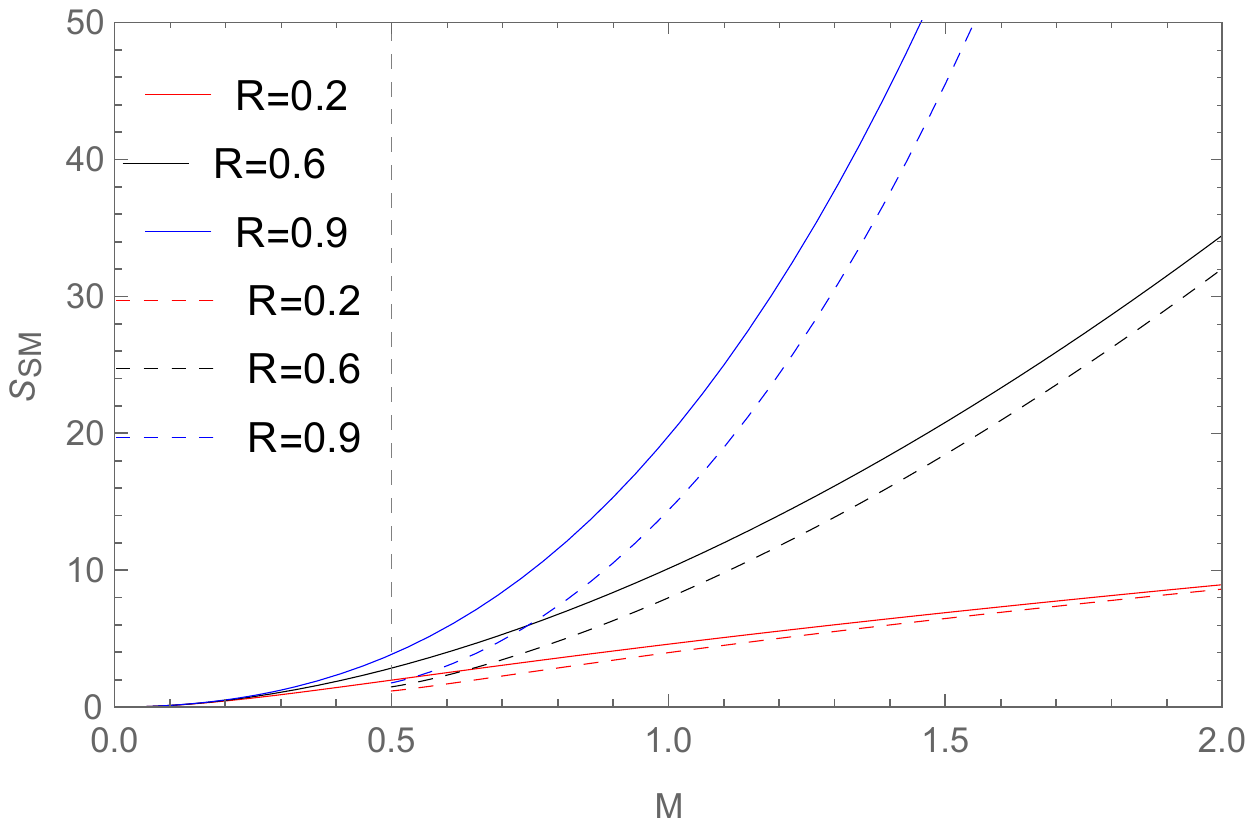}
    \caption{Plot of the Sharma-Mittal entropy for $\lambda=0.7$. Dashed lines correspond to a GUP case.}
    \label{fig:ssm}
\end{figure}
Since SM entropy is the generalization of the Tsallis and R\'enyi entropy, the behavior of the temperature and the entropy are similar to that of $S_B$ and $S_R$ and $T_H$ and $T_R$ for different values of Sharma-Mittal parameter $R$. Also, the black hole does not evaporate in this case as well, and the evaporation process stops at $M_r$, leaving the final state of the black hole as a remnant having finite entropy and temperature. The plots of SM entropy and temperature are given in Figs. \ref{fig:ssm} and \ref{fig:tsm}.

\subsubsection{Heat Capacity for Sharma-Mittal Black Holes}

\begin{figure}[hbt!]
    \centering
    \includegraphics[scale=0.5]{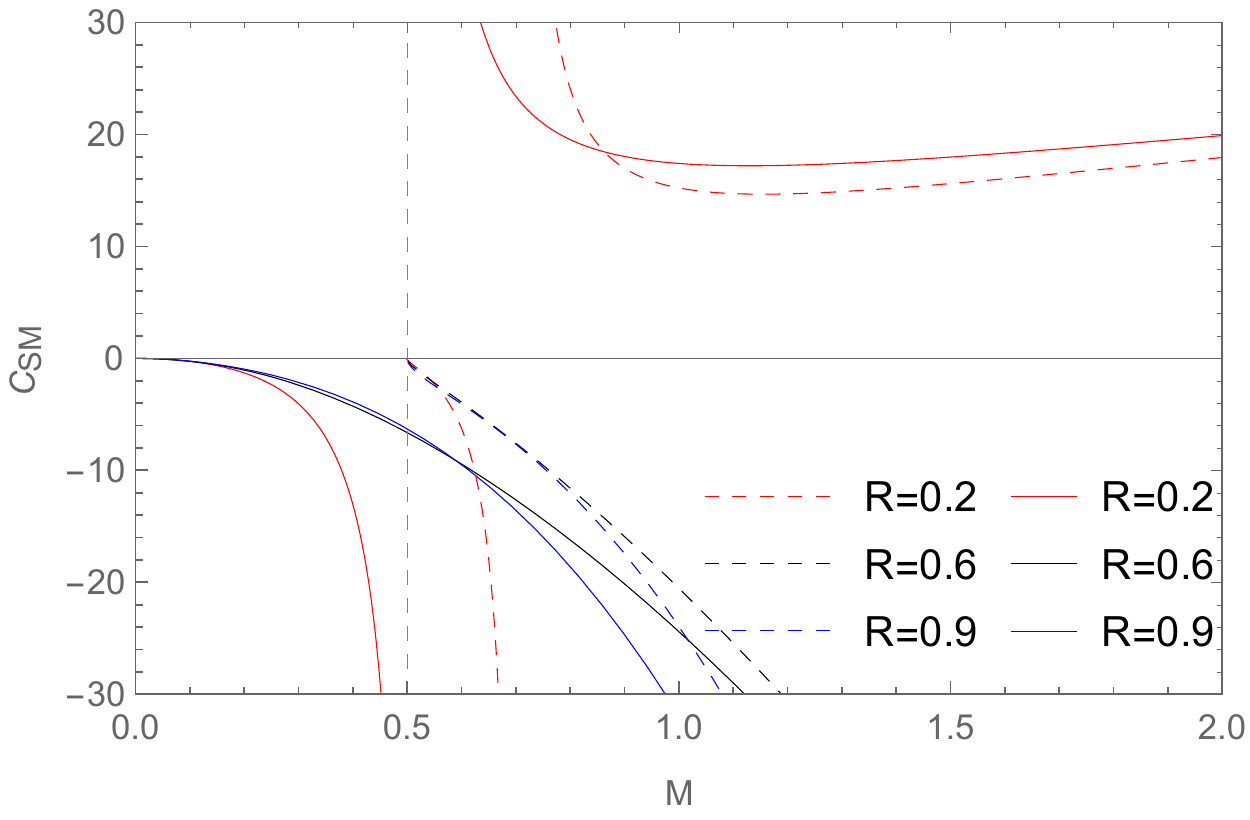}
    \caption{Heat capacity $C_{SM}$ for Sharma-Mittal entropy for $\lambda = 0.7$. Dashed lines correspond to a GUP case.}
    \label{fig:csm}
\end{figure}

By following the previous subsections, we can calculate the heat capacity $C_{SM}$ for the SM black holes as
\begin{eqnarray}
C_{SM}= \frac{C_{Sc} (1+\frac{\lambda}{k_B} S_B)^{\frac{R}{\lambda}}}{(1+\frac{\lambda}{k_B} S_B)-\frac{\lambda}{k_B} C_{Sc}\left(\frac{R}{\lambda}-1\right)} \;,
\label{csm}
\end{eqnarray}
and for the GUP SM black holes case, it reads as
\begin{eqnarray}
C_{SMgup}= \frac{C_{GUP} (1+\frac{\lambda}{k_B} S_{GUP})^{\frac{R}{\lambda}}}{(1+\frac{\lambda}{k_B} S_{GUP})- \frac{\lambda}{k_B} C_{GUP}\left(\frac{R}{\lambda}-1\right)}\; .
\label{csmgup}
\end{eqnarray}
The plots of (\ref{csm}) and (\ref{csmgup}) are given in Fig. \ref{fig:csm}. Similarly as for the R\'enyi case, we define the characteristic length scale $\mathcal{L}_{SM}$ in terms of $\lambda$ and $R$ by employing the singular point of $C_{SM}$. For the non-GUP case, we have such a singular point for
\begin{eqnarray}
\lambda=\frac{R C_{Sc}-k_B}{C_{Sc}+S_{B}}\;.
\label{lsm1}
\end{eqnarray}
From (\ref{lsm1}), we can easily define the following characteristic relation by solving it for $M$, which reads
\be
\mathcal{L}_{SM}=2l_p\sqrt{\pi(\lambda-2R)} ,
\ee 
where $\mathcal{L}_{SM}=GM_c/c^2$, and the mass scale $M_c$ is defined as 
\be
M_c=\frac{m_p}{2\sqrt{\pi(\lambda-2R)}}.
\ee
Similarly, one can define $\mathcal{L}_{SMgup}$ for the GUP case by using the following singular point at
\begin{eqnarray}
\lambda=\frac{R C_{GUP}-k_B}{C_{GUP}+S_{GUP}}\;\;,
\label{lsm2}
\end{eqnarray}
and solve it for $M$. Since the analytic solution is not possible, one could use a perturbative approach to find the GUP corrections to $\mathcal{L}_{SM}$ up to the first order in $\alpha$.
Note that $R\rightarrow0$ limit yields the $\mathcal{L}_{R}$ for the R\'enyi case. For $\lambda-2R>0$ and $M>M_c$, the heat capacity is positive for both non-GUP and GUP cases, and for $M<M_c$, the heat capacity is negative for both non-GUP and GUP cases.

\subsubsection{Sparsity of the Sharma-Mittal Radiation}

\begin{figure}[hbt!]
    \centering
    \includegraphics[scale=0.5]{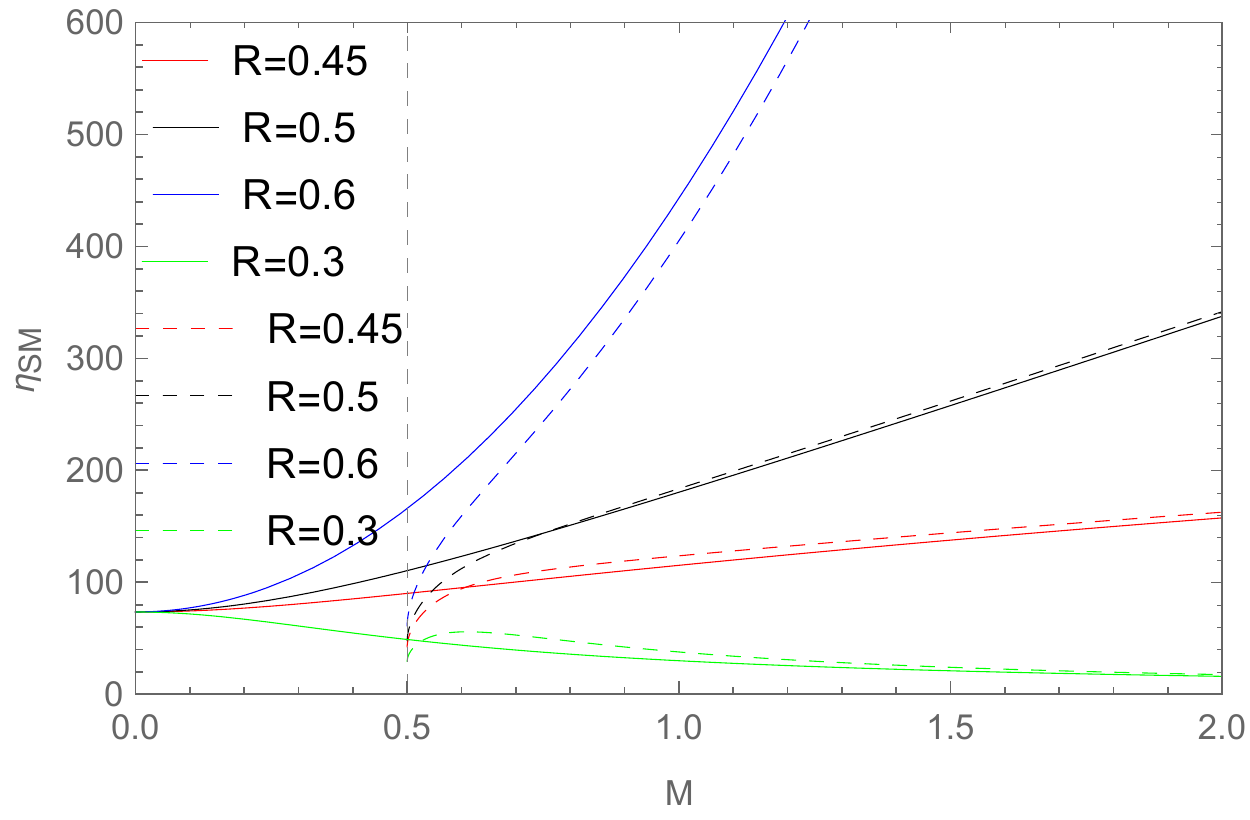}
     \caption{Sparsity profile for Sharma-Mittal entropy for $\lambda=0.4$. Dashed lines correspond to a GUP case.}
    \label{fig:etasm}
\end{figure}

The sparsity profile $\eta_{SM}$ can be derived by applying the Sharma-Mittal temperature to (\ref{Eta}), and reads
\begin{eqnarray}
\eta_{SM}=\eta_H (1+\frac{\lambda}{k_B} S_B)^{2(\frac{R}{\lambda}-1)} ,
\label{etasm}
\end{eqnarray}
and for the GUP case, substituting equations (\ref{tsmgup}) and (\ref{Agup}) in (\ref{Eta}), the GUP modified sparsity profile for the Sharma-Mittal radiation reads as
\begin{eqnarray}
\eta_{SMgup}=\eta_{GUP} (1+\frac{\lambda}{k_B} S_{GUP})^{2(\frac{R}{\lambda}-1)} .
\label{etasmgup}
\end{eqnarray}
The plots of the sparsity profile for SM (\ref{etasm}) and SM GUP (\ref{etasmgup}) cases are given in Fig. \ref{fig:etasm}. 
The behavior of the sparsity profile again depends on the Sharma-Mittal parameter $R$ in addition to the nonextensive parameter $\lambda$ and also the GUP parameter $\alpha$ in the case of GUP corrections. For the values of $\lambda$ and $R$, which satisfy the inequality $\lambda>R$, the sparsity of the Sharma-Mittal radiation behaves like the sparsity of the R\'enyi radiation for both non-GUP and GUP cases. This means that, initially, the Sharma-Mittal radiation is not sparse, and at the end of the evaporation, its value approaches the value of Hawking's case, i.e., $\eta_H$, for the non-GUP case. At the $R \rightarrow 0$ limit we obtain the sparsity profile of the Rènyi entropy. For $R>\lambda$, initially, the Sharma-Mittal sparsity profile is higher than $\eta_H$ and its value exactly approaches $\eta_H$ at the end of the evaporation, while for the case of GUP, it approaches to some finite value less than $\eta_H$. It is interesting to note that, for $\alpha>0$, the GUP modified sparsity parameter is always less than the standard Hawking case.

\subsection{Kaniadakis Entropy}

Kaniadakis entropy \cite{Kaniadakis:2002zz,Drepanou:2021jiv} is a type of nonextensive entropy that results from the Lorentz transformation of special relativity. It is a single parameter deformation of Gibbs entropy in which the standard Gibbs entropy is generalized to the relativistic regime with the help of a new parameter $K$ that is connected to the dimensionless rest energy of the various parts of a multibody relativistic system. The Kaniadakis entropy $S_K$ is defined as
\begin{eqnarray}
    S_K=k_B\log_K \Omega \label{SK}
\end{eqnarray}
where 
\begin{eqnarray} \label{logK}
    \log_K (\Omega) = \frac{\Omega^K-\Omega^{-K}}{2 K} \;.
\end{eqnarray}
Considering $S_B=k_B\ln \Omega$, which means that the number of microstates $\Omega$ for a black hole is proportional to $e^{S_B/k_B}$, the above equation (\ref{SK}) can be written in the following form
\begin{eqnarray}
S_K=\frac{k_B}{K}\sinh \left[K \frac{S_B}{k_B}\right],
\end{eqnarray}
where we have used equation (\ref{logK}) for the $\sinh x$ function and used the relation $\Omega=e^{S_B/k_B}$. Replacing $S_B$ with $S_{GUP}$, the GUP modified Kaniadakis entropy $S_{KGUP}$ reads as
\begin{eqnarray}
S_{KGUP}=\frac{k_B}{K}\sinh \left[K \frac{S_{GUP}}{k_B}\right].
\end{eqnarray}
\begin{figure}[hbt!]
    \includegraphics[scale=0.5]{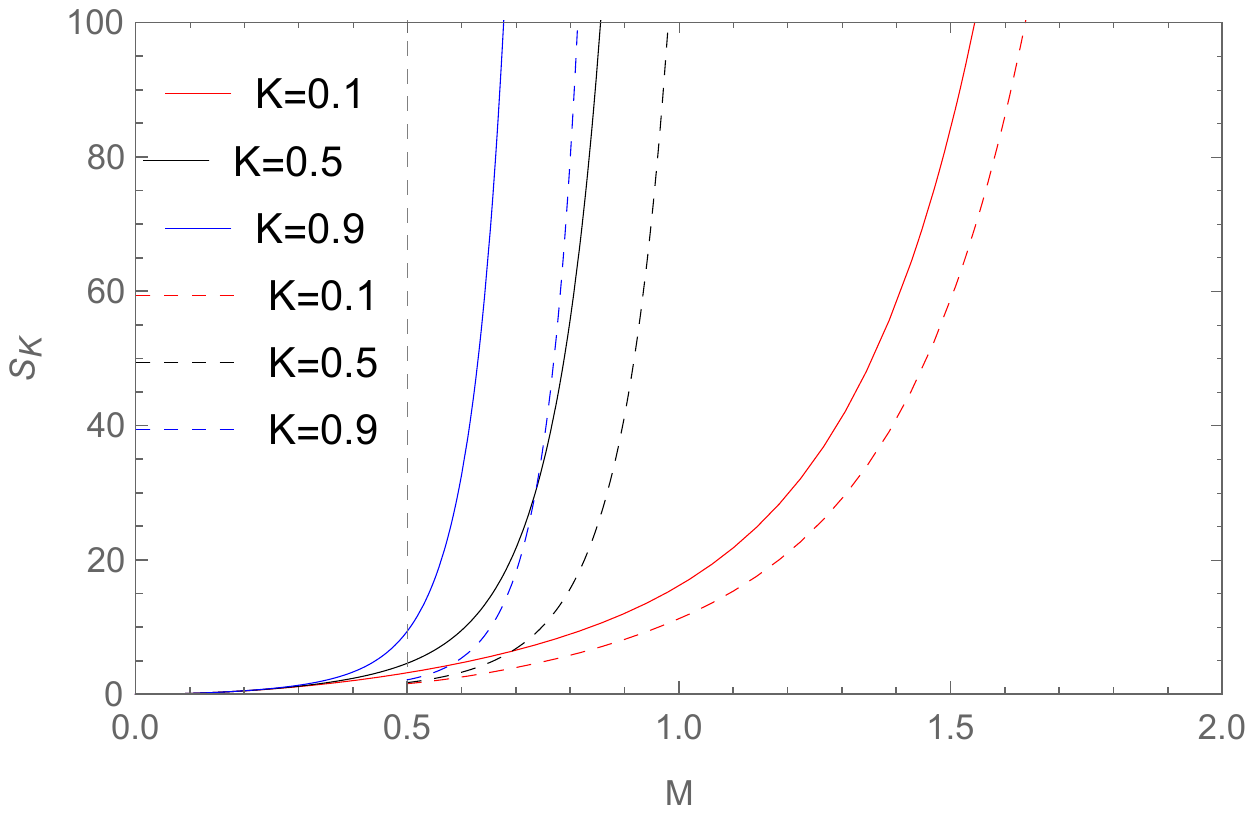}
    \caption{Kaniadakis Entropy $S_K$ vs mass $M$. Dashed lines correspond to a GUP case.}
    \label{fig:skaniadakis}
\end{figure}
Note that, in the limit $K\rightarrow0$,  $S_K$ reduces to Gibbs entropy. In Fig. (\ref{fig:skaniadakis}), one can see the characteristic form of sine hyperbolic $(\sinh)$ function for different small values of $K$ which shows the similar behaviour like the Bekenstein entropy. As expected, for the GUP case, black holes do not evaporate completely and the final state of the black hole is a remnant like for the case of standard GUP modified Bekenstein-Hawking case. Furthermore, as $K$ increases, the entropy increases sharply.
By using the Clausius relation, the corresponding Kaniadakis black black hole temperature $T_K$ reads as 
\be \label{T_K}
T_K= T_H \,\sech \left[K \frac{S_B}{k_B}\right],
\ee
and the GUP modified Kaniadakis temperature $T_{KGUP}$ can be written as
\be \label{T_KGUP}
T_{Kgup}= T_{GUP}\, \sech \left[K \frac{S_{GUP}}{k_B}\right].
\ee
By using (\ref{T_K}) and (\ref{T_KGUP}), one can write the following inverse temperature parameters $\beta_K$ as follows
\be
k_B \beta_K = k_B\beta \cosh \left[K \frac{S_B}{k_B}\right] ,
\ee
and for the GUP case, $\beta_{KGUP}$ reads
\begin{eqnarray}
k_B \beta_{Kgup} = k_B \beta_{GUP} \cosh \left[K \frac{S_{GUP}}{k_B}\right],
\end{eqnarray}
which can further be used to find the heat capacities for Kaniadiakis black holes.
\begin{figure}[hbt!]
    \includegraphics[scale=0.5]{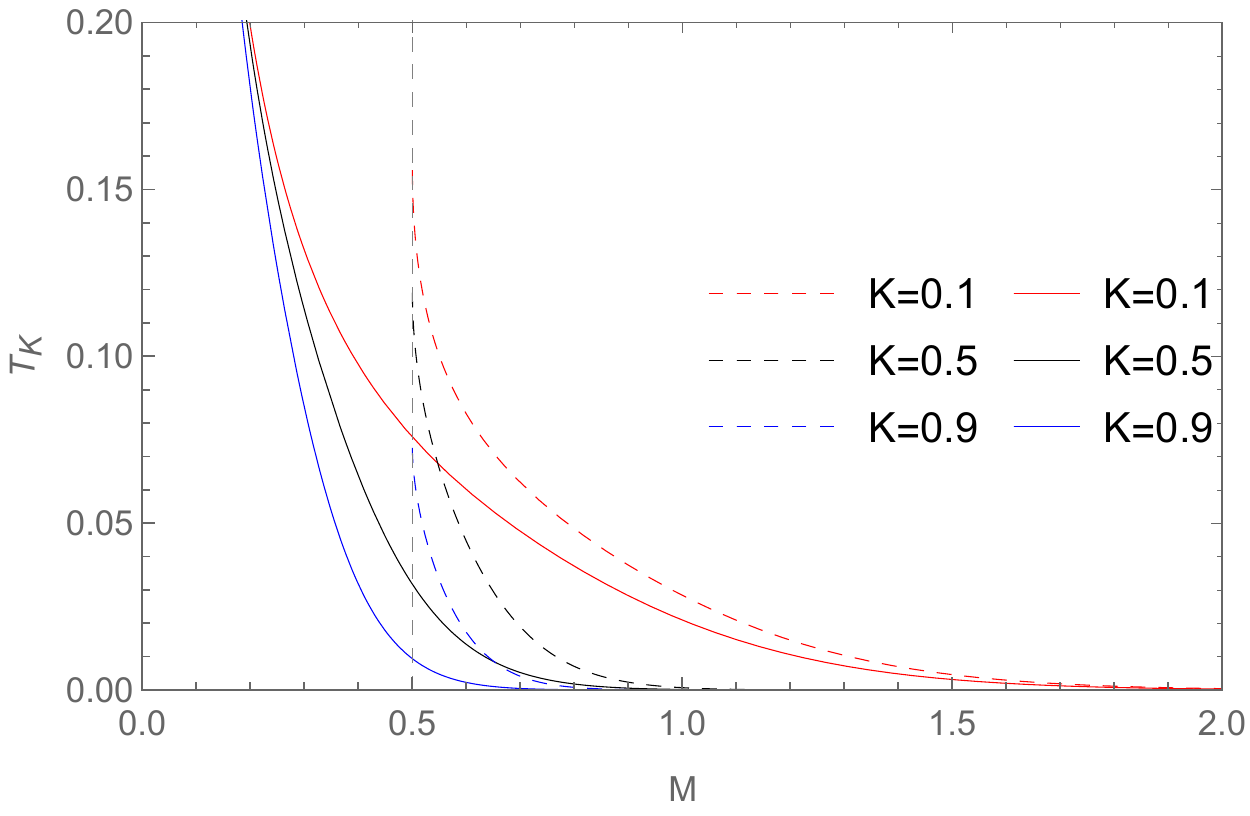}
    \caption{Kaniadakis temprature $T_K$ vs mass. Dashed lines correspond to a GUP case.}
    \label{fig:tkaniadakis}
\end{figure}
Fig. (\ref{fig:tkaniadakis}) shows that Kaniadakis temperature behaves as Hawking temperature with a slight change depending on the parameter $K$. For the GUP case, it stops at some finite value, when $M$ approaches to $M_r$ during the final stages of the black hole evaporation process.

\subsubsection{Heat capacity for Kaniadakis Black Holes}

The heat capacities for Kaniadakis entropy can be calculated by following the previous subsections. For the non-GUP case, the heat capacity $C_K$ for Kaniadakis black hole reads as 
\begin{eqnarray}
C_K= C_{Sc} \frac{\cosh^2[K \frac{S_B}{k_B}]}{\cosh[K \frac{S_B}{k_B}]-C_{Sc} \sinh[K \frac{S_B}{k_B}]} ,
\end{eqnarray}
and for the GUP modified heat capacity, $C_{Kgup}$, it can written as
\begin{eqnarray}
C_{Kgup}= C_{GUP} \frac{\cosh^2[K \frac{S_{GUP}}{k_B}]}{\cosh[K \frac{S_{GUP}}{k_B}]-C_{GUP} \sinh[K \frac{S_{GUP}}{k_B}]}. \;\;\;
\end{eqnarray}
\begin{figure}[hbt!]
    \includegraphics[scale=0.5]{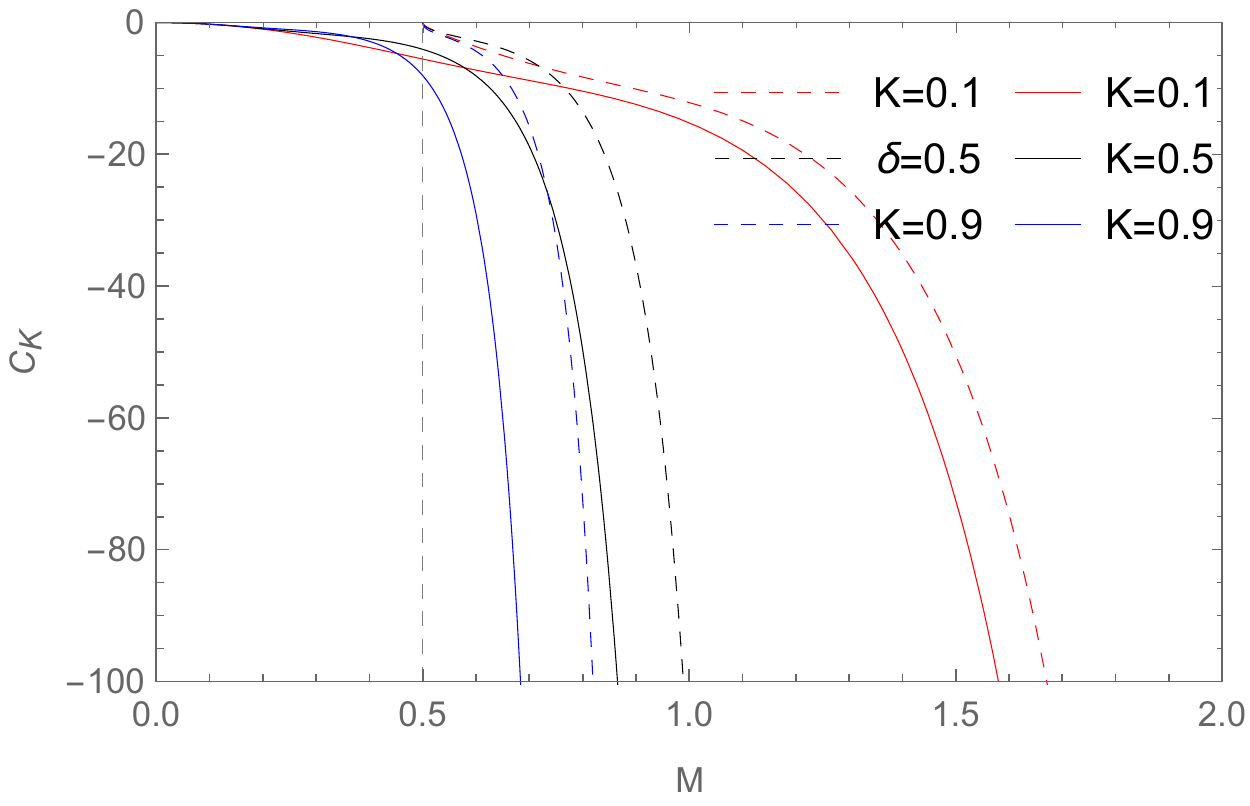}
    \caption{Kaniadakis heat capacity $C_K$ vs mass $M$. Dashed lines correspond to a GUP case.}
    \label{fig:ckaniadakis}
\end{figure}
\begin{figure}[hbt!]
    \centering
    \includegraphics[scale=0.5]{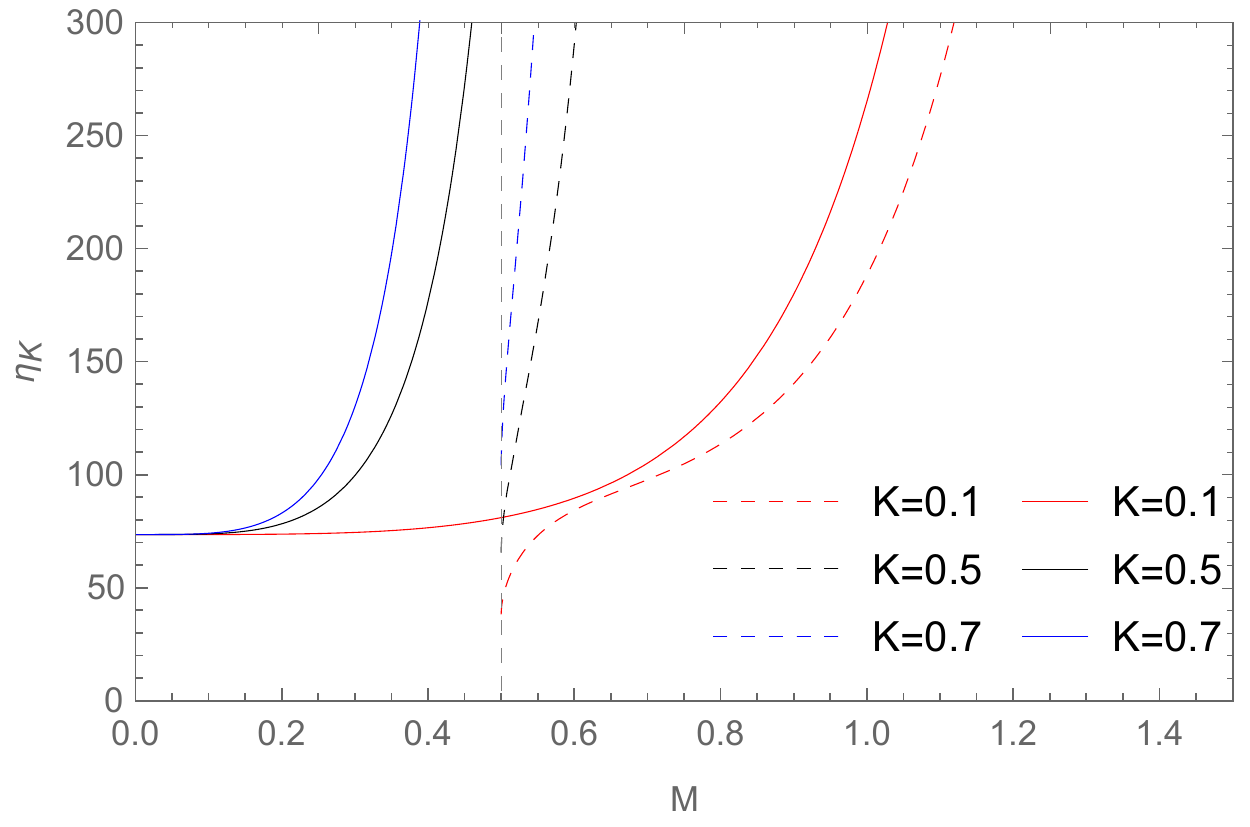}
    \caption{Sparsity profile $\eta_K$ for Kaniadakis radiation vs mass $M$ of Kaniadakis black hole. Dashed lines correspond to a GUP case.}
    \label{fig:etakaniadakis}
\end{figure}
From Fig. (\ref{fig:ckaniadakis}), one can easily notice the negative heat capacities for all values of $K$. This means that Kaniadakis black holes are thermodynamically unstable for all $M$.

\subsubsection{Sparsity profile of the Kaniadakis Radiation}

The sparsity profile $\eta_K$ for the Kaniadakis radiation can be derived by applying (\ref{T_K}) into (\ref{Eta}), and reads 
\begin{eqnarray}
\eta_K = \eta_H \cosh^2 \left(K \frac{S_B}{k_B}\right) ,
\end{eqnarray}
and for the GUP modified sparsity profile $\eta_{KGUP}$, we apply (\ref{T_KGUP}) and (\ref{Agup}) into (\ref{Eta}), to obtain
\begin{eqnarray}
\eta_{KGUP} = \eta_{GUP} \cosh^2 \left(K \frac{S_{GUP}}{k_B}\right) .
\end{eqnarray}
From Fig. (\ref{fig:etakaniadakis}), the sparsity parameter for the Kaniadakis case is always high from the beginning of the evaporation process as compared to the standard Bekenstein Hawking case. However, for the non-GUP case, $\eta_K$ approaches to the value of $\eta_H$ at the end of the evaporation. For the GUP case, again, it approaches to some finite value of sparsity when $M$ approaches $M_r$, which is always less than the sparsity profile $\eta_H$. Furthermore, we see that increasing value of $K$ directly results in sparser Kaniadakis radiation. 

\subsection{Barrow entropy} 

Barrow entropy \cite{Barrow:2020tzx} is an entropic form that has no statistical roots, but is closely tied to black hole horizon geometry. It is proposed to replace the smooth black hole horizon with a fractal of spheres known as a sphereflake. This structure is distinguished by its fractal dimension $d_f$, where $3\geq d_f\geq2$, and results in an effective horizon area of $r_+{}^{d_f}$, where $r_+$ is the horizon radius. As a result, in this scenario, the horizon area is modified, yielding Barrow entropy  as below $S_{Barrow}$ 
\begin{eqnarray}
S_{Barrow}=k_B \left( \frac{A}{A_p}\right)^{1+\frac{\Delta}{2}}
\label{barrowarea}
\end{eqnarray}
where $A$ is the horizon area, $A_p$ is the Planck area, and $\Delta$ is the parameter directly tied to the fractal dimension $d_f$ through $\Delta=d_f-2$. In this form, $\Delta$ can take values between 0 and 1, and $\Delta\rightarrow1$ limit yields maximally fractal structure, where the horizon area effectively behaves like a $3-$dimensional volume, while $\Delta\rightarrow0$ limit yields the well-known Bekenstein area law where no fractalization occurs. Although Barrow entropy offers a different picture in the geometrical sense, in its essence, it has the same form as Tsallis-Cirto entropy. We can see that they are equivalent by making the following parametrization in Tsallis-Cirto entropy \cite{Abreu:2020wbz}
\begin{eqnarray}
\delta \rightarrow 1+\frac{\Delta}{2}
\label{barrowdelta}
\end{eqnarray}
Thus, qualitatively, both entropic forms yield the same temperatures and heat capacities as a function of black hole mass. Similarly, the Tsallis-Cirto entropy limit $\Delta=1$ ($\delta=3/2$ for $S_\delta$) yields an extensive, but still nonadditive entropy for black holes.

\section{Summary and discussion}
\label{summary}

We have investigated the nonextensive thermodynamics of black holes, the impact of the generalized uncertainty principle on nonextensive thermodynamics quantities, and the sparsity and GUP-modified sparsity of the radiation in the nonextensive scenario. We have found that all nonextensive black hole entropies and associated temperatures have finite values at the end of the black hole evaporation process due to GUP modifications, indicating the existence of a remnant at the end of the evaporation. This means that black holes do not evaporate fully in the nonextensive setup as well. We have also investigated the sparsity profiles in each nonextensive configuration. Despite the fact that the behavior of the sparsity parameter varies for each nonextensive scenario, GUP consistently lowers the radiation sparsity in all circumstances toward the end of the evaporation process. Even though multiple nonextensive scenarios have the same temperatures and entropic profiles, we have demonstrated that the sparsity parameter can be used to {\it distinguish} between them.


We have introduced GUP and GUP-corrected thermodynamic parameters and have revised otherwise well-known GUP corrected quantities to a better form in which the two crucial limits - the extensivity limit for $\lambda \to 0$ and the HUP limit for $\alpha \to 0$ - are easily identified. Even though GUP corrections on R\'enyi entropy in black hole thermodynamics have been researched in the literature, we presented a full discussion of it in order to help readers distinguish between various sorts of nonextensive scenarios. Additionally, we have provided non-perturbative results for each quantity, with a focus on the R\'enyi sparsity parameter, which rises (as shown by the "bump" in Fig. (\ref{fig:etarenyi})) before the value of the remnant mass. This is because it is assumed that the area can change as a result of the GUP-modified Bekenstein entropy, which is explicitly shown in (\ref{etagup}). This indicates that $A_{GUP}$ as well as $T_{GUP}$ have an impact on the sparsity parameter. Furthermore, we have introduced black hole mass scale $M_c=m_p/2\sqrt{\pi\lambda}$ for the nonextensive parameter $\lambda$ for the R\'enyi black hole quantities and we defined corresponding characteristic length for $\lambda$ in terms of $M_c$, i.e. $\mathcal{L}_R=GM_c/c^2=2l_p\sqrt{\pi \lambda}$. We have shown that, for $M>M_c$, the heat capacity is positive and hence black holes in R\'enyi scenario are thermodynamically stable, while for $M<M_c$, the heat capacity is negative and $S_R$ and $T_R$ behave like Bekenstein entropy $S_B$ and Hawking temperature $T_H$, hence unstable black holes. 

Similarly, we have also analyzed the thermodynamic black hole quantities associated with Tsallis-Cirto black hole entropy. Particularly, we have focused on GUP corrections and the sparsity of the Tsallis-Cirto radiation. We have shown that, when GUP corrections are included, Tsallis-Cirto entropy and associated temperature have a finite value, and this proves that the final state of the black hole is also a remnant with finite entropy and temperature. It is interesting to note that the Tsallis-Cirto parameter $\delta$ plays a significant role. We have found that, for $\delta > 1/2$, Tsallis-Cirto entropy and temperature behave similarly to Bekenstein entropy and Hawking temperature, and hence have negative heat capacity. For the GUP case, Tsallis-Cirto temperature behaves like R\'enyi temperature and has positive heat capacity for $\delta<1/2$. This means that, in this framework, we must have $\delta<1/2$ for thermodynamic stability of black holes. In this way, we have shown that the Tsallis-Cirto sparsity parameter is very high during the start of the evaporation for $\delta>1$, but it approaches zero at the the end of the black hole evaporation. On the contrary, for $\delta<1$, we have shown that the Tsallis-Cirto radiation is not sparse during the start of the evaporation, but at the end of the evaporation, the sparsity parameter becomes infinite and hence shows the highly sparse Tsallis-Cirto radiation. The behavior of the GUP case is initially the same as that of the non-GUP case, but as the mass approaches the order of Planck mass, i.e., $M_r$, the Tsallis-Cirto sparsity parameter for each case reduces to some finite values. It should be noted that all of these finite sparsity parameter values are less than the sparsity parameter $\eta_H$ for the standard Hawking case.

We have also shown that the behavior of the temperature and the entropy for the Sharma-Mittal case is comparable to that of $S_B$ and $S_R$ and $T_H$ and $T_R$ for different values of the Sharma-Mittal parameter $R$ since the Sharma-Mittal entropy is the extension of the Tsallis and R\'enyi entropy.
Also, in this instance, the black hole does not evaporate, and the evaporation process stops at $M_r$, leaving the black hole in its ultimate state as a remnant of mass $M_r$ with finite entropy and temperature. We have analysed the sparsity of the Sharma-Mittal radiation and compared it with the standard Hawking case. We have found that the sparsity of the Sharma-Mittal radiation behaves similarly to the R\'enyi radiation in both non-GUP and GUP instances for values of $\lambda$ and $R$ that fulfill the condition $\lambda-2R>0$. This indicates that the Sharma-Mittal radiation is initially not sparse and that by the end of the evaporation, its value approaches that of Hawking's scenario, or $\eta_H$, for the non-GUP case. When $M$ approaches $M_r$ for the GUP case, the Sharma-Mittal sparsity parameter approaches a finite value that is smaller than $\eta_H$. For the case, $R>\lambda$, we have shown that the Sharma-Mittal sparsity parameter is initially larger than $\eta_H$ and its value exactly approaches $\eta_H$ by the end of the evaporation whereas for the case of GUP, it approaches a finite value that is smaller than $\eta_H$. It is noteworthy to notice that, for $\alpha>0$, the GUP modified sparsity parameter is always lower than the standard Hawking case. Moreover, we have also introduced the characteristic mass scale, $M_c=m_p/2\sqrt{\pi(\lambda-2R})$, for the Sharma-Mittal scenario and also, defined the corresponding characteristic length scale $\mathcal{L}_{SM}=GM_c/c^2=2l_p\sqrt{\pi(\lambda-2R)}$. We have shown that, for $M>M_c$ with $\lambda-2R>0$, the black holes are thermodynamically stable in the Sharma-Mittal scenario for both GUP and non-GUP cases, while for $M<M_c$, black holes are thermodynamically unstable.
 
We have also examined the Kaniadakis thermodynamic black hole quantities, and the results demonstrate that, with a little variation depending on the parameter $K$, Kaniadakis entropy and temperature behave similarly to Bekenstein entropy and Hawking temperature. In the case of the GUP, both quantities reach a finite value as black hole mass approaches $M_r$ during the late stages of the black hole evaporation process. It results in negative heat capacity for all values of $K$, indicating that Kaniadakis black holes are thermodynamically {\it unstable} for all values of black hole mass.
Furthermore, in contrast to the typical Hawking example, the sparsity parameter for the Kaniadakis instance is consistently high from the beginning of the evaporation process. For the non-GUP example, however, $\eta_K$ approaches the value of $\eta_H$ at the end of the evaporation. In the GUP situation, it approaches some finite value of sparsity when $M$ approaches $M_r$, which is always smaller than the sparsity parameter $\eta_H$. Additionally, it is clear that a rise in the value of $K$ causes the Kaniadakis radiation to become sparser.  

Our short look onto the Barrow entropy has proven its equivalence (though in a restricted range of parameters) to the Tsallis-Cirto entropy. In view of that, all the discussion of thermodynamical quantities for Barrow entropy should be the same as for Tsallis-Cirto.

 The main assumption of a nonextensive setup is based on considering Bekenstein entropy as Tsallis entropy. Therefore in calculations for sparsities, a Planckian distribution is assumed for all nonextensive entropies. Therefore a more in-depth study of sparsities of nonextensive statistics can be done by considering corresponding proper statistics. However, Planck distribution is the only methodological way to obtain the temperature as surface gravity. Therefore assuming a different statistical distribution(such as q-distribution for Tsallis statistics) would again be an educated guess at best. Thus, current calculations for sparsities are adequate for qualitative phenomenological assessment. 
 
 It is also worth mentioning that we have consequently defined in the paper the temperatures which were related to the appropriate entropies. It is the fact that some of the temperatures and entropies are generalising others (like Sharma-Mittal which generalises Tsallis and R\'enyi. An open issue remains as which of these temperatures and entropies have firm physical relevance. This problem will be addressed elsewhere.

Finally, it is important to emphasize that our conclusions mainly apply to the simplest spherically symmetric Schwarzschild black holes and may not be appropriate for physically more advanced (rotating, hairy, multidimensional etc.) objects automatically.

\section*{Acknowledgments}
The work of I.C. and M.P.D. was supported by the Polish National Science Centre grant No. DEC-2020/39/O/ST2/02323. 

\bibliography{ref}{}

\begin{thebibliography}{108}%
\makeatletter
\providecommand \@ifxundefined [1]{%
 \@ifx{#1\undefined}
}%
\providecommand \@ifnum [1]{%
 \ifnum #1\expandafter \@firstoftwo
 \else \expandafter \@secondoftwo
 \fi
}%
\providecommand \@ifx [1]{%
 \ifx #1\expandafter \@firstoftwo
 \else \expandafter \@secondoftwo
 \fi
}%
\providecommand \natexlab [1]{#1}%
\providecommand \enquote  [1]{``#1''}%
\providecommand \bibnamefont  [1]{#1}%
\providecommand \bibfnamefont [1]{#1}%
\providecommand \citenamefont [1]{#1}%
\providecommand \href@noop [0]{\@secondoftwo}%
\providecommand \href [0]{\begingroup \@sanitize@url \@href}%
\providecommand \@href[1]{\@@startlink{#1}\@@href}%
\providecommand \@@href[1]{\endgroup#1\@@endlink}%
\providecommand \@sanitize@url [0]{\catcode `\\12\catcode `\$12\catcode
  `\&12\catcode `\#12\catcode `\^12\catcode `\_12\catcode `\%12\relax}%
\providecommand \@@startlink[1]{}%
\providecommand \@@endlink[0]{}%
\providecommand \url  [0]{\begingroup\@sanitize@url \@url }%
\providecommand \@url [1]{\endgroup\@href {#1}{\urlprefix }}%
\providecommand \urlprefix  [0]{URL }%
\providecommand \Eprint [0]{\href }%
\providecommand \doibase [0]{http://dx.doi.org/}%
\providecommand \selectlanguage [0]{\@gobble}%
\providecommand \bibinfo  [0]{\@secondoftwo}%
\providecommand \bibfield  [0]{\@secondoftwo}%
\providecommand \translation [1]{[#1]}%
\providecommand \BibitemOpen [0]{}%
\providecommand \bibitemStop [0]{}%
\providecommand \bibitemNoStop [0]{.\EOS\space}%
\providecommand \EOS [0]{\spacefactor3000\relax}%
\providecommand \BibitemShut  [1]{\csname bibitem#1\endcsname}%
\let\auto@bib@innerbib\@empty
\bibitem [{\citenamefont {Hawking}(1974)}]{Hawking:1974rv}%
  \BibitemOpen
  \bibfield  {author} {\bibinfo {author} {\bibfnamefont {S.~W.}\ \bibnamefont
  {Hawking}},\ }\href {\doibase 10.1038/248030a0} {\bibfield  {journal}
  {\bibinfo  {journal} {Nature}\ }\textbf {\bibinfo {volume} {248}},\ \bibinfo
  {pages} {30} (\bibinfo {year} {1974})}\BibitemShut {NoStop}%
\bibitem [{\citenamefont {Bekenstein}(1973)}]{Bekenstein:1973ur}%
  \BibitemOpen
  \bibfield  {author} {\bibinfo {author} {\bibfnamefont {J.~D.}\ \bibnamefont
  {Bekenstein}},\ }\href {\doibase 10.1103/PhysRevD.7.2333} {\bibfield
  {journal} {\bibinfo  {journal} {Phys. Rev. D}\ }\textbf {\bibinfo {volume}
  {7}},\ \bibinfo {pages} {2333} (\bibinfo {year} {1973})}\BibitemShut
  {NoStop}%
\bibitem [{\citenamefont {Hawking}(1976{\natexlab{a}})}]{Hawking:1976ra}%
  \BibitemOpen
  \bibfield  {author} {\bibinfo {author} {\bibfnamefont {S.~W.}\ \bibnamefont
  {Hawking}},\ }\href {\doibase 10.1103/PhysRevD.14.2460} {\bibfield  {journal}
  {\bibinfo  {journal} {Phys. Rev. D}\ }\textbf {\bibinfo {volume} {14}},\
  \bibinfo {pages} {2460} (\bibinfo {year} {1976}{\natexlab{a}})}\BibitemShut
  {NoStop}%
\bibitem [{\citenamefont {Chen}\ \emph {et~al.}(2015)\citenamefont {Chen},
  \citenamefont {Ong},\ and\ \citenamefont {Yeom}}]{Chen:2014jwq}%
  \BibitemOpen
  \bibfield  {author} {\bibinfo {author} {\bibfnamefont {P.}~\bibnamefont
  {Chen}}, \bibinfo {author} {\bibfnamefont {Y.~C.}\ \bibnamefont {Ong}}, \
  and\ \bibinfo {author} {\bibfnamefont {D.-h.}\ \bibnamefont {Yeom}},\ }\href
  {\doibase 10.1016/j.physrep.2015.10.007} {\bibfield  {journal} {\bibinfo
  {journal} {Phys. Rept.}\ }\textbf {\bibinfo {volume} {603}},\ \bibinfo
  {pages} {1} (\bibinfo {year} {2015})},\ \Eprint
  {http://arxiv.org/abs/1412.8366} {arXiv:1412.8366 [gr-qc]} \BibitemShut
  {NoStop}%
\bibitem [{\citenamefont {Unruh}\ and\ \citenamefont
  {Wald}(2017)}]{Unruh:2017uaw}%
  \BibitemOpen
  \bibfield  {author} {\bibinfo {author} {\bibfnamefont {W.~G.}\ \bibnamefont
  {Unruh}}\ and\ \bibinfo {author} {\bibfnamefont {R.~M.}\ \bibnamefont
  {Wald}},\ }\href {\doibase 10.1088/1361-6633/aa778e} {\bibfield  {journal}
  {\bibinfo  {journal} {Rept. Prog. Phys.}\ }\textbf {\bibinfo {volume} {80}},\
  \bibinfo {pages} {092002} (\bibinfo {year} {2017})},\ \Eprint
  {http://arxiv.org/abs/1703.02140} {arXiv:1703.02140 [hep-th]} \BibitemShut
  {NoStop}%
\bibitem [{\citenamefont {Bardeen}\ \emph {et~al.}(1973)\citenamefont
  {Bardeen}, \citenamefont {Carter},\ and\ \citenamefont
  {Hawking}}]{Bardeen:1973gs}%
  \BibitemOpen
  \bibfield  {author} {\bibinfo {author} {\bibfnamefont {J.~M.}\ \bibnamefont
  {Bardeen}}, \bibinfo {author} {\bibfnamefont {B.}~\bibnamefont {Carter}}, \
  and\ \bibinfo {author} {\bibfnamefont {S.~W.}\ \bibnamefont {Hawking}},\
  }\href {\doibase 10.1007/BF01645742} {\bibfield  {journal} {\bibinfo
  {journal} {Commun. Math. Phys.}\ }\textbf {\bibinfo {volume} {31}},\ \bibinfo
  {pages} {161} (\bibinfo {year} {1973})}\BibitemShut {NoStop}%
\bibitem [{\citenamefont {Gibbons}\ and\ \citenamefont
  {Perry}(1978)}]{Gibbons:1976pt}%
  \BibitemOpen
  \bibfield  {author} {\bibinfo {author} {\bibfnamefont {G.~W.}\ \bibnamefont
  {Gibbons}}\ and\ \bibinfo {author} {\bibfnamefont {M.~J.}\ \bibnamefont
  {Perry}},\ }\href {\doibase 10.1098/rspa.1978.0022} {\bibfield  {journal}
  {\bibinfo  {journal} {Proc. Roy. Soc. Lond. A}\ }\textbf {\bibinfo {volume}
  {358}},\ \bibinfo {pages} {467} (\bibinfo {year} {1978})}\BibitemShut
  {NoStop}%
\bibitem [{\citenamefont {Hawking}\ and\ \citenamefont
  {Page}(1983)}]{Hawking:1982dh}%
  \BibitemOpen
  \bibfield  {author} {\bibinfo {author} {\bibfnamefont {S.~W.}\ \bibnamefont
  {Hawking}}\ and\ \bibinfo {author} {\bibfnamefont {D.~N.}\ \bibnamefont
  {Page}},\ }\href {\doibase 10.1007/BF01208266} {\bibfield  {journal}
  {\bibinfo  {journal} {Commun. Math. Phys.}\ }\textbf {\bibinfo {volume}
  {87}},\ \bibinfo {pages} {577} (\bibinfo {year} {1983})}\BibitemShut
  {NoStop}%
\bibitem [{\citenamefont {Hawking}(1976{\natexlab{b}})}]{Hawkingnew1}%
  \BibitemOpen
  \bibfield  {author} {\bibinfo {author} {\bibfnamefont {S.~W.}\ \bibnamefont
  {Hawking}},\ }\href {\doibase 10.1103/PhysRevD.13.191} {\bibfield  {journal}
  {\bibinfo  {journal} {Phys. Rev. D}\ }\textbf {\bibinfo {volume} {13}},\
  \bibinfo {pages} {191} (\bibinfo {year} {1976}{\natexlab{b}})}\BibitemShut
  {NoStop}%
\bibitem [{\citenamefont {Hawking}(1975)}]{Hawking:1975vcx}%
  \BibitemOpen
  \bibfield  {author} {\bibinfo {author} {\bibfnamefont {S.~W.}\ \bibnamefont
  {Hawking}},\ }\href {\doibase 10.1007/BF02345020} {\bibfield  {journal}
  {\bibinfo  {journal} {Commun. Math. Phys.}\ }\textbf {\bibinfo {volume}
  {43}},\ \bibinfo {pages} {199} (\bibinfo {year} {1975})},\ \bibinfo {note}
  {[Erratum: Commun.Math.Phys. 46, 206 (1976)]}\BibitemShut {NoStop}%
\bibitem [{\citenamefont {Jacobson}(1995)}]{Jacobson:1995ab}%
  \BibitemOpen
  \bibfield  {author} {\bibinfo {author} {\bibfnamefont {T.}~\bibnamefont
  {Jacobson}},\ }\href {\doibase 10.1103/PhysRevLett.75.1260} {\bibfield
  {journal} {\bibinfo  {journal} {Phys. Rev. Lett.}\ }\textbf {\bibinfo
  {volume} {75}},\ \bibinfo {pages} {1260} (\bibinfo {year} {1995})},\ \Eprint
  {http://arxiv.org/abs/gr-qc/9504004} {arXiv:gr-qc/9504004} \BibitemShut
  {NoStop}%
\bibitem [{\citenamefont {Verlinde}(2011)}]{Verlinde:2010hp}%
  \BibitemOpen
  \bibfield  {author} {\bibinfo {author} {\bibfnamefont {E.~P.}\ \bibnamefont
  {Verlinde}},\ }\href {\doibase 10.1007/JHEP04(2011)029} {\bibfield  {journal}
  {\bibinfo  {journal} {JHEP}\ }\textbf {\bibinfo {volume} {04}},\ \bibinfo
  {pages} {029} (\bibinfo {year} {2011})},\ \Eprint
  {http://arxiv.org/abs/1001.0785} {arXiv:1001.0785 [hep-th]} \BibitemShut
  {NoStop}%
\bibitem [{\citenamefont {Padmanabhan}(2010)}]{Padmanabhan:2009kr}%
  \BibitemOpen
  \bibfield  {author} {\bibinfo {author} {\bibfnamefont {T.}~\bibnamefont
  {Padmanabhan}},\ }\href {\doibase 10.1142/S021773231003313X} {\bibfield
  {journal} {\bibinfo  {journal} {Mod. Phys. Lett. A}\ }\textbf {\bibinfo
  {volume} {25}},\ \bibinfo {pages} {1129} (\bibinfo {year} {2010})},\ \Eprint
  {http://arxiv.org/abs/0912.3165} {arXiv:0912.3165 [gr-qc]} \BibitemShut
  {NoStop}%
\bibitem [{\citenamefont {Kubiznak}\ and\ \citenamefont
  {Mann}(2015)}]{Kubiznak:2014zwa}%
  \BibitemOpen
  \bibfield  {author} {\bibinfo {author} {\bibfnamefont {D.}~\bibnamefont
  {Kubiznak}}\ and\ \bibinfo {author} {\bibfnamefont {R.~B.}\ \bibnamefont
  {Mann}},\ }\href {\doibase 10.1139/cjp-2014-0465} {\bibfield  {journal}
  {\bibinfo  {journal} {Can. J. Phys.}\ }\textbf {\bibinfo {volume} {93}},\
  \bibinfo {pages} {999} (\bibinfo {year} {2015})},\ \Eprint
  {http://arxiv.org/abs/1404.2126} {arXiv:1404.2126 [gr-qc]} \BibitemShut
  {NoStop}%
\bibitem [{\citenamefont {Cvetic}\ \emph {et~al.}(2011)\citenamefont {Cvetic},
  \citenamefont {Gibbons}, \citenamefont {Kubiznak},\ and\ \citenamefont
  {Pope}}]{Cvetic:2010jb}%
  \BibitemOpen
  \bibfield  {author} {\bibinfo {author} {\bibfnamefont {M.}~\bibnamefont
  {Cvetic}}, \bibinfo {author} {\bibfnamefont {G.~W.}\ \bibnamefont {Gibbons}},
  \bibinfo {author} {\bibfnamefont {D.}~\bibnamefont {Kubiznak}}, \ and\
  \bibinfo {author} {\bibfnamefont {C.~N.}\ \bibnamefont {Pope}},\ }\href
  {\doibase 10.1103/PhysRevD.84.024037} {\bibfield  {journal} {\bibinfo
  {journal} {Phys. Rev. D}\ }\textbf {\bibinfo {volume} {84}},\ \bibinfo
  {pages} {024037} (\bibinfo {year} {2011})},\ \Eprint
  {http://arxiv.org/abs/1012.2888} {arXiv:1012.2888 [hep-th]} \BibitemShut
  {NoStop}%
\bibitem [{\citenamefont {Caldarelli}\ \emph {et~al.}(2000)\citenamefont
  {Caldarelli}, \citenamefont {Cognola},\ and\ \citenamefont
  {Klemm}}]{Caldarelli:1999xj}%
  \BibitemOpen
  \bibfield  {author} {\bibinfo {author} {\bibfnamefont {M.~M.}\ \bibnamefont
  {Caldarelli}}, \bibinfo {author} {\bibfnamefont {G.}~\bibnamefont {Cognola}},
  \ and\ \bibinfo {author} {\bibfnamefont {D.}~\bibnamefont {Klemm}},\ }\href
  {\doibase 10.1088/0264-9381/17/2/310} {\bibfield  {journal} {\bibinfo
  {journal} {Class. Quant. Grav.}\ }\textbf {\bibinfo {volume} {17}},\ \bibinfo
  {pages} {399} (\bibinfo {year} {2000})},\ \Eprint
  {http://arxiv.org/abs/hep-th/9908022} {arXiv:hep-th/9908022} \BibitemShut
  {NoStop}%
\bibitem [{\citenamefont {Cai}\ and\ \citenamefont {Kim}(2005)}]{Cai:2005ra}%
  \BibitemOpen
  \bibfield  {author} {\bibinfo {author} {\bibfnamefont {R.-G.}\ \bibnamefont
  {Cai}}\ and\ \bibinfo {author} {\bibfnamefont {S.~P.}\ \bibnamefont {Kim}},\
  }\href {\doibase 10.1088/1126-6708/2005/02/050} {\bibfield  {journal}
  {\bibinfo  {journal} {JHEP}\ }\textbf {\bibinfo {volume} {02}},\ \bibinfo
  {pages} {050} (\bibinfo {year} {2005})},\ \Eprint
  {http://arxiv.org/abs/hep-th/0501055} {arXiv:hep-th/0501055} \BibitemShut
  {NoStop}%
\bibitem [{\citenamefont {Davies}(1977)}]{Davies:1977bgr}%
  \BibitemOpen
  \bibfield  {author} {\bibinfo {author} {\bibfnamefont {P.~C.~W.}\
  \bibnamefont {Davies}},\ }\href {\doibase 10.1098/rspa.1977.0047} {\bibfield
  {journal} {\bibinfo  {journal} {Proc. Roy. Soc. Lond. A}\ }\textbf {\bibinfo
  {volume} {353}},\ \bibinfo {pages} {499} (\bibinfo {year}
  {1977})}\BibitemShut {NoStop}%
\bibitem [{\citenamefont {Dolan}(2012)}]{Dolan:2012jh}%
  \BibitemOpen
  \bibfield  {author} {\bibinfo {author} {\bibfnamefont {B.~P.}\ \bibnamefont
  {Dolan}},\ }\enquote {\bibinfo {title} {{Where Is the PdV in the First Law of
  Black Hole Thermodynamics?}}}\ \ (\bibinfo  {publisher} {INTECH},\ \bibinfo
  {year} {2012})\ \Eprint {http://arxiv.org/abs/1209.1272} {arXiv:1209.1272
  [gr-qc]} \BibitemShut {NoStop}%
\bibitem [{\citenamefont {Easson}\ \emph {et~al.}(2011)\citenamefont {Easson},
  \citenamefont {Frampton},\ and\ \citenamefont {Smoot}}]{Easson:2010av}%
  \BibitemOpen
  \bibfield  {author} {\bibinfo {author} {\bibfnamefont {D.~A.}\ \bibnamefont
  {Easson}}, \bibinfo {author} {\bibfnamefont {P.~H.}\ \bibnamefont
  {Frampton}}, \ and\ \bibinfo {author} {\bibfnamefont {G.~F.}\ \bibnamefont
  {Smoot}},\ }\href {\doibase 10.1016/j.physletb.2010.12.025} {\bibfield
  {journal} {\bibinfo  {journal} {Phys. Lett. B}\ }\textbf {\bibinfo {volume}
  {696}},\ \bibinfo {pages} {273} (\bibinfo {year} {2011})},\ \Eprint
  {http://arxiv.org/abs/1002.4278} {arXiv:1002.4278 [hep-th]} \BibitemShut
  {NoStop}%
\bibitem [{\citenamefont {Hawking}(1971)}]{PhysRevLett.26.1344}%
  \BibitemOpen
  \bibfield  {author} {\bibinfo {author} {\bibfnamefont {S.~W.}\ \bibnamefont
  {Hawking}},\ }\href {\doibase 10.1103/PhysRevLett.26.1344} {\bibfield
  {journal} {\bibinfo  {journal} {Phys. Rev. Lett.}\ }\textbf {\bibinfo
  {volume} {26}},\ \bibinfo {pages} {1344} (\bibinfo {year}
  {1971})}\BibitemShut {NoStop}%
\bibitem [{\citenamefont {Landsberg}(1999)}]{soton29487}%
  \BibitemOpen
  \bibfield  {author} {\bibinfo {author} {\bibfnamefont {P.}~\bibnamefont
  {Landsberg}},\ }\href {https://eprints.soton.ac.uk/29487/} {\bibfield
  {journal} {\bibinfo  {journal} {Brazilian Journal of Physics}\ }\textbf
  {\bibinfo {volume} {29}},\ \bibinfo {pages} {46} (\bibinfo {year}
  {1999})}\BibitemShut {NoStop}%
\bibitem [{\citenamefont {Swendsen}(2011)}]{Swendsen2011}%
  \BibitemOpen
  \bibfield  {author} {\bibinfo {author} {\bibfnamefont {R.~H.}\ \bibnamefont
  {Swendsen}},\ }\href {\doibase 10.1119/1.3536633} {\bibfield  {journal}
  {\bibinfo  {journal} {American Journal of Physics}\ }\textbf {\bibinfo
  {volume} {79}},\ \bibinfo {pages} {342} (\bibinfo {year} {2011})},\ \Eprint
  {http://arxiv.org/abs/https://doi.org/10.1119/1.3536633}
  {https://doi.org/10.1119/1.3536633} \BibitemShut {NoStop}%
\bibitem [{\citenamefont {Mannaerts}(2014)}]{Mannaerts_2014}%
  \BibitemOpen
  \bibfield  {author} {\bibinfo {author} {\bibfnamefont {S.~H.}\ \bibnamefont
  {Mannaerts}},\ }\href {\doibase 10.1088/0143-0807/35/3/035017} {\bibfield
  {journal} {\bibinfo  {journal} {European Journal of Physics}\ }\textbf
  {\bibinfo {volume} {35}},\ \bibinfo {pages} {035017} (\bibinfo {year}
  {2014})}\BibitemShut {NoStop}%
\bibitem [{\citenamefont {Tsallis}(1988)}]{Tsallis:1987eu}%
  \BibitemOpen
  \bibfield  {author} {\bibinfo {author} {\bibfnamefont {C.}~\bibnamefont
  {Tsallis}},\ }\href {\doibase 10.1007/BF01016429} {\bibfield  {journal}
  {\bibinfo  {journal} {J. Statist. Phys.}\ }\textbf {\bibinfo {volume} {52}},\
  \bibinfo {pages} {479} (\bibinfo {year} {1988})}\BibitemShut {NoStop}%
\bibitem [{\citenamefont {Tsallis}\ \emph {et~al.}(1998)\citenamefont
  {Tsallis}, \citenamefont {Mendes},\ and\ \citenamefont
  {Plastino}}]{TSALLIS1998534}%
  \BibitemOpen
  \bibfield  {author} {\bibinfo {author} {\bibfnamefont {C.}~\bibnamefont
  {Tsallis}}, \bibinfo {author} {\bibfnamefont {R.}~\bibnamefont {Mendes}}, \
  and\ \bibinfo {author} {\bibfnamefont {A.}~\bibnamefont {Plastino}},\ }\href
  {\doibase https://doi.org/10.1016/S0378-4371(98)00437-3} {\bibfield
  {journal} {\bibinfo  {journal} {Physica A: Statistical Mechanics and its
  Applications}\ }\textbf {\bibinfo {volume} {261}},\ \bibinfo {pages} {534}
  (\bibinfo {year} {1998})}\BibitemShut {NoStop}%
\bibitem [{\citenamefont {Tsallis}(2009)}]{tsallisbook}%
  \BibitemOpen
  \bibfield  {author} {\bibinfo {author} {\bibfnamefont {C.}~\bibnamefont
  {Tsallis}},\ }\href {\doibase 10.1007/978-0-387-85359-8} {\emph {\bibinfo
  {title} {Introduction to Nonextensive Statistical Mechanics: Approaching a
  Complex World}}}\ (\bibinfo  {publisher} {Springer New York, NY},\ \bibinfo
  {year} {2009})\BibitemShut {NoStop}%
\bibitem [{\citenamefont {Abe}\ \emph {et~al.}(2001)\citenamefont {Abe},
  \citenamefont {Mart{\i}nez}, \citenamefont {Pennini},\ and\ \citenamefont
  {Plastino}}]{Abe_2001}%
  \BibitemOpen
  \bibfield  {author} {\bibinfo {author} {\bibfnamefont {S.}~\bibnamefont
  {Abe}}, \bibinfo {author} {\bibfnamefont {S.}~\bibnamefont {Mart{\i}nez}},
  \bibinfo {author} {\bibfnamefont {F.}~\bibnamefont {Pennini}}, \ and\
  \bibinfo {author} {\bibfnamefont {A.}~\bibnamefont {Plastino}},\ }\href
  {\doibase 10.1016/s0375-9601(01)00127-x} {\bibfield  {journal} {\bibinfo
  {journal} {Physics Letters A}\ }\textbf {\bibinfo {volume} {281}},\ \bibinfo
  {pages} {126} (\bibinfo {year} {2001})}\BibitemShut {NoStop}%
\bibitem [{\citenamefont {Abe}\ and\ \citenamefont
  {Rajagopal}(2001)}]{Abe_2001a}%
  \BibitemOpen
  \bibfield  {author} {\bibinfo {author} {\bibfnamefont {S.}~\bibnamefont
  {Abe}}\ and\ \bibinfo {author} {\bibfnamefont {A.~K.}\ \bibnamefont
  {Rajagopal}},\ }\href {\doibase 10.1209/epl/i2001-00373-4} {\bibfield
  {journal} {\bibinfo  {journal} {Europhysics Letters ({EPL})}\ }\textbf
  {\bibinfo {volume} {55}},\ \bibinfo {pages} {6} (\bibinfo {year}
  {2001})}\BibitemShut {NoStop}%
\bibitem [{\citenamefont {Abe}(2001)}]{Abe_2001c}%
  \BibitemOpen
  \bibfield  {author} {\bibinfo {author} {\bibfnamefont {S.}~\bibnamefont
  {Abe}},\ }\href {\doibase 10.1103/physreve.63.061105} {\bibfield  {journal}
  {\bibinfo  {journal} {Physical Review E}\ }\textbf {\bibinfo {volume} {63}}
  (\bibinfo {year} {2001}),\ 10.1103/physreve.63.061105}\BibitemShut {NoStop}%
\bibitem [{\citenamefont {Bir{\'{o}}}\ and\ \citenamefont
  {V{\'{a}}n}(2011)}]{Bir__2011}%
  \BibitemOpen
  \bibfield  {author} {\bibinfo {author} {\bibfnamefont {T.~S.}\ \bibnamefont
  {Bir{\'{o}}}}\ and\ \bibinfo {author} {\bibfnamefont {P.}~\bibnamefont
  {V{\'{a}}n}},\ }\href {\doibase 10.1103/physreve.83.061147} {\bibfield
  {journal} {\bibinfo  {journal} {Physical Review E}\ }\textbf {\bibinfo
  {volume} {83}} (\bibinfo {year} {2011}),\
  10.1103/physreve.83.061147}\BibitemShut {NoStop}%
\bibitem [{\citenamefont {Nauenberg}(2003)}]{PhysRevE.67.036114}%
  \BibitemOpen
  \bibfield  {author} {\bibinfo {author} {\bibfnamefont {M.}~\bibnamefont
  {Nauenberg}},\ }\href {\doibase 10.1103/PhysRevE.67.036114} {\bibfield
  {journal} {\bibinfo  {journal} {Phys. Rev. E}\ }\textbf {\bibinfo {volume}
  {67}},\ \bibinfo {pages} {036114} (\bibinfo {year} {2003})}\BibitemShut
  {NoStop}%
\bibitem [{\citenamefont {Bir\'o}\ and\ \citenamefont
  {V\'an}(2011)}]{PhysRevE.83.061147}%
  \BibitemOpen
  \bibfield  {author} {\bibinfo {author} {\bibfnamefont {T.~S.}\ \bibnamefont
  {Bir\'o}}\ and\ \bibinfo {author} {\bibfnamefont {P.}~\bibnamefont {V\'an}},\
  }\href {\doibase 10.1103/PhysRevE.83.061147} {\bibfield  {journal} {\bibinfo
  {journal} {Phys. Rev. E}\ }\textbf {\bibinfo {volume} {83}},\ \bibinfo
  {pages} {061147} (\bibinfo {year} {2011})}\BibitemShut {NoStop}%
\bibitem [{\citenamefont {Parvan}\ and\ \citenamefont
  {Biro}(2005)}]{Parvan:2004vn}%
  \BibitemOpen
  \bibfield  {author} {\bibinfo {author} {\bibfnamefont {A.~S.}\ \bibnamefont
  {Parvan}}\ and\ \bibinfo {author} {\bibfnamefont {T.~S.}\ \bibnamefont
  {Biro}},\ }\href {\doibase 10.1016/j.physleta.2005.04.036} {\bibfield
  {journal} {\bibinfo  {journal} {Phys. Lett. A}\ }\textbf {\bibinfo {volume}
  {340}},\ \bibinfo {pages} {375} (\bibinfo {year} {2005})},\ \Eprint
  {http://arxiv.org/abs/hep-ph/0407131} {arXiv:hep-ph/0407131} \BibitemShut
  {NoStop}%
\bibitem [{\citenamefont {Tsallis}\ and\ \citenamefont
  {Cirto}(2013)}]{Tsallis:2012js}%
  \BibitemOpen
  \bibfield  {author} {\bibinfo {author} {\bibfnamefont {C.}~\bibnamefont
  {Tsallis}}\ and\ \bibinfo {author} {\bibfnamefont {L.~J.~L.}\ \bibnamefont
  {Cirto}},\ }\href {\doibase 10.1140/epjc/s10052-013-2487-6} {\bibfield
  {journal} {\bibinfo  {journal} {Eur. Phys. J. C}\ }\textbf {\bibinfo {volume}
  {73}},\ \bibinfo {pages} {2487} (\bibinfo {year} {2013})},\ \Eprint
  {http://arxiv.org/abs/1202.2154} {arXiv:1202.2154 [cond-mat.stat-mech]}
  \BibitemShut {NoStop}%
\bibitem [{\citenamefont {Bir\'o}\ and\ \citenamefont
  {Czinner}(2013)}]{Biro:2013cra}%
  \BibitemOpen
  \bibfield  {author} {\bibinfo {author} {\bibfnamefont {T.~S.}\ \bibnamefont
  {Bir\'o}}\ and\ \bibinfo {author} {\bibfnamefont {V.~G.}\ \bibnamefont
  {Czinner}},\ }\href {\doibase 10.1016/j.physletb.2013.09.032} {\bibfield
  {journal} {\bibinfo  {journal} {Phys. Lett. B}\ }\textbf {\bibinfo {volume}
  {726}},\ \bibinfo {pages} {861} (\bibinfo {year} {2013})},\ \Eprint
  {http://arxiv.org/abs/1309.4261} {arXiv:1309.4261 [gr-qc]} \BibitemShut
  {NoStop}%
\bibitem [{\citenamefont {Czinner}(2015)}]{Czinner:2015ena}%
  \BibitemOpen
  \bibfield  {author} {\bibinfo {author} {\bibfnamefont {V.~G.}\ \bibnamefont
  {Czinner}},\ }\href {\doibase 10.1142/S0218271815420158} {\bibfield
  {journal} {\bibinfo  {journal} {Int. J. Mod. Phys. D}\ }\textbf {\bibinfo
  {volume} {24}},\ \bibinfo {pages} {1542015} (\bibinfo {year}
  {2015})}\BibitemShut {NoStop}%
\bibitem [{\citenamefont {Czinner}\ and\ \citenamefont
  {Iguchi}(2017{\natexlab{a}})}]{Czinner:2017bwc}%
  \BibitemOpen
  \bibfield  {author} {\bibinfo {author} {\bibfnamefont {V.~G.}\ \bibnamefont
  {Czinner}}\ and\ \bibinfo {author} {\bibfnamefont {H.}~\bibnamefont
  {Iguchi}},\ }\href {\doibase 10.3390/universe3010014} {\bibfield  {journal}
  {\bibinfo  {journal} {Universe}\ }\textbf {\bibinfo {volume} {3}},\ \bibinfo
  {pages} {14} (\bibinfo {year} {2017}{\natexlab{a}})}\BibitemShut {NoStop}%
\bibitem [{\citenamefont {Czinner}\ and\ \citenamefont
  {Iguchi}(2016)}]{Czinner:2015eyk}%
  \BibitemOpen
  \bibfield  {author} {\bibinfo {author} {\bibfnamefont {V.~G.}\ \bibnamefont
  {Czinner}}\ and\ \bibinfo {author} {\bibfnamefont {H.}~\bibnamefont
  {Iguchi}},\ }\href {\doibase 10.1016/j.physletb.2015.11.061} {\bibfield
  {journal} {\bibinfo  {journal} {Phys. Lett. B}\ }\textbf {\bibinfo {volume}
  {752}},\ \bibinfo {pages} {306} (\bibinfo {year} {2016})},\ \Eprint
  {http://arxiv.org/abs/1511.06963} {arXiv:1511.06963 [gr-qc]} \BibitemShut
  {NoStop}%
\bibitem [{\citenamefont {Czinner}\ and\ \citenamefont
  {Iguchi}(2017{\natexlab{b}})}]{Czinner:2017tjq}%
  \BibitemOpen
  \bibfield  {author} {\bibinfo {author} {\bibfnamefont {V.~G.}\ \bibnamefont
  {Czinner}}\ and\ \bibinfo {author} {\bibfnamefont {H.}~\bibnamefont
  {Iguchi}},\ }\href {\doibase 10.1140/epjc/s10052-017-5453-x} {\bibfield
  {journal} {\bibinfo  {journal} {Eur. Phys. J. C}\ }\textbf {\bibinfo {volume}
  {77}},\ \bibinfo {pages} {892} (\bibinfo {year} {2017}{\natexlab{b}})},\
  \Eprint {http://arxiv.org/abs/1702.05341} {arXiv:1702.05341 [gr-qc]}
  \BibitemShut {NoStop}%
\bibitem [{\citenamefont {Tsallis}(2019)}]{Tsallis:2019giw}%
  \BibitemOpen
  \bibfield  {author} {\bibinfo {author} {\bibfnamefont {C.}~\bibnamefont
  {Tsallis}},\ }\href {\doibase 10.3390/e22010017} {\bibfield  {journal}
  {\bibinfo  {journal} {Entropy}\ }\textbf {\bibinfo {volume} {22}},\ \bibinfo
  {pages} {17} (\bibinfo {year} {2019})}\BibitemShut {NoStop}%
\bibitem [{\citenamefont {Alonso-Serrano}\ \emph {et~al.}(2021)\citenamefont
  {Alonso-Serrano}, \citenamefont {Dabrowski},\ and\ \citenamefont
  {Gohar}}]{Alonso-Serrano:2020hpb}%
  \BibitemOpen
  \bibfield  {author} {\bibinfo {author} {\bibfnamefont {A.}~\bibnamefont
  {Alonso-Serrano}}, \bibinfo {author} {\bibfnamefont {M.~P.}\ \bibnamefont
  {Dabrowski}}, \ and\ \bibinfo {author} {\bibfnamefont {H.}~\bibnamefont
  {Gohar}},\ }\href {\doibase 10.1103/PhysRevD.103.026021} {\bibfield
  {journal} {\bibinfo  {journal} {Phys. Rev. D}\ }\textbf {\bibinfo {volume}
  {103}},\ \bibinfo {pages} {026021} (\bibinfo {year} {2021})},\ \Eprint
  {http://arxiv.org/abs/2009.02129} {arXiv:2009.02129 [gr-qc]} \BibitemShut
  {NoStop}%
\bibitem [{\citenamefont {R\'enyi}(1959)}]{Renyi1}%
  \BibitemOpen
  \bibfield  {author} {\bibinfo {author} {\bibfnamefont {A.}~\bibnamefont
  {R\'enyi}},\ }\href {\doibase https://doi.org/10.1007/BF02063299} {\bibfield
  {journal} {\bibinfo  {journal} {Acta Mathematica Academiae Scientiarum
  Hungaricae}\ }\textbf {\bibinfo {volume} {10}},\ \bibinfo {pages} {193–215}
  (\bibinfo {year} {1959})}\BibitemShut {NoStop}%
\bibitem [{\citenamefont {Sharma}\ and\ \citenamefont {Mittal}(1977)}]{SM}%
  \BibitemOpen
  \bibfield  {author} {\bibinfo {author} {\bibfnamefont {B.~D.}\ \bibnamefont
  {Sharma}}\ and\ \bibinfo {author} {\bibfnamefont {D.~P.}\ \bibnamefont
  {Mittal}},\ }\href@noop {} {\bibfield  {journal} {\bibinfo  {journal}
  {J.Comb.Inf.Syst.Sci.}\ }\textbf {\bibinfo {volume} {2}},\ \bibinfo {pages}
  {122} (\bibinfo {year} {1977})}\BibitemShut {NoStop}%
\bibitem [{\citenamefont {Sharma}\ and\ \citenamefont
  {Mittal}()}]{sharma1975new}%
  \BibitemOpen
  \bibfield  {author} {\bibinfo {author} {\bibfnamefont {B.~D.}\ \bibnamefont
  {Sharma}}\ and\ \bibinfo {author} {\bibfnamefont {D.~P.}\ \bibnamefont
  {Mittal}},\ }\href@noop {} {\bibfield  {journal} {\bibinfo  {journal} {J.
  Math. Sci}\ }\textbf {\bibinfo {volume} {10}},\ \bibinfo {pages}
  {28}}\BibitemShut {NoStop}%
\bibitem [{\citenamefont {Kaniadakis}(2002)}]{Kaniadakis:2002zz}%
  \BibitemOpen
  \bibfield  {author} {\bibinfo {author} {\bibfnamefont {G.}~\bibnamefont
  {Kaniadakis}},\ }\href {\doibase 10.1103/PhysRevE.66.056125} {\bibfield
  {journal} {\bibinfo  {journal} {Phys. Rev. E}\ }\textbf {\bibinfo {volume}
  {66}},\ \bibinfo {pages} {056125} (\bibinfo {year} {2002})},\ \Eprint
  {http://arxiv.org/abs/cond-mat/0210467} {arXiv:cond-mat/0210467} \BibitemShut
  {NoStop}%
\bibitem [{\citenamefont {Kaniadakis}(2005)}]{Kaniadakis:2005zk}%
  \BibitemOpen
  \bibfield  {author} {\bibinfo {author} {\bibfnamefont {G.}~\bibnamefont
  {Kaniadakis}},\ }\href {\doibase 10.1103/PhysRevE.72.036108} {\bibfield
  {journal} {\bibinfo  {journal} {Phys. Rev. E}\ }\textbf {\bibinfo {volume}
  {72}},\ \bibinfo {pages} {036108} (\bibinfo {year} {2005})},\ \Eprint
  {http://arxiv.org/abs/cond-mat/0507311} {arXiv:cond-mat/0507311} \BibitemShut
  {NoStop}%
\bibitem [{\citenamefont {Barrow}(2020)}]{Barrow:2020tzx}%
  \BibitemOpen
  \bibfield  {author} {\bibinfo {author} {\bibfnamefont {J.~D.}\ \bibnamefont
  {Barrow}},\ }\href {\doibase https://doi.org/10.1016/j.physletb.2020.135643}
  {\bibfield  {journal} {\bibinfo  {journal} {Physics Letters B}\ }\textbf
  {\bibinfo {volume} {808}},\ \bibinfo {pages} {135643} (\bibinfo {year}
  {2020})}\BibitemShut {NoStop}%
\bibitem [{\citenamefont {Nojiri}\ \emph {et~al.}(2021)\citenamefont {Nojiri},
  \citenamefont {Odintsov},\ and\ \citenamefont {Faraoni}}]{Nojiri:2021czz}%
  \BibitemOpen
  \bibfield  {author} {\bibinfo {author} {\bibfnamefont {S.}~\bibnamefont
  {Nojiri}}, \bibinfo {author} {\bibfnamefont {S.~D.}\ \bibnamefont
  {Odintsov}}, \ and\ \bibinfo {author} {\bibfnamefont {V.}~\bibnamefont
  {Faraoni}},\ }\href {\doibase 10.1103/PhysRevD.104.084030} {\bibfield
  {journal} {\bibinfo  {journal} {Phys. Rev. D}\ }\textbf {\bibinfo {volume}
  {104}},\ \bibinfo {pages} {084030} (\bibinfo {year} {2021})},\ \Eprint
  {http://arxiv.org/abs/2109.05315} {arXiv:2109.05315 [gr-qc]} \BibitemShut
  {NoStop}%
\bibitem [{\citenamefont {Nojiri}\ \emph
  {et~al.}(2022{\natexlab{a}})\citenamefont {Nojiri}, \citenamefont
  {Odintsov},\ and\ \citenamefont {Faraoni}}]{Nojiri:2022aof}%
  \BibitemOpen
  \bibfield  {author} {\bibinfo {author} {\bibfnamefont {S.}~\bibnamefont
  {Nojiri}}, \bibinfo {author} {\bibfnamefont {S.~D.}\ \bibnamefont
  {Odintsov}}, \ and\ \bibinfo {author} {\bibfnamefont {V.}~\bibnamefont
  {Faraoni}},\ }\href {\doibase 10.1103/PhysRevD.105.044042} {\bibfield
  {journal} {\bibinfo  {journal} {Phys. Rev. D}\ }\textbf {\bibinfo {volume}
  {105}},\ \bibinfo {pages} {044042} (\bibinfo {year} {2022}{\natexlab{a}})},\
  \Eprint {http://arxiv.org/abs/2201.02424} {arXiv:2201.02424 [gr-qc]}
  \BibitemShut {NoStop}%
\bibitem [{\citenamefont {Nojiri}\ \emph
  {et~al.}(2022{\natexlab{b}})\citenamefont {Nojiri}, \citenamefont
  {Odintsov},\ and\ \citenamefont {Faraoni}}]{Nojiri:2022sfd}%
  \BibitemOpen
  \bibfield  {author} {\bibinfo {author} {\bibfnamefont {S.}~\bibnamefont
  {Nojiri}}, \bibinfo {author} {\bibfnamefont {S.~D.}\ \bibnamefont
  {Odintsov}}, \ and\ \bibinfo {author} {\bibfnamefont {V.}~\bibnamefont
  {Faraoni}},\ }\href {\doibase 10.1142/S0219887822502103} {\bibfield
  {journal} {\bibinfo  {journal} {Int. J. Geom. Meth. Mod. Phys.}\ }\textbf
  {\bibinfo {volume} {19}},\ \bibinfo {pages} {2250210} (\bibinfo {year}
  {2022}{\natexlab{b}})},\ \Eprint {http://arxiv.org/abs/2207.07905}
  {arXiv:2207.07905 [gr-qc]} \BibitemShut {NoStop}%
\bibitem [{\citenamefont {Nojiri}\ \emph
  {et~al.}(2022{\natexlab{c}})\citenamefont {Nojiri}, \citenamefont
  {Odintsov},\ and\ \citenamefont {Paul}}]{Nojiri:2021jxf}%
  \BibitemOpen
  \bibfield  {author} {\bibinfo {author} {\bibfnamefont {S.}~\bibnamefont
  {Nojiri}}, \bibinfo {author} {\bibfnamefont {S.~D.}\ \bibnamefont
  {Odintsov}}, \ and\ \bibinfo {author} {\bibfnamefont {T.}~\bibnamefont
  {Paul}},\ }\href {\doibase 10.1016/j.physletb.2021.136844} {\bibfield
  {journal} {\bibinfo  {journal} {Phys. Lett. B}\ }\textbf {\bibinfo {volume}
  {825}},\ \bibinfo {pages} {136844} (\bibinfo {year} {2022}{\natexlab{c}})},\
  \Eprint {http://arxiv.org/abs/2112.10159} {arXiv:2112.10159 [gr-qc]}
  \BibitemShut {NoStop}%
\bibitem [{\citenamefont {Promsiri}\ \emph {et~al.}(2020)\citenamefont
  {Promsiri}, \citenamefont {Hirunsirisawat},\ and\ \citenamefont
  {Liewrian}}]{Promsiri:2020jga}%
  \BibitemOpen
  \bibfield  {author} {\bibinfo {author} {\bibfnamefont {C.}~\bibnamefont
  {Promsiri}}, \bibinfo {author} {\bibfnamefont {E.}~\bibnamefont
  {Hirunsirisawat}}, \ and\ \bibinfo {author} {\bibfnamefont {W.}~\bibnamefont
  {Liewrian}},\ }\href {\doibase 10.1103/PhysRevD.102.064014} {\bibfield
  {journal} {\bibinfo  {journal} {Phys. Rev. D}\ }\textbf {\bibinfo {volume}
  {102}},\ \bibinfo {pages} {064014} (\bibinfo {year} {2020})},\ \Eprint
  {http://arxiv.org/abs/2003.12986} {arXiv:2003.12986 [hep-th]} \BibitemShut
  {NoStop}%
\bibitem [{\citenamefont {Promsiri}\ \emph {et~al.}(2021)\citenamefont
  {Promsiri}, \citenamefont {Hirunsirisawat},\ and\ \citenamefont
  {Liewrian}}]{Promsiri:2021hhv}%
  \BibitemOpen
  \bibfield  {author} {\bibinfo {author} {\bibfnamefont {C.}~\bibnamefont
  {Promsiri}}, \bibinfo {author} {\bibfnamefont {E.}~\bibnamefont
  {Hirunsirisawat}}, \ and\ \bibinfo {author} {\bibfnamefont {W.}~\bibnamefont
  {Liewrian}},\ }\href {\doibase 10.1103/PhysRevD.104.064004} {\bibfield
  {journal} {\bibinfo  {journal} {Phys. Rev. D}\ }\textbf {\bibinfo {volume}
  {104}},\ \bibinfo {pages} {064004} (\bibinfo {year} {2021})},\ \Eprint
  {http://arxiv.org/abs/2106.02406} {arXiv:2106.02406 [hep-th]} \BibitemShut
  {NoStop}%
\bibitem [{\citenamefont {Tannukij}\ \emph {et~al.}(2020)\citenamefont
  {Tannukij}, \citenamefont {Wongjun}, \citenamefont {Hirunsirisawat},
  \citenamefont {Deesuwan},\ and\ \citenamefont {Promsiri}}]{Tannukij:2020njz}%
  \BibitemOpen
  \bibfield  {author} {\bibinfo {author} {\bibfnamefont {L.}~\bibnamefont
  {Tannukij}}, \bibinfo {author} {\bibfnamefont {P.}~\bibnamefont {Wongjun}},
  \bibinfo {author} {\bibfnamefont {E.}~\bibnamefont {Hirunsirisawat}},
  \bibinfo {author} {\bibfnamefont {T.}~\bibnamefont {Deesuwan}}, \ and\
  \bibinfo {author} {\bibfnamefont {C.}~\bibnamefont {Promsiri}},\ }\href
  {\doibase 10.1140/epjp/s13360-020-00517-2} {\bibfield  {journal} {\bibinfo
  {journal} {Eur. Phys. J. Plus}\ }\textbf {\bibinfo {volume} {135}},\ \bibinfo
  {pages} {500} (\bibinfo {year} {2020})},\ \Eprint
  {http://arxiv.org/abs/2002.00377} {arXiv:2002.00377 [gr-qc]} \BibitemShut
  {NoStop}%
\bibitem [{\citenamefont {Nakarachinda}\ \emph {et~al.}(2021)\citenamefont
  {Nakarachinda}, \citenamefont {Hirunsirisawat}, \citenamefont {Tannukij},\
  and\ \citenamefont {Wongjun}}]{Nakarachinda:2021jxd}%
  \BibitemOpen
  \bibfield  {author} {\bibinfo {author} {\bibfnamefont {R.}~\bibnamefont
  {Nakarachinda}}, \bibinfo {author} {\bibfnamefont {E.}~\bibnamefont
  {Hirunsirisawat}}, \bibinfo {author} {\bibfnamefont {L.}~\bibnamefont
  {Tannukij}}, \ and\ \bibinfo {author} {\bibfnamefont {P.}~\bibnamefont
  {Wongjun}},\ }\href {\doibase 10.1103/PhysRevD.104.064003} {\bibfield
  {journal} {\bibinfo  {journal} {Phys. Rev. D}\ }\textbf {\bibinfo {volume}
  {104}},\ \bibinfo {pages} {064003} (\bibinfo {year} {2021})},\ \Eprint
  {http://arxiv.org/abs/2106.02838} {arXiv:2106.02838 [gr-qc]} \BibitemShut
  {NoStop}%
\bibitem [{\citenamefont {\c{C}imdiker}\ \emph {et~al.}(2022)\citenamefont
  {\c{C}imdiker}, \citenamefont {Dabrowski},\ and\ \citenamefont
  {Gohar}}]{Cimdiker:2022ics}%
  \BibitemOpen
  \bibfield  {author} {\bibinfo {author} {\bibfnamefont {I.}~\bibnamefont
  {\c{C}imdiker}}, \bibinfo {author} {\bibfnamefont {M.~P.}\ \bibnamefont
  {Dabrowski}}, \ and\ \bibinfo {author} {\bibfnamefont {H.}~\bibnamefont
  {Gohar}},\ }\href@noop {} {\  (\bibinfo {year} {2022})},\ \Eprint
  {http://arxiv.org/abs/2208.04473} {arXiv:2208.04473 [gr-qc]} \BibitemShut
  {NoStop}%
\bibitem [{\citenamefont {Promsiri}\ \emph {et~al.}(2022)\citenamefont
  {Promsiri}, \citenamefont {Hirunsirisawat},\ and\ \citenamefont
  {Nakarachinda}}]{Promsiri:2022qin}%
  \BibitemOpen
  \bibfield  {author} {\bibinfo {author} {\bibfnamefont {C.}~\bibnamefont
  {Promsiri}}, \bibinfo {author} {\bibfnamefont {E.}~\bibnamefont
  {Hirunsirisawat}}, \ and\ \bibinfo {author} {\bibfnamefont {R.}~\bibnamefont
  {Nakarachinda}},\ }\href {\doibase 10.1103/PhysRevD.105.124049} {\bibfield
  {journal} {\bibinfo  {journal} {Phys. Rev. D}\ }\textbf {\bibinfo {volume}
  {105}},\ \bibinfo {pages} {124049} (\bibinfo {year} {2022})},\ \Eprint
  {http://arxiv.org/abs/2204.13023} {arXiv:2204.13023 [hep-th]} \BibitemShut
  {NoStop}%
\bibitem [{\citenamefont {Nakarachinda}\ \emph {et~al.}(2022)\citenamefont
  {Nakarachinda}, \citenamefont {Promsiri}, \citenamefont {Tannukij},\ and\
  \citenamefont {Wongjun}}]{Nakarachinda:2022gsb}%
  \BibitemOpen
  \bibfield  {author} {\bibinfo {author} {\bibfnamefont {R.}~\bibnamefont
  {Nakarachinda}}, \bibinfo {author} {\bibfnamefont {C.}~\bibnamefont
  {Promsiri}}, \bibinfo {author} {\bibfnamefont {L.}~\bibnamefont {Tannukij}},
  \ and\ \bibinfo {author} {\bibfnamefont {P.}~\bibnamefont {Wongjun}},\
  }\href@noop {} {\  (\bibinfo {year} {2022})},\ \Eprint
  {http://arxiv.org/abs/2211.05989} {arXiv:2211.05989 [gr-qc]} \BibitemShut
  {NoStop}%
\bibitem [{\citenamefont {Saridakis}(2020)}]{Saridakis:2020zol}%
  \BibitemOpen
  \bibfield  {author} {\bibinfo {author} {\bibfnamefont {E.~N.}\ \bibnamefont
  {Saridakis}},\ }\href {\doibase 10.1103/PhysRevD.102.123525} {\bibfield
  {journal} {\bibinfo  {journal} {Phys. Rev. D}\ }\textbf {\bibinfo {volume}
  {102}},\ \bibinfo {pages} {123525} (\bibinfo {year} {2020})},\ \Eprint
  {http://arxiv.org/abs/2005.04115} {arXiv:2005.04115 [gr-qc]} \BibitemShut
  {NoStop}%
\bibitem [{\citenamefont {Dabrowski}\ and\ \citenamefont
  {Salzano}(2020)}]{Dabrowski:2020atl}%
  \BibitemOpen
  \bibfield  {author} {\bibinfo {author} {\bibfnamefont {M.~P.}\ \bibnamefont
  {Dabrowski}}\ and\ \bibinfo {author} {\bibfnamefont {V.}~\bibnamefont
  {Salzano}},\ }\href {\doibase 10.1103/PhysRevD.102.064047} {\bibfield
  {journal} {\bibinfo  {journal} {Phys. Rev. D}\ }\textbf {\bibinfo {volume}
  {102}},\ \bibinfo {pages} {064047} (\bibinfo {year} {2020})},\ \Eprint
  {http://arxiv.org/abs/2009.08306} {arXiv:2009.08306 [astro-ph.CO]}
  \BibitemShut {NoStop}%
\bibitem [{\citenamefont {Nojiri}\ \emph
  {et~al.}(2022{\natexlab{d}})\citenamefont {Nojiri}, \citenamefont
  {Odintsov},\ and\ \citenamefont {Faraoni}}]{Nojiri:2022ljp}%
  \BibitemOpen
  \bibfield  {author} {\bibinfo {author} {\bibfnamefont {S.}~\bibnamefont
  {Nojiri}}, \bibinfo {author} {\bibfnamefont {S.~D.}\ \bibnamefont
  {Odintsov}}, \ and\ \bibinfo {author} {\bibfnamefont {V.}~\bibnamefont
  {Faraoni}},\ }\href@noop {} {\  (\bibinfo {year} {2022}{\natexlab{d}})},\
  \Eprint {http://arxiv.org/abs/2208.10235} {arXiv:2208.10235 [gr-qc]}
  \BibitemShut {NoStop}%
\bibitem [{\citenamefont {Komatsu}(2017)}]{Komatsu:2016vof}%
  \BibitemOpen
  \bibfield  {author} {\bibinfo {author} {\bibfnamefont {N.}~\bibnamefont
  {Komatsu}},\ }\href {\doibase 10.1140/epjc/s10052-017-4800-2} {\bibfield
  {journal} {\bibinfo  {journal} {Eur. Phys. J. C}\ }\textbf {\bibinfo {volume}
  {77}},\ \bibinfo {pages} {229} (\bibinfo {year} {2017})},\ \Eprint
  {http://arxiv.org/abs/1611.04084} {arXiv:1611.04084 [gr-qc]} \BibitemShut
  {NoStop}%
\bibitem [{\citenamefont {Komatsu}\ and\ \citenamefont
  {Kimura}(2016)}]{Komatsu:2015nkb}%
  \BibitemOpen
  \bibfield  {author} {\bibinfo {author} {\bibfnamefont {N.}~\bibnamefont
  {Komatsu}}\ and\ \bibinfo {author} {\bibfnamefont {S.}~\bibnamefont
  {Kimura}},\ }\href {\doibase 10.1103/PhysRevD.93.043530} {\bibfield
  {journal} {\bibinfo  {journal} {Phys. Rev. D}\ }\textbf {\bibinfo {volume}
  {93}},\ \bibinfo {pages} {043530} (\bibinfo {year} {2016})},\ \Eprint
  {http://arxiv.org/abs/1511.04364} {arXiv:1511.04364 [gr-qc]} \BibitemShut
  {NoStop}%
\bibitem [{\citenamefont {Nunes}\ \emph {et~al.}(2016)\citenamefont {Nunes},
  \citenamefont {Barboza}, \citenamefont {Abreu},\ and\ \citenamefont
  {Neto}}]{Nunes:2015xsa}%
  \BibitemOpen
  \bibfield  {author} {\bibinfo {author} {\bibfnamefont {R.~C.}\ \bibnamefont
  {Nunes}}, \bibinfo {author} {\bibfnamefont {E.~M.}\ \bibnamefont {Barboza},
  \bibfnamefont {Jr.}}, \bibinfo {author} {\bibfnamefont {E.~M.~C.}\
  \bibnamefont {Abreu}}, \ and\ \bibinfo {author} {\bibfnamefont {J.~A.}\
  \bibnamefont {Neto}},\ }\href {\doibase 10.1088/1475-7516/2016/08/051}
  {\bibfield  {journal} {\bibinfo  {journal} {JCAP}\ }\textbf {\bibinfo
  {volume} {08}},\ \bibinfo {pages} {051} (\bibinfo {year} {2016})},\ \Eprint
  {http://arxiv.org/abs/1509.05059} {arXiv:1509.05059 [gr-qc]} \BibitemShut
  {NoStop}%
\bibitem [{\citenamefont {Liu}(2022)}]{Liu:2022snq}%
  \BibitemOpen
  \bibfield  {author} {\bibinfo {author} {\bibfnamefont {Y.}~\bibnamefont
  {Liu}},\ }\href {\doibase 10.1140/epjc/s10052-022-10744-9} {\bibfield
  {journal} {\bibinfo  {journal} {Eur. Phys. J. C}\ }\textbf {\bibinfo {volume}
  {82}},\ \bibinfo {pages} {762} (\bibinfo {year} {2022})},\ \Eprint
  {http://arxiv.org/abs/2203.01814} {arXiv:2203.01814 [gr-qc]} \BibitemShut
  {NoStop}%
\bibitem [{\citenamefont {Majhi}(2017)}]{Majhi:2017zao}%
  \BibitemOpen
  \bibfield  {author} {\bibinfo {author} {\bibfnamefont {A.}~\bibnamefont
  {Majhi}},\ }\href {\doibase 10.1016/j.physletb.2017.10.043} {\bibfield
  {journal} {\bibinfo  {journal} {Phys. Lett. B}\ }\textbf {\bibinfo {volume}
  {775}},\ \bibinfo {pages} {32} (\bibinfo {year} {2017})},\ \Eprint
  {http://arxiv.org/abs/1703.09355} {arXiv:1703.09355 [gr-qc]} \BibitemShut
  {NoStop}%
\bibitem [{\citenamefont {Luciano}\ and\ \citenamefont
  {Blasone}(2021)}]{Luciano:2021mto}%
  \BibitemOpen
  \bibfield  {author} {\bibinfo {author} {\bibfnamefont {G.~G.}\ \bibnamefont
  {Luciano}}\ and\ \bibinfo {author} {\bibfnamefont {M.}~\bibnamefont
  {Blasone}},\ }\href {\doibase 10.1103/PhysRevD.104.045004} {\bibfield
  {journal} {\bibinfo  {journal} {Phys. Rev. D}\ }\textbf {\bibinfo {volume}
  {104}},\ \bibinfo {pages} {045004} (\bibinfo {year} {2021})},\ \Eprint
  {http://arxiv.org/abs/2104.00395} {arXiv:2104.00395 [hep-th]} \BibitemShut
  {NoStop}%
\bibitem [{\citenamefont {Di~Gennaro}\ and\ \citenamefont
  {Ong}(2022)}]{DiGennaro:2022ykp}%
  \BibitemOpen
  \bibfield  {author} {\bibinfo {author} {\bibfnamefont {S.}~\bibnamefont
  {Di~Gennaro}}\ and\ \bibinfo {author} {\bibfnamefont {Y.~C.}\ \bibnamefont
  {Ong}},\ }\href {\doibase 10.3390/universe8100541} {\bibfield  {journal}
  {\bibinfo  {journal} {Universe}\ }\textbf {\bibinfo {volume} {8}},\ \bibinfo
  {pages} {541} (\bibinfo {year} {2022})},\ \Eprint
  {http://arxiv.org/abs/2205.09311} {arXiv:2205.09311 [gr-qc]} \BibitemShut
  {NoStop}%
\bibitem [{\citenamefont {Di~Gennaro}\ \emph {et~al.}(2022)\citenamefont
  {Di~Gennaro}, \citenamefont {Xu},\ and\ \citenamefont
  {Ong}}]{DiGennaro:2022grw}%
  \BibitemOpen
  \bibfield  {author} {\bibinfo {author} {\bibfnamefont {S.}~\bibnamefont
  {Di~Gennaro}}, \bibinfo {author} {\bibfnamefont {H.}~\bibnamefont {Xu}}, \
  and\ \bibinfo {author} {\bibfnamefont {Y.~C.}\ \bibnamefont {Ong}},\ }\href
  {\doibase 10.1140/epjc/s10052-022-11040-2} {\bibfield  {journal} {\bibinfo
  {journal} {Eur. Phys. J. C}\ }\textbf {\bibinfo {volume} {82}},\ \bibinfo
  {pages} {1066} (\bibinfo {year} {2022})},\ \Eprint
  {http://arxiv.org/abs/2207.09271} {arXiv:2207.09271 [gr-qc]} \BibitemShut
  {NoStop}%
\bibitem [{\citenamefont {Asghari}\ and\ \citenamefont
  {Sheykhi}(2022)}]{Asghari:2021bqa}%
  \BibitemOpen
  \bibfield  {author} {\bibinfo {author} {\bibfnamefont {M.}~\bibnamefont
  {Asghari}}\ and\ \bibinfo {author} {\bibfnamefont {A.}~\bibnamefont
  {Sheykhi}},\ }\href {\doibase 10.1140/epjc/s10052-022-10262-8} {\bibfield
  {journal} {\bibinfo  {journal} {Eur. Phys. J. C}\ }\textbf {\bibinfo {volume}
  {82}},\ \bibinfo {pages} {388} (\bibinfo {year} {2022})},\ \Eprint
  {http://arxiv.org/abs/2110.00059} {arXiv:2110.00059 [gr-qc]} \BibitemShut
  {NoStop}%
\bibitem [{\citenamefont {Abreu}\ and\ \citenamefont
  {Neto}(2022)}]{Abreu:2022pil}%
  \BibitemOpen
  \bibfield  {author} {\bibinfo {author} {\bibfnamefont {E.~M.~C.}\
  \bibnamefont {Abreu}}\ and\ \bibinfo {author} {\bibfnamefont {J.~A.}\
  \bibnamefont {Neto}},\ }\href {\doibase 10.1016/j.physletb.2022.137565}
  {\bibfield  {journal} {\bibinfo  {journal} {Phys. Lett. B}\ }\textbf
  {\bibinfo {volume} {835}},\ \bibinfo {pages} {137565} (\bibinfo {year}
  {2022})},\ \Eprint {http://arxiv.org/abs/2207.13652} {arXiv:2207.13652
  [gr-qc]} \BibitemShut {NoStop}%
\bibitem [{\citenamefont {Sayahian~Jahromi}\ \emph {et~al.}(2018)\citenamefont
  {Sayahian~Jahromi}, \citenamefont {Moosavi}, \citenamefont {Moradpour},
  \citenamefont {Morais~Gra\c{c}a}, \citenamefont {Lobo}, \citenamefont
  {Salako},\ and\ \citenamefont {Jawad}}]{SayahianJahromi:2018irq}%
  \BibitemOpen
  \bibfield  {author} {\bibinfo {author} {\bibfnamefont {A.}~\bibnamefont
  {Sayahian~Jahromi}}, \bibinfo {author} {\bibfnamefont {S.~A.}\ \bibnamefont
  {Moosavi}}, \bibinfo {author} {\bibfnamefont {H.}~\bibnamefont {Moradpour}},
  \bibinfo {author} {\bibfnamefont {J.~P.}\ \bibnamefont {Morais~Gra\c{c}a}},
  \bibinfo {author} {\bibfnamefont {I.~P.}\ \bibnamefont {Lobo}}, \bibinfo
  {author} {\bibfnamefont {I.~G.}\ \bibnamefont {Salako}}, \ and\ \bibinfo
  {author} {\bibfnamefont {A.}~\bibnamefont {Jawad}},\ }\href {\doibase
  10.1016/j.physletb.2018.02.052} {\bibfield  {journal} {\bibinfo  {journal}
  {Phys. Lett. B}\ }\textbf {\bibinfo {volume} {780}},\ \bibinfo {pages} {21}
  (\bibinfo {year} {2018})},\ \Eprint {http://arxiv.org/abs/1802.07722}
  {arXiv:1802.07722 [gr-qc]} \BibitemShut {NoStop}%
\bibitem [{\citenamefont {Drepanou}\ \emph {et~al.}(2022)\citenamefont
  {Drepanou}, \citenamefont {Lymperis}, \citenamefont {Saridakis},\ and\
  \citenamefont {Yesmakhanova}}]{Drepanou:2021jiv}%
  \BibitemOpen
  \bibfield  {author} {\bibinfo {author} {\bibfnamefont {N.}~\bibnamefont
  {Drepanou}}, \bibinfo {author} {\bibfnamefont {A.}~\bibnamefont {Lymperis}},
  \bibinfo {author} {\bibfnamefont {E.~N.}\ \bibnamefont {Saridakis}}, \ and\
  \bibinfo {author} {\bibfnamefont {K.}~\bibnamefont {Yesmakhanova}},\ }\href
  {\doibase 10.1140/epjc/s10052-022-10415-9} {\bibfield  {journal} {\bibinfo
  {journal} {Eur. Phys. J. C}\ }\textbf {\bibinfo {volume} {82}},\ \bibinfo
  {pages} {449} (\bibinfo {year} {2022})},\ \Eprint
  {http://arxiv.org/abs/2109.09181} {arXiv:2109.09181 [gr-qc]} \BibitemShut
  {NoStop}%
\bibitem [{\citenamefont {Carlip}(2001)}]{Carlip:2001wq}%
  \BibitemOpen
  \bibfield  {author} {\bibinfo {author} {\bibfnamefont {S.}~\bibnamefont
  {Carlip}},\ }\href {\doibase 10.1088/0034-4885/64/8/301} {\bibfield
  {journal} {\bibinfo  {journal} {Rept. Prog. Phys.}\ }\textbf {\bibinfo
  {volume} {64}},\ \bibinfo {pages} {885} (\bibinfo {year} {2001})},\ \Eprint
  {http://arxiv.org/abs/gr-qc/0108040} {arXiv:gr-qc/0108040} \BibitemShut
  {NoStop}%
\bibitem [{\citenamefont {Konishi}\ \emph {et~al.}(1990)\citenamefont
  {Konishi}, \citenamefont {Paffuti},\ and\ \citenamefont
  {Provero}}]{Konishi:1989wk}%
  \BibitemOpen
  \bibfield  {author} {\bibinfo {author} {\bibfnamefont {K.}~\bibnamefont
  {Konishi}}, \bibinfo {author} {\bibfnamefont {G.}~\bibnamefont {Paffuti}}, \
  and\ \bibinfo {author} {\bibfnamefont {P.}~\bibnamefont {Provero}},\ }\href
  {\doibase 10.1016/0370-2693(90)91927-4} {\bibfield  {journal} {\bibinfo
  {journal} {Phys. Lett. B}\ }\textbf {\bibinfo {volume} {234}},\ \bibinfo
  {pages} {276} (\bibinfo {year} {1990})}\BibitemShut {NoStop}%
\bibitem [{\citenamefont {Adler}\ and\ \citenamefont
  {Santiago}(1999)}]{Adler:1999bu}%
  \BibitemOpen
  \bibfield  {author} {\bibinfo {author} {\bibfnamefont {R.~J.}\ \bibnamefont
  {Adler}}\ and\ \bibinfo {author} {\bibfnamefont {D.~I.}\ \bibnamefont
  {Santiago}},\ }\href {\doibase 10.1142/S0217732399001462} {\bibfield
  {journal} {\bibinfo  {journal} {Mod. Phys. Lett. A}\ }\textbf {\bibinfo
  {volume} {14}},\ \bibinfo {pages} {1371} (\bibinfo {year} {1999})},\ \Eprint
  {http://arxiv.org/abs/gr-qc/9904026} {arXiv:gr-qc/9904026} \BibitemShut
  {NoStop}%
\bibitem [{\citenamefont {Rovelli}(1996)}]{Rovelli:1996dv}%
  \BibitemOpen
  \bibfield  {author} {\bibinfo {author} {\bibfnamefont {C.}~\bibnamefont
  {Rovelli}},\ }\href {\doibase 10.1103/PhysRevLett.77.3288} {\bibfield
  {journal} {\bibinfo  {journal} {Phys. Rev. Lett.}\ }\textbf {\bibinfo
  {volume} {77}},\ \bibinfo {pages} {3288} (\bibinfo {year} {1996})},\ \Eprint
  {http://arxiv.org/abs/gr-qc/9603063} {arXiv:gr-qc/9603063} \BibitemShut
  {NoStop}%
\bibitem [{\citenamefont {Meissner}(2004)}]{Meissner:2004ju}%
  \BibitemOpen
  \bibfield  {author} {\bibinfo {author} {\bibfnamefont {K.~A.}\ \bibnamefont
  {Meissner}},\ }\href {\doibase 10.1088/0264-9381/21/22/015} {\bibfield
  {journal} {\bibinfo  {journal} {Class. Quant. Grav.}\ }\textbf {\bibinfo
  {volume} {21}},\ \bibinfo {pages} {5245} (\bibinfo {year} {2004})},\ \Eprint
  {http://arxiv.org/abs/gr-qc/0407052} {arXiv:gr-qc/0407052} \BibitemShut
  {NoStop}%
\bibitem [{\citenamefont {Scardigli}(1999)}]{Scardigli:1999jh}%
  \BibitemOpen
  \bibfield  {author} {\bibinfo {author} {\bibfnamefont {F.}~\bibnamefont
  {Scardigli}},\ }\href {\doibase 10.1016/S0370-2693(99)00167-7} {\bibfield
  {journal} {\bibinfo  {journal} {Phys. Lett. B}\ }\textbf {\bibinfo {volume}
  {452}},\ \bibinfo {pages} {39} (\bibinfo {year} {1999})},\ \Eprint
  {http://arxiv.org/abs/hep-th/9904025} {arXiv:hep-th/9904025} \BibitemShut
  {NoStop}%
\bibitem [{\citenamefont {Hossenfelder}(2013)}]{Hossenfelder:2012jw}%
  \BibitemOpen
  \bibfield  {author} {\bibinfo {author} {\bibfnamefont {S.}~\bibnamefont
  {Hossenfelder}},\ }\href {\doibase 10.12942/lrr-2013-2} {\bibfield  {journal}
  {\bibinfo  {journal} {Living Rev. Rel.}\ }\textbf {\bibinfo {volume} {16}},\
  \bibinfo {pages} {2} (\bibinfo {year} {2013})},\ \Eprint
  {http://arxiv.org/abs/1203.6191} {arXiv:1203.6191 [gr-qc]} \BibitemShut
  {NoStop}%
\bibitem [{\citenamefont {Maggiore}(1993)}]{Maggiore:1993kv}%
  \BibitemOpen
  \bibfield  {author} {\bibinfo {author} {\bibfnamefont {M.}~\bibnamefont
  {Maggiore}},\ }\href {\doibase 10.1016/0370-2693(93)90785-G} {\bibfield
  {journal} {\bibinfo  {journal} {Phys. Lett. B}\ }\textbf {\bibinfo {volume}
  {319}},\ \bibinfo {pages} {83} (\bibinfo {year} {1993})},\ \Eprint
  {http://arxiv.org/abs/hep-th/9309034} {arXiv:hep-th/9309034} \BibitemShut
  {NoStop}%
\bibitem [{\citenamefont {Giddings}(1992)}]{Giddings:1992hh}%
  \BibitemOpen
  \bibfield  {author} {\bibinfo {author} {\bibfnamefont {S.~B.}\ \bibnamefont
  {Giddings}},\ }\href {\doibase 10.1103/PhysRevD.46.1347} {\bibfield
  {journal} {\bibinfo  {journal} {Phys. Rev. D}\ }\textbf {\bibinfo {volume}
  {46}},\ \bibinfo {pages} {1347} (\bibinfo {year} {1992})},\ \Eprint
  {http://arxiv.org/abs/hep-th/9203059} {arXiv:hep-th/9203059} \BibitemShut
  {NoStop}%
\bibitem [{\citenamefont {Page}(1976{\natexlab{a}})}]{Page:1976df}%
  \BibitemOpen
  \bibfield  {author} {\bibinfo {author} {\bibfnamefont {D.~N.}\ \bibnamefont
  {Page}},\ }\href {\doibase 10.1103/PhysRevD.13.198} {\bibfield  {journal}
  {\bibinfo  {journal} {Phys. Rev. D}\ }\textbf {\bibinfo {volume} {13}},\
  \bibinfo {pages} {198} (\bibinfo {year} {1976}{\natexlab{a}})}\BibitemShut
  {NoStop}%
\bibitem [{\citenamefont {Page}(1976{\natexlab{b}})}]{Page:1976ki}%
  \BibitemOpen
  \bibfield  {author} {\bibinfo {author} {\bibfnamefont {D.~N.}\ \bibnamefont
  {Page}},\ }\href {\doibase 10.1103/PhysRevD.14.3260} {\bibfield  {journal}
  {\bibinfo  {journal} {Phys. Rev. D}\ }\textbf {\bibinfo {volume} {14}},\
  \bibinfo {pages} {3260} (\bibinfo {year} {1976}{\natexlab{b}})}\BibitemShut
  {NoStop}%
\bibitem [{\citenamefont {Page}(1977)}]{Page:1977um}%
  \BibitemOpen
  \bibfield  {author} {\bibinfo {author} {\bibfnamefont {D.~N.}\ \bibnamefont
  {Page}},\ }\href {\doibase 10.1103/PhysRevD.16.2402} {\bibfield  {journal}
  {\bibinfo  {journal} {Phys. Rev. D}\ }\textbf {\bibinfo {volume} {16}},\
  \bibinfo {pages} {2402} (\bibinfo {year} {1977})}\BibitemShut {NoStop}%
\bibitem [{\citenamefont {Schuster}(2018)}]{Schuster:2018lmz}%
  \BibitemOpen
  \bibfield  {author} {\bibinfo {author} {\bibfnamefont {S.}~\bibnamefont
  {Schuster}},\ }\emph {\bibinfo {title} {{Black Hole Evaporation: Sparsity in
  Analogue and General Relativistic Space-Times}}},\ \href@noop {} {Ph.D.
  thesis},\ \bibinfo  {school} {Victoria U., Wellington} (\bibinfo {year}
  {2018}),\ \Eprint {http://arxiv.org/abs/1901.05648} {arXiv:1901.05648
  [gr-qc]} \BibitemShut {NoStop}%
\bibitem [{\citenamefont {Gray}\ \emph {et~al.}(2016)\citenamefont {Gray},
  \citenamefont {Schuster}, \citenamefont {Van-Brunt},\ and\ \citenamefont
  {Visser}}]{Gray:2015pma}%
  \BibitemOpen
  \bibfield  {author} {\bibinfo {author} {\bibfnamefont {F.}~\bibnamefont
  {Gray}}, \bibinfo {author} {\bibfnamefont {S.}~\bibnamefont {Schuster}},
  \bibinfo {author} {\bibfnamefont {A.}~\bibnamefont {Van-Brunt}}, \ and\
  \bibinfo {author} {\bibfnamefont {M.}~\bibnamefont {Visser}},\ }\href
  {\doibase 10.1088/0264-9381/33/11/115003} {\bibfield  {journal} {\bibinfo
  {journal} {Class. Quant. Grav.}\ }\textbf {\bibinfo {volume} {33}},\ \bibinfo
  {pages} {115003} (\bibinfo {year} {2016})},\ \Eprint
  {http://arxiv.org/abs/1506.03975} {arXiv:1506.03975 [gr-qc]} \BibitemShut
  {NoStop}%
\bibitem [{\citenamefont {Schuster}(2021)}]{Schuster:2019xvp}%
  \BibitemOpen
  \bibfield  {author} {\bibinfo {author} {\bibfnamefont {S.}~\bibnamefont
  {Schuster}},\ }\href {\doibase 10.1088/1361-6382/abd144} {\bibfield
  {journal} {\bibinfo  {journal} {Class. Quant. Grav.}\ }\textbf {\bibinfo
  {volume} {38}},\ \bibinfo {pages} {047002} (\bibinfo {year} {2021})},\
  \Eprint {http://arxiv.org/abs/1910.07256} {arXiv:1910.07256 [gr-qc]}
  \BibitemShut {NoStop}%
\bibitem [{\citenamefont {Paul}\ and\ \citenamefont
  {Majhi}(2017)}]{Paul:2016xvb}%
  \BibitemOpen
  \bibfield  {author} {\bibinfo {author} {\bibfnamefont {A.}~\bibnamefont
  {Paul}}\ and\ \bibinfo {author} {\bibfnamefont {B.~R.}\ \bibnamefont
  {Majhi}},\ }\href {\doibase 10.1142/S0217751X17500889} {\bibfield  {journal}
  {\bibinfo  {journal} {Int. J. Mod. Phys. A}\ }\textbf {\bibinfo {volume}
  {32}},\ \bibinfo {pages} {1750088} (\bibinfo {year} {2017})},\ \Eprint
  {http://arxiv.org/abs/1601.07310} {arXiv:1601.07310 [gr-qc]} \BibitemShut
  {NoStop}%
\bibitem [{\citenamefont {Alonso-Serrano}\ \emph
  {et~al.}(2018{\natexlab{a}})\citenamefont {Alonso-Serrano}, \citenamefont
  {D\c{a}browski},\ and\ \citenamefont {Gohar}}]{PhysRevD.97.044029}%
  \BibitemOpen
  \bibfield  {author} {\bibinfo {author} {\bibfnamefont {A.}~\bibnamefont
  {Alonso-Serrano}}, \bibinfo {author} {\bibfnamefont {M.~P.}\ \bibnamefont
  {D\c{a}browski}}, \ and\ \bibinfo {author} {\bibfnamefont {H.}~\bibnamefont
  {Gohar}},\ }\href {\doibase 10.1103/PhysRevD.97.044029} {\bibfield  {journal}
  {\bibinfo  {journal} {Phys. Rev. D}\ }\textbf {\bibinfo {volume} {97}},\
  \bibinfo {pages} {044029} (\bibinfo {year} {2018}{\natexlab{a}})}\BibitemShut
  {NoStop}%
\bibitem [{\citenamefont {Alonso-Serrano}\ \emph
  {et~al.}(2018{\natexlab{b}})\citenamefont {Alonso-Serrano}, \citenamefont
  {Dabrowski},\ and\ \citenamefont {Gohar}}]{Alonso-Serrano:2018mfo}%
  \BibitemOpen
  \bibfield  {author} {\bibinfo {author} {\bibfnamefont {A.}~\bibnamefont
  {Alonso-Serrano}}, \bibinfo {author} {\bibfnamefont {M.~P.}\ \bibnamefont
  {Dabrowski}}, \ and\ \bibinfo {author} {\bibfnamefont {H.}~\bibnamefont
  {Gohar}},\ }\href {\doibase 10.1142/S0218271818470284} {\bibfield  {journal}
  {\bibinfo  {journal} {Int. J. Mod. Phys. D}\ }\textbf {\bibinfo {volume}
  {27}},\ \bibinfo {pages} {1847028} (\bibinfo {year} {2018}{\natexlab{b}})},\
  \Eprint {http://arxiv.org/abs/1805.07690} {arXiv:1805.07690 [gr-qc]}
  \BibitemShut {NoStop}%
\bibitem [{\citenamefont {Ong}(2018{\natexlab{a}})}]{Ong:2018syk}%
  \BibitemOpen
  \bibfield  {author} {\bibinfo {author} {\bibfnamefont {Y.~C.}\ \bibnamefont
  {Ong}},\ }\href {\doibase 10.1007/JHEP10(2018)195} {\bibfield  {journal}
  {\bibinfo  {journal} {JHEP}\ }\textbf {\bibinfo {volume} {10}},\ \bibinfo
  {pages} {195} (\bibinfo {year} {2018}{\natexlab{a}})},\ \Eprint
  {http://arxiv.org/abs/1806.03691} {arXiv:1806.03691 [gr-qc]} \BibitemShut
  {NoStop}%
\bibitem [{\citenamefont {Feng}\ \emph {et~al.}(2020)\citenamefont {Feng},
  \citenamefont {Zhou}, \citenamefont {Zhou},\ and\ \citenamefont
  {Feng}}]{Feng:2018jqf}%
  \BibitemOpen
  \bibfield  {author} {\bibinfo {author} {\bibfnamefont {Z.-W.}\ \bibnamefont
  {Feng}}, \bibinfo {author} {\bibfnamefont {X.}~\bibnamefont {Zhou}}, \bibinfo
  {author} {\bibfnamefont {S.-Q.}\ \bibnamefont {Zhou}}, \ and\ \bibinfo
  {author} {\bibfnamefont {D.-D.}\ \bibnamefont {Feng}},\ }\href {\doibase
  10.1016/j.aop.2020.168144} {\bibfield  {journal} {\bibinfo  {journal} {Annals
  Phys.}\ }\textbf {\bibinfo {volume} {416}},\ \bibinfo {pages} {168144}
  (\bibinfo {year} {2020})},\ \Eprint {http://arxiv.org/abs/1808.09958}
  {arXiv:1808.09958 [gr-qc]} \BibitemShut {NoStop}%
\bibitem [{\citenamefont {Amati}\ \emph {et~al.}(1989)\citenamefont {Amati},
  \citenamefont {Ciafaloni},\ and\ \citenamefont {Veneziano}}]{Amati:1988tn}%
  \BibitemOpen
  \bibfield  {author} {\bibinfo {author} {\bibfnamefont {D.}~\bibnamefont
  {Amati}}, \bibinfo {author} {\bibfnamefont {M.}~\bibnamefont {Ciafaloni}}, \
  and\ \bibinfo {author} {\bibfnamefont {G.}~\bibnamefont {Veneziano}},\ }\href
  {\doibase 10.1016/0370-2693(89)91366-X} {\bibfield  {journal} {\bibinfo
  {journal} {Phys. Lett. B}\ }\textbf {\bibinfo {volume} {216}},\ \bibinfo
  {pages} {41} (\bibinfo {year} {1989})}\BibitemShut {NoStop}%
\bibitem [{\citenamefont {Kempf}\ \emph {et~al.}(1995)\citenamefont {Kempf},
  \citenamefont {Mangano},\ and\ \citenamefont {Mann}}]{Kempf:1994su}%
  \BibitemOpen
  \bibfield  {author} {\bibinfo {author} {\bibfnamefont {A.}~\bibnamefont
  {Kempf}}, \bibinfo {author} {\bibfnamefont {G.}~\bibnamefont {Mangano}}, \
  and\ \bibinfo {author} {\bibfnamefont {R.~B.}\ \bibnamefont {Mann}},\ }\href
  {\doibase 10.1103/PhysRevD.52.1108} {\bibfield  {journal} {\bibinfo
  {journal} {Phys. Rev. D}\ }\textbf {\bibinfo {volume} {52}},\ \bibinfo
  {pages} {1108} (\bibinfo {year} {1995})},\ \Eprint
  {http://arxiv.org/abs/hep-th/9412167} {arXiv:hep-th/9412167} \BibitemShut
  {NoStop}%
\bibitem [{\citenamefont {Schiller}(2005)}]{schiller2005general}%
  \BibitemOpen
  \bibfield  {author} {\bibinfo {author} {\bibfnamefont {C.}~\bibnamefont
  {Schiller}},\ }\href@noop {} {\bibfield  {journal} {\bibinfo  {journal}
  {International Journal of Theoretical Physics}\ }\textbf {\bibinfo {volume}
  {44}},\ \bibinfo {pages} {1629} (\bibinfo {year} {2005})}\BibitemShut
  {NoStop}%
\bibitem [{\citenamefont {Barrow}\ and\ \citenamefont
  {Gibbons}(2014)}]{BG2014}%
  \BibitemOpen
  \bibfield  {author} {\bibinfo {author} {\bibfnamefont {J.~D.}\ \bibnamefont
  {Barrow}}\ and\ \bibinfo {author} {\bibfnamefont {G.~W.}\ \bibnamefont
  {Gibbons}},\ }\href {\doibase 10.1093/mnras/stu2378} {\bibfield  {journal}
  {\bibinfo  {journal} {Monthly Notices of the Royal Astronomical Society}\
  }\textbf {\bibinfo {volume} {446}},\ \bibinfo {pages} {3874} (\bibinfo {year}
  {2014})},\ \Eprint
  {http://arxiv.org/abs/https://academic.oup.com/mnras/article-pdf/446/4/3874/9388063/stu2378.pdf}
  {https://academic.oup.com/mnras/article-pdf/446/4/3874/9388063/stu2378.pdf}
  \BibitemShut {NoStop}%
\bibitem [{\citenamefont {Dabrowski}\ and\ \citenamefont
  {Gohar}(2015)}]{dabrowski2015abolishing}%
  \BibitemOpen
  \bibfield  {author} {\bibinfo {author} {\bibfnamefont {M.~P.}\ \bibnamefont
  {Dabrowski}}\ and\ \bibinfo {author} {\bibfnamefont {H.}~\bibnamefont
  {Gohar}},\ }\href@noop {} {\bibfield  {journal} {\bibinfo  {journal} {Physics
  Letters B}\ }\textbf {\bibinfo {volume} {748}},\ \bibinfo {pages} {428}
  (\bibinfo {year} {2015})}\BibitemShut {NoStop}%
\bibitem [{\citenamefont {Ong}(2018{\natexlab{b}})}]{Ong:2018xna}%
  \BibitemOpen
  \bibfield  {author} {\bibinfo {author} {\bibfnamefont {Y.~C.}\ \bibnamefont
  {Ong}},\ }\href {\doibase 10.1016/j.physletb.2018.08.065} {\bibfield
  {journal} {\bibinfo  {journal} {Phys. Lett. B}\ }\textbf {\bibinfo {volume}
  {785}},\ \bibinfo {pages} {217} (\bibinfo {year} {2018}{\natexlab{b}})},\
  \Eprint {http://arxiv.org/abs/1809.00442} {arXiv:1809.00442 [gr-qc]}
  \BibitemShut {NoStop}%
\bibitem [{\citenamefont {Gao}\ and\ \citenamefont {Zhan}(2016)}]{Gao:2016fmk}%
  \BibitemOpen
  \bibfield  {author} {\bibinfo {author} {\bibfnamefont {D.}~\bibnamefont
  {Gao}}\ and\ \bibinfo {author} {\bibfnamefont {M.}~\bibnamefont {Zhan}},\
  }\href {\doibase 10.1103/PhysRevA.94.013607} {\bibfield  {journal} {\bibinfo
  {journal} {Phys. Rev. A}\ }\textbf {\bibinfo {volume} {94}},\ \bibinfo
  {pages} {013607} (\bibinfo {year} {2016})},\ \Eprint
  {http://arxiv.org/abs/1607.04353} {arXiv:1607.04353 [gr-qc]} \BibitemShut
  {NoStop}%
\bibitem [{\citenamefont {Feng}\ \emph {et~al.}(2017)\citenamefont {Feng},
  \citenamefont {Yang}, \citenamefont {Li},\ and\ \citenamefont
  {Zu}}]{Feng:2016tyt}%
  \BibitemOpen
  \bibfield  {author} {\bibinfo {author} {\bibfnamefont {Z.-W.}\ \bibnamefont
  {Feng}}, \bibinfo {author} {\bibfnamefont {S.-Z.}\ \bibnamefont {Yang}},
  \bibinfo {author} {\bibfnamefont {H.-L.}\ \bibnamefont {Li}}, \ and\ \bibinfo
  {author} {\bibfnamefont {X.-T.}\ \bibnamefont {Zu}},\ }\href {\doibase
  10.1016/j.physletb.2017.02.043} {\bibfield  {journal} {\bibinfo  {journal}
  {Phys. Lett. B}\ }\textbf {\bibinfo {volume} {768}},\ \bibinfo {pages} {81}
  (\bibinfo {year} {2017})},\ \Eprint {http://arxiv.org/abs/1610.08549}
  {arXiv:1610.08549 [hep-ph]} \BibitemShut {NoStop}%
\bibitem [{\citenamefont {Bosso}\ \emph {et~al.}(2018)\citenamefont {Bosso},
  \citenamefont {Das},\ and\ \citenamefont {Mann}}]{Bosso:2018ckz}%
  \BibitemOpen
  \bibfield  {author} {\bibinfo {author} {\bibfnamefont {P.}~\bibnamefont
  {Bosso}}, \bibinfo {author} {\bibfnamefont {S.}~\bibnamefont {Das}}, \ and\
  \bibinfo {author} {\bibfnamefont {R.~B.}\ \bibnamefont {Mann}},\ }\href
  {\doibase 10.1016/j.physletb.2018.08.061} {\bibfield  {journal} {\bibinfo
  {journal} {Phys. Lett. B}\ }\textbf {\bibinfo {volume} {785}},\ \bibinfo
  {pages} {498} (\bibinfo {year} {2018})},\ \Eprint
  {http://arxiv.org/abs/1804.03620} {arXiv:1804.03620 [gr-qc]} \BibitemShut
  {NoStop}%
\bibitem [{\citenamefont {Gao}\ \emph {et~al.}(2017)\citenamefont {Gao},
  \citenamefont {Wang},\ and\ \citenamefont {Zhan}}]{Gao:2017zch}%
  \BibitemOpen
  \bibfield  {author} {\bibinfo {author} {\bibfnamefont {D.}~\bibnamefont
  {Gao}}, \bibinfo {author} {\bibfnamefont {J.}~\bibnamefont {Wang}}, \ and\
  \bibinfo {author} {\bibfnamefont {M.}~\bibnamefont {Zhan}},\ }\href {\doibase
  10.1103/PhysRevA.95.042106} {\bibfield  {journal} {\bibinfo  {journal} {Phys.
  Rev. A}\ }\textbf {\bibinfo {volume} {95}},\ \bibinfo {pages} {042106}
  (\bibinfo {year} {2017})},\ \Eprint {http://arxiv.org/abs/1704.02037}
  {arXiv:1704.02037 [gr-qc]} \BibitemShut {NoStop}%
\bibitem [{\citenamefont {Giardino}\ and\ \citenamefont
  {Salzano}(2021)}]{Giardino:2020myz}%
  \BibitemOpen
  \bibfield  {author} {\bibinfo {author} {\bibfnamefont {S.}~\bibnamefont
  {Giardino}}\ and\ \bibinfo {author} {\bibfnamefont {V.}~\bibnamefont
  {Salzano}},\ }\href {\doibase 10.1140/epjc/s10052-021-08914-2} {\bibfield
  {journal} {\bibinfo  {journal} {Eur. Phys. J. C}\ }\textbf {\bibinfo {volume}
  {81}},\ \bibinfo {pages} {110} (\bibinfo {year} {2021})},\ \Eprint
  {http://arxiv.org/abs/2006.01580} {arXiv:2006.01580 [gr-qc]} \BibitemShut
  {NoStop}%
\bibitem [{\citenamefont {Jizba}\ \emph {et~al.}(2010)\citenamefont {Jizba},
  \citenamefont {Kleinert},\ and\ \citenamefont {Scardigli}}]{Jizba:2009qf}%
  \BibitemOpen
  \bibfield  {author} {\bibinfo {author} {\bibfnamefont {P.}~\bibnamefont
  {Jizba}}, \bibinfo {author} {\bibfnamefont {H.}~\bibnamefont {Kleinert}}, \
  and\ \bibinfo {author} {\bibfnamefont {F.}~\bibnamefont {Scardigli}},\ }\href
  {\doibase 10.1103/PhysRevD.81.084030} {\bibfield  {journal} {\bibinfo
  {journal} {Phys. Rev. D}\ }\textbf {\bibinfo {volume} {81}},\ \bibinfo
  {pages} {084030} (\bibinfo {year} {2010})},\ \Eprint
  {http://arxiv.org/abs/0912.2253} {arXiv:0912.2253 [hep-th]} \BibitemShut
  {NoStop}%
\bibitem [{\citenamefont {Masi}(2005)}]{MASI2005217}%
  \BibitemOpen
  \bibfield  {author} {\bibinfo {author} {\bibfnamefont {M.}~\bibnamefont
  {Masi}},\ }\href {\doibase https://doi.org/10.1016/j.physleta.2005.01.094}
  {\bibfield  {journal} {\bibinfo  {journal} {Physics Letters A}\ }\textbf
  {\bibinfo {volume} {338}},\ \bibinfo {pages} {217} (\bibinfo {year}
  {2005})}\BibitemShut {NoStop}%
\bibitem [{\citenamefont {Abreu}\ and\ \citenamefont
  {Neto}(2020)}]{Abreu:2020wbz}%
  \BibitemOpen
  \bibfield  {author} {\bibinfo {author} {\bibfnamefont {E.~M.~C.}\
  \bibnamefont {Abreu}}\ and\ \bibinfo {author} {\bibfnamefont {J.~A.}\
  \bibnamefont {Neto}},\ }\href {\doibase 10.1016/j.physletb.2020.135805}
  {\bibfield  {journal} {\bibinfo  {journal} {Phys. Lett. B}\ }\textbf
  {\bibinfo {volume} {810}},\ \bibinfo {pages} {135805} (\bibinfo {year}
  {2020})},\ \Eprint {http://arxiv.org/abs/2009.10133} {arXiv:2009.10133
  [gr-qc]} \BibitemShut {NoStop}%
\end{thebibliography}%
\bibliographystyle{apsrev4-1}

\end{document}